\documentclass[preprint,11pt]{elsarticle}

\usepackage[T1]{fontenc}
\usepackage{amsmath}
\usepackage{amssymb}
\usepackage{epsfig}
\usepackage{xfrac}
\usepackage{graphicx}
\usepackage{bm}
\usepackage{slashed}
\usepackage{cleveref}
\crefname{figure}{Fig.}{Figs.}
\crefname{equation}{Eq.}{Eqs.}
\usepackage{natbib}
\usepackage{booktabs}
\usepackage{caption}
\usepackage{siunitx}

\numberwithin{equation}{section}
\numberwithin{table}{section}
\numberwithin{figure}{section}
\biboptions{square,numbers,comma,sort&compress}

\newcommand{\be}{\begin{equation}}
\newcommand{\ee}{\end{equation}}
\newcommand{\bea}{\begin{eqnarray}}
\newcommand{\eea}{\end{eqnarray}}
\newcommand{\nn}{\nonumber}
\newcommand{\degree}{\mbox{$^\circ$}}

\def\g2{\gamma\gamma}
\def\ubar{\bar{u}}
\def\gegm{G_E/G_M} 
\def\text{\rm}
\newcommand{\eps}{\varepsilon}

\def\etal{{\em et al.}}

\renewcommand{\Re}{\operatorname{Re}}
\renewcommand{\Im}{\operatorname{Im}}

\begin{document}
  
\journal{Progress in Particle and Nuclear Physics}

\begin{frontmatter}
 
\title{Two-photon exchange in elastic electron-proton scattering}

\author[L1]{A.~Afanasev~\fnref{myfootnote1}}
\author[L2,L3]{P.~G.~Blunden~\fnref{myfootnote2}}
\author[L4]{D.~Hasell~\fnref{myfootnote4}}
\author[L5]{B.~A.~Raue~\fnref{myfootnote5}}

\address[L1]{George Washington University, Washington, D.C., U.S.A.}
\address[L2]{University of Manitoba, Winnipeg, MB, Canada}
\address[L3]{Jefferson Lab, Newport News, VA, U.S.A.}
\address[L4]{Massachusetts Institute of Technology, Cambridge, MA, U.S.A.}
\address[L5]{Florida International University, Miami, FL, U.S.A.}

\fntext[myfootnote1]{afanas@gwu.edu}
\fntext[myfootnote2]{blunden@physics.umanitoba.ca}
\fntext[myfootnote4]{hasell@mit.edu}
\fntext[myfootnote5]{baraue@fiu.edu}

\begin{abstract}
We review recent theoretical and experimental progress on the role of two-photon
exchange (TPE) in electron-proton scattering at low to moderate momentum
transfers. We make a detailed comparison and analysis of the results of
competing experiments on the ratio of $e^+p$ to $e^-p$ elastic scattering cross
sections, and of the theoretical calculations describing them. A summary of the
current experimental situation is provided, along with an outlook for future
experiments.
\end{abstract}

\begin{keyword}
Two-photon exchange \sep Radiative corrections \sep Form factors\\
{\em Preprint:} JLAB-THY-17-2404
\end{keyword}

\end{frontmatter}

\newpage
\thispagestyle{empty}
\tableofcontents

\newpage
\section{Introduction}\label{sec:intro}

Electron scattering has been a primary experimental tool in the study of hadron
physics for many decades. The electromagnetic interaction is well understood,
and the pointlike nature of electrons makes it an ideal tool to probe the
internal structure of hadrons. Furthermore, the relatively small value of the
electromagnetic coupling, $\alpha\sim 1/137$, makes the electromagnetic
interaction amenable to a perturbative treatment in the context of quantum field
theory.

Much of our information on the structure of the proton comes from unpolarized
measurements of the inclusive electron-proton scattering cross section. These
measurements determine the proton electric ($G_E^p$) and magnetic ($G_M^p$) form
factors, which are fundamental observables characterizing the internal structure
of the proton. Specifically, the quantities $(G_E^p)^2$ and $(G_M^p)^2$ can be
extracted from the angular dependence of the unpolarized electron scattering
cross section.

More recently, polarized beams, polarized targets, and measurements of the
recoil polarization of the target proton have been used to provide additional
information on the spin structure of the proton, and to improve our knowledge of
proton form factors. Polarization measurements have proven to be a crucial
ingredient in studies of proton form factors over the past two decades. These
experiments access the ratio $G_E^p/G_M^p$ directly from the ratio of transverse
to longitudinal nuclear polarization measurements.

In what has become known colloquially as ``the proton form factor puzzle'', a
comparison of the form factor ratio extracted from both types of experiments
revealed a significant discrepancy in kinematic regions where both techniques
provide precise measurements. Because these, and essentially all other electron
scattering measurements, are analyzed in the framework of the one-photon
exchange (OPE) or Born approximation, this discrepancy led to a reexamination of
the possible role played by radiative corrections to the electron scattering
cross sections. For electron scattering, radiative corrections must be applied
to measured cross sections in order to extract an equivalent OPE form.  Although
these radiative corrections are large, they are generally model-independent and
well understood. In particular, the standard radiative corrections are
independent of hadronic structure.

Attempts to reconcile the unpolarized and polarized measurements have mostly
focussed on improved treatments of these radiative corrections. Of particular
interest, and the subject of this review, are considerations of two-photon
exchange (TPE) effects beyond the minimal model-independent terms incorporated
into the standard radiative corrections.  The challenge in calculating these TPE
contributions is that they are not independent of hadronic structure. The
challenge in measuring them directly is that they are most prominent at high
momentum transfer and backward scattering angles, where the cross section is
suppressed. Early measurements and calculations suggested that TPE effects are a
few percent correction to cross sections, consistent with the expectation that
they are of order ${\cal O}(\alpha)$ compared to the OPE approximation. However,
there is now convincing evidence that these corrections can nevertheless be
extremely important in specific circumstances.

Over the past 15 years there has been a significant investment, on both the
theoretical and experimental fronts, to studying TPE in electromagnetic
processes. Many of these efforts have been the subject of previous reviews, such
as the 2007 review by Carlson and Vanderhaeghen~\cite{Carlson:2007sp}, and 2011
review by Arrington, Blunden and Melnitchouk~\cite{Arrington:2011dn}. Since the
2011 review~\cite{Arrington:2011dn} there has been significant progress in
theoretical calculations, which we highlight here. In addition, results have
recently been reported from the VEPP-3, CLAS, and OLYMPUS experiments, which
were designed to directly measure TPE effects from the ratio of $e^+p$ to $e^-p$
elastic scattering cross sections.

Interest in TPE effects has been furthered by the so-called ``proton radius
problem''~\cite{Pohl:2013yb}. Briefly, the proton radius extracted from electron
scattering and atomic hydrogen spectroscopy measurements disagrees by several
standard deviations from the proton radius extracted by spectroscopy on muonic
atoms. Two-photon exchange is one contributor to the energy shift in atomic
systems. This is described in a recent review by Carlson~\cite{Carlson:2015jba},
and we don't address it further in this review.

The outline of this review is as follows. Section~\ref{sec:theory} provides a
theoretical overview. This includes the relevant electron scattering formalism
in sections~\ref{ssec:kinematics} and \ref{ssec:BornFF}. Section~\ref{ssec:TPE}
describes the formalism and calculations of TPE corrections in unpolarized
electron scattering, summarizing the older work but focussing on recent
improvements in the past five years. Two-photon exchange for spin-polarization
effects are described in \cref{ssec:spin}.

Section~\ref{sec:expt} focusses on experimental measurements. In particular, the
recent VEPP-3, CLAS, and OLYMPUS experiments, which look for direct evidence of
TPE effects by measuring the ratio of $e^+p$ to $e^-p$ elastic cross sections,
are each described in some detail. A comparison and analysis of the results of
these experiments is made in \cref{ssec:compEM}. Conclusions and the outlook for
both theory and experiment are given in \cref{sec:conclusion}.

\section{Theoretical overview}\label{sec:theory}

\subsection{Kinematics and definitions}\label{ssec:kinematics}

In this section we define the general kinematics of elastic electron--nucleon
scattering, and present amplitudes and cross sections in the OPE or Born
approximation.

\begin{figure}[t]
\centering
\includegraphics[width=0.7\textwidth]{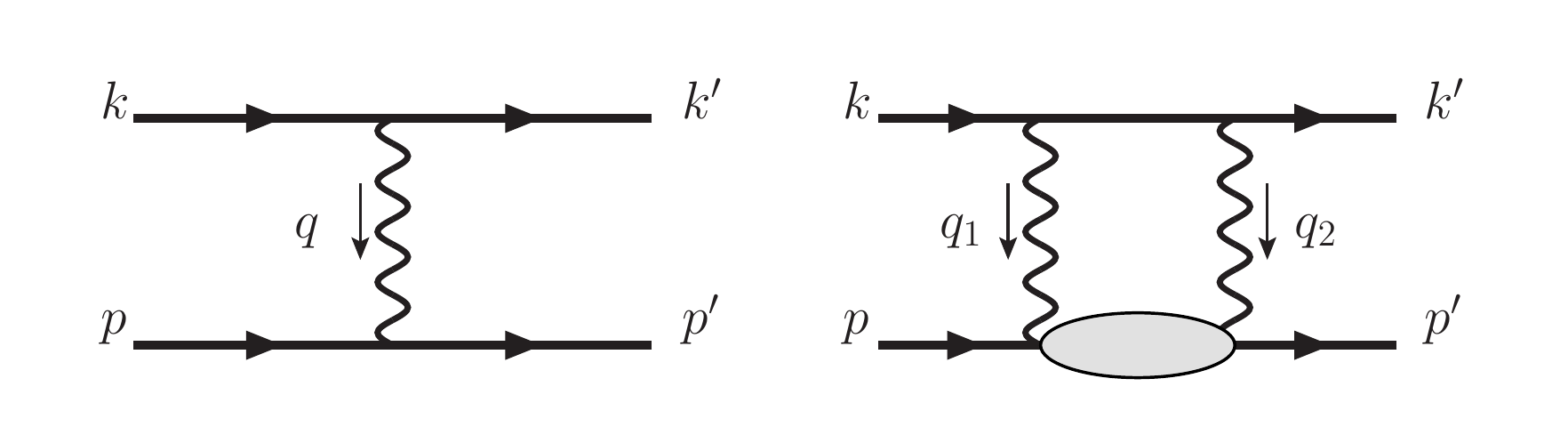}
\caption{Contributions to elastic electron--nucleon scattering from (a) one-photon
	exchange (OPE), and (b) two-photon exchange (TPE) amplitudes, with
	particle momenta as indicated. For TPE we have $q_1+q_2=q$.
	Only the $s$-channel ``box'' diagram is drawn. The ``crossed-box'' contribution,
	which can be obtained by applying crossing symmetry $s\to u$, is implied.
	}
\label{fig:OPETPE}
\end{figure}

For the elastic scattering process $e N \to e N$ (see Fig.~\ref{fig:OPETPE}),
the four-momenta of the initial and final electrons (mass $m_e$) are labelled by
$k$ and $k'$, with corresponding energies $E$ and $E'$, and of the initial and
final nucleons (mass $M$) by $p$ and $p'$, respectively. The four-momentum
transfer from the electron to the nucleon is given by $q = p'-p = k-k'$, with
$Q^2 \equiv -q^2 > 0$. One can express the elastic cross section in terms of any
two of the Mandelstam variables $s$ (total electron--nucleon invariant mass
squared), $t$, and $u$, where
\be
s=(k+p)^2=(k'+p')^2\,,\quad
t=(k-k')^2=q^2\,,\quad
u=(p-k')^2=(p'-k)^2\,,
\label{eq:Mandelstam}
\ee
with the constraint $s + t + u = 2 M^2 + 2 m_e^2$. The electron mass $m_e$ can
generally be ignored at the kinematics of interest here. In particular, there
are no mass singularities in the limit $m_e \to 0$ in either the OPE amplitude
or the {\em total} TPE amplitude.

Conventionally, the elastic scattering cross section is defined in terms of $Q^2$
and the electron scattering angle, $\theta_e$, or equivalently, any two of the
dimensionless quantities
\be
\tau = \frac{Q^2}{4M^2}\, ,\qquad
\eps = \frac{\nu^2 - \tau (1+\tau)}{\nu^2 + \tau (1+\tau)}
=\frac{2 \left(M^4-s u\right)}{s^2+u^2-2 M^4}\,,\qquad \nu = \frac{k\cdot p}{M^2} -\tau\,.
\ee
The inverse relationships are also useful:
\be
\nu = \frac{s-u}{4 M^2} = \sqrt\frac{\tau (1+\tau)(1+\eps)}{1-\eps}\, .
\ee
In the target rest frame we have the relations
\be
\eps = \left(1 + 2 (1+\tau) \tan^2{\frac{\theta_e}{2}}\right)^{-1}\, ,
\qquad \tau=\frac{E-E'}{2M}\, ,\qquad \nu=\frac{E+E'}{2M}\, ,
\ee
where $\eps$ is identified with the relative flux of longitudinal virtual
photons, and $E$ $(E')$ is the energy of the incident (scattered) electron.

In the Born (OPE) approximation the electron--nucleon scattering invariant
amplitude can be written as
\be
{\cal M}_\gamma = -\frac{e^2}{q^2}\, j_{\gamma \mu}\, J_\gamma^\mu\, ,
\label{eq:Mg}
\ee
where $e$ is the electric charge, and the matrix elements of the electromagnetic
leptonic and hadronic currents are given in terms of the lepton ($u_e$) and
nucleon ($u_N$) spinors by
\be
j_{\gamma \mu} = \ubar_e(k')\, \gamma_\mu\, u_e(k)\,, \qquad
J_\gamma^\mu = \ubar_N(p')\, \Gamma_\gamma^\mu(q)\, u_N(p)\,.
\ee
The electromagnetic hadron current operator $\Gamma_\gamma^\mu$ is parametrized
by the Dirac ($F_1$) and Pauli ($F_2$) form factors as
\be
\Gamma_\gamma^\mu(q) = F_1(Q^2)\, \gamma^\mu\ 
+\ F_2(Q^2)\, \frac{i \sigma^{\mu\nu} q_\nu}{2 M}\, .
\label{eq:Jg}
\ee
In terms of the amplitude ${\cal M}_\gamma$, the differential Born cross section
is given by
\be
\frac{d\sigma}{d\Omega }
= \left( \frac{\alpha}{4 M Q^2 } \frac{E'}{E} \right)^2
    \left| {\cal M}_\gamma \right|^2 
 = \frac{\sigma_{\rm Mott}}{\eps (1+\tau)}\, \sigma_R\, ,\qquad
 \sigma_{\rm Mott}
= \frac{\alpha^2 E' \cos^2(\theta_e/2)}{4 E^3 \sin^4(\theta_e/2) }\, ,
\label{eq:sigma0}
\ee
where $\alpha = e^2/4\pi$ is the electromagnetic fine structure constant, and
the $\sigma_{\rm Mott}$ is the cross section for scattering from a point particle.
In our convention, the reduced Born cross section $\sigma_R$ is given by
\be
\sigma_R = \eps\, G_E^2(Q^2)\ + \tau\, G_M^2(Q^2)\, ,
\label{eq:sigmaR}
\ee
where the Sachs electric and magnetic form factors $G_{E,M}(Q^2)$ are defined in
terms of the Dirac and Pauli form factors as
\be
G_E(Q^2) = F_1(Q^2) - \tau F_2(Q^2)\, ,\qquad
G_M(Q^2) = F_1(Q^2) + F_2(Q^2)\, .
\label{eq:GEMdef}
\ee
The form factors are normalized such that $G_E^{p\,(n)}(0)=1\,(0)$ and
$G_M^{p\,(n)}(0)=\mu_{p\,(n)}$ for the proton (neutron), where
$\mu_{p\,(n)} = 2.793\,(-1.913)$ is the proton (neutron) magnetic moment.

\subsection{Experimental measurements of proton form factors}
\label{ssec:BornFF}

For many decades the standard experimental technique for extracting proton form
factors has been the Rosenbluth, or longitudinal-transverse (LT), separation
method~\cite{Rosenbluth:1950yq}.  The method requires applying a number of
standard radiative corrections~\cite{Tsai:1961zz, Mo:1968cg, Maximon:2000hm} to
the measured cross section to extract a reduced cross section $\sigma_R$
equivalent to the OPE form given in Eq.~(\ref{eq:sigmaR}). The standard radiative
corrections are large, but generally speaking they are independent of hadron
structure.  Analyzing $\sigma_R$ as a function of the longitudinal photon
polarization $\eps$ at fixed $Q^2$ allows one to extract $G_M^2(Q^2)$ from the
$\eps$-intercept, and $G_E^2(Q^2)$ from the slope in $\eps$. Because of the
$\eps/\tau$ weighting of $G_E^2$ relative to $G_M^2$, the contribution from the
electric form factor to the cross section is suppressed at large $Q^2$. The
proton form factor ratios extracted via the Rosenbluth technique have generally
been consistent with $Q^2$ scaling
$|G_E| \approx |G_M/\mu_p|$~\cite{Walker:1993vj, Christy:2004rc, Qattan:2004ht}.

An alternative method of extracting the form factors, known as the polarization
transfer (PT) technique~\cite{Akhiezer:1974em, Arnold:1980zj}, utilizes polarization
degrees of freedom to increase the sensitivity to the electric form factor at large
$Q^2$.  Here, longitudinally polarized electrons are scattered from an unpolarized
proton target, with the polarization of the recoiling proton detected,
$\vec{e}+p\to e+\vec{p}$.

In the Born approximation the elastic cross section for scattering a
longitudinally polarized electron of helicity $h$ with a recoil proton polarized
longitudinally ({\em i.e.} in the direction of motion) is given by
\be
\frac{d\sigma^{(L)}}{d\Omega}
= h\ \sigma_{\rm Mott}\
   \frac{E + E'}{M} \sqrt\frac{\tau}{1+\tau} \tan^2\frac{\theta_e}{2}\
    G_M^2\, .
\label{eq:sigL}
\ee
For a recoil proton detected with a polarization transverse to the proton
momentum, but still in the scattering plane, the cross section is
\be
\frac{d\sigma^{(T)}}{d\Omega }
= h\ \sigma_{\rm Mott}\
    2 \sqrt\frac{\tau}{1+\tau} \tan\frac{\theta_e}{2}\
    G_E\, G_M\, .
\label{eq:sigT}
\ee
Taking the ratio of the transverse to longitudinal proton cross sections then
yields the ratio of the electric to magnetic proton form factors,
\be
R=-\mu_p \sqrt\frac{\tau (1+\eps)}{2 \eps}\, \frac{P_T}{P_L}
= -\mu_p \frac{E+E'}{2 M} \tan\frac{\theta_e}{2}\ \frac{P_T}{P_L}
= \mu_p \frac{G_E}{G_M}\, ,
\label{eq:poltrans}
\ee
where $P_L$ and $P_T$ are the polarizations of the recoil proton longitudinal
and transverse to the proton momentum in the scattering plane, respectively. The
polarization transfer normal to the scattering plane, $P_N$, vanishes in the OPE
approximation, but it can be non-zero in general.
\begin{figure}[t]
\begin{center}
\includegraphics[width=0.5\textwidth]{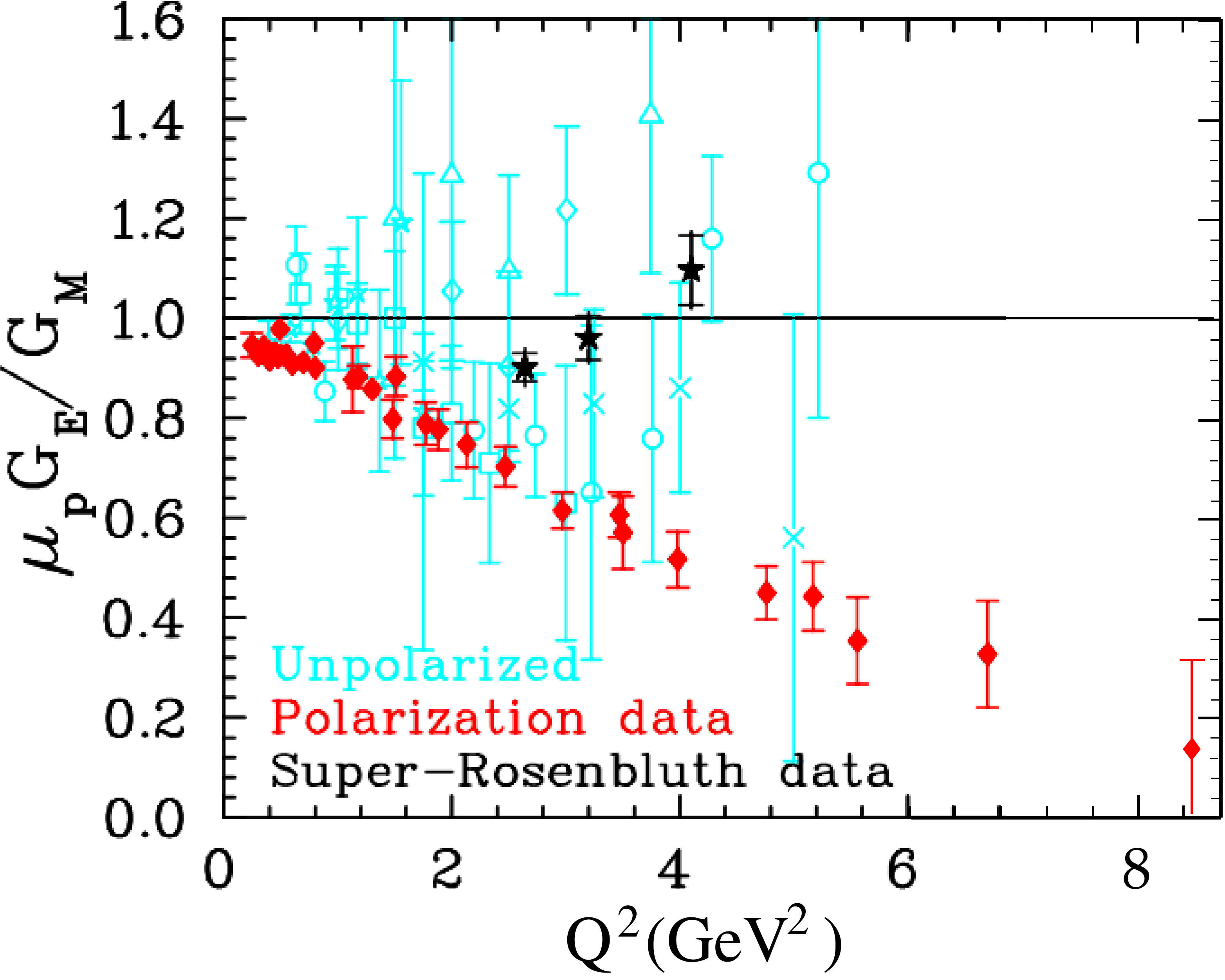}
\caption{Ratio $R=\mu_p G_E^p/G_M^p$ as extracted
	using Rosenbluth separation~\cite{Arrington:2003df} (open cyan points), 
	polarization transfer measurements~\cite{Punjabi:2005wq,
	Puckett:2010ac, Puckett:2011xg, Zhan:2011ji, Ron:2011rd} (filled red diamonds),
	and ``Super-Rosenbluth''
	measurements~\cite{Qattan:2004ht} (black stars).
	Figure taken from Ref.~\cite{Rimal:2016toz}.}
\label{fig:GEGM}
\end{center}
\end{figure}

In a series of experiments at Jefferson Lab~\cite{Jones:1999rz, Gayou:2001qt,
Gayou:2001qd, Punjabi:2005wq, MacLachlan:2006vw, Ron:2007vr, Puckett:2010ac,
Zhan:2011ji, Puckett:2011xg, Ron:2011rd} beginning in the late 1990's, the
polarization transfer (PT) technique was used to accurately determine the
ratio $\gegm$ up to $Q^2 = 8.5$~GeV$^2$. In addition, there have been
complementary measurements using polarized targets at
MIT-Bates~\cite{Crawford:2006rz} and Jefferson Lab~\cite{Jones:2006kf}. The
results, illustrated in Fig.~\ref{fig:GEGM}, are in striking contrast to the ratio
obtained via LT or Rosenbluth separations, showing an approximately linear
decrease of $R$ with $Q^2$.

The discrepancy between the LT and PT measurements of $\gegm$ has stimulated
considerable theoretical and experimental activity over the past 15 years. 
Attempts to reconcile the measurements have mostly focussed on improved
treatments of radiative corrections, particularly those associated with the
model-dependent hadronic terms that arise in TPE. These terms can lead to
additional $\eps$-dependence of the cross section. In the following sections we
discuss theoretical efforts to better understand the discrepancy, as well as the
impact of these calculations on other observables ({\em e.g.} single-spin
asymmetries).

\subsection{Two-photon exchange}
\label{ssec:TPE}

The first quantitative calculation to address the $G_E/G_M$ discrepancy was made
by Blunden \etal~\cite{Blunden:2003sp}, who computed the effect on $\gegm$ from
TPE, incorporating explicitly the nucleon's substructure through hadronic form
factors. A number of other studies have followed, examining TPE in a variety of
frameworks, and exploring reactions beyond elastic $ep$ scattering.
In a parallel effort, Guichon and Vanderhaeghen~\cite{Guichon:2003qm} provided a
generalized formalism for elastic scattering, allowing for possible TPE
contributions through three ``generalized form factors'' denoted as
$\tilde{F}_1$, $\tilde{F}_2$, and $\tilde{F}_3$.  They demonstrated that
it was natural to have TPE contributions that could significantly change the LT
extraction of $\gegm$ with minimal impact on the PT measurements.
In this section we review these efforts, paying particular attention to recent
progress in conventional hadronic-level calculations, which are most applicable
to data analysis at low to moderate $Q^2$ values. Theoretical progress on TPE
calculations at higher $Q^2$, using partonic degrees of freedom, is briefly
discussed in the final subsection.

\subsubsection{General formulation}
\label{ssec:formulation}

Using the kinematics of Fig.~\ref{fig:OPETPE}, the contribution to the TPE box
amplitude from an intermediate hadronic state $R$ of invariant mass $M_R$ can be
written in the general form~\cite{Blunden:2003sp, Kondratyuk:2005kk}
\be
{\cal M}_{\gamma\gamma}^{\rm box}
= -ie^4 \int \frac{d^4 q_1}{(2\pi)^4}\
    \frac{L_{\mu\nu} H_R^{\mu\nu}}{(q_1^2-\lambda^2)(q_2^2-\lambda^2)}\,,
\label{eq:Mggbox}
\ee
with $q_2=q-q_1$, and an infinitesimal photon mass $\lambda$ is introduced to
regulate any infrared divergences. The leptonic and hadronic tensors are given by
\begin{eqnarray}
L_{\mu\nu}
&=& \ubar_e(k')\, \gamma_\mu\, S_F(k-q_1,m_e)\, \gamma_\nu\, u_e(k)\,,\\
H^{\mu\nu}
&=&\ubar_N(p')\, \Gamma_{R\to\gamma N}^{\mu\alpha}(p+q_1,-q_2)\,
     S_{\alpha\beta}(p+q_1,M_R)\,
		 \Gamma_{\gamma N\to R}^{\beta\nu}(p+q_1,q_1)\, u_N(p)\,.
\end{eqnarray}
The electron propagator is
\be
S_F(k,m_e)
= \frac{(\slashed{k} + m_e)}{k^2 - m_e^2 + i\epsilon }\, .
\ee
The hadronic transition current operator $\gamma N\to R$ is written in a general
form $\Gamma_{\gamma N\to R}^{\alpha\mu}(p_R,q)$ that allows for a possible
dependence on the {\em incoming} momentum $q$ of the photon and the {\em
outgoing} momentum $p_R$ of the hadron, while $\mu$ and $\alpha$ are Lorentz
indices, and $S_{\alpha\beta}(p_R,M_R)$ is the hadronic state propagator. One
can obtain the crossed-box term directly from the box term by applying
crossing symmetry. For example, in the unpolarized case, we have
\be
{\cal M}_{\gamma\gamma}^{\rm xbox}(u,t) = -{\cal M}_{\gamma\gamma}^{\rm box}(s,t)
\vert_{s\to u}\,.
\ee
In general, ${\cal M}_{\gamma\gamma}^{\rm box}(s,t)$ has both real and imaginary
parts, whereas ${\cal M}_{\gamma\gamma}^{\rm xbox}(u,t)$ is purely real. The
imaginary part is of interest for calculations of the normal spin asymmetry,
which is discussed in section~(\ref{ssec:spin}) of this Review. The imaginary
part is also useful as an alternative method to calculating the real part
through the use of dispersion relations and analyticity. This is discussed in
further detail in section~(\ref{ssec:dispersive}).

The relative correction to the reduced Born cross section,
Eq.~(\ref{eq:sigmaR}), due to the interference of the one- and two-photon
exchange amplitudes shown in Fig.~\ref{fig:OPETPE}, is given by
\be
\delta_{\rm TPE}
= \frac{2 \Re \left( {\cal M}_\gamma^* {\cal M}_{\gamma\gamma} \right)}
     {\left| {\cal M}_\gamma \right|^2 }\,.
\label{eq:delta_TPE}
\ee

Within the framework of the simplest hadronic models, analytic evaluation of
$\delta_{\rm TPE}$ is made possible by writing the
transition form factors at the $\gamma$-hadron vertices as a sum and/or product
of monopole form factors~\cite{Blunden:2003sp, Blunden:2005ew}, which are
typically fit to empirical transition form factors over a suitable range in
space-like four-momentum transfer.  Four-dimensional integrals over the momentum
in the one-loop box diagram can then be expressed in terms of Passarino-Veltman
scalar functions $A_0$, $B_0$, $C_0$, and $D_0$~\cite{tHooft:1978jhc,
Passarino:1978jh}, which can be evaluated numerically using packages like
LoopTools~\cite{Hahn:1998yk}. 

\subsubsection{Model-independent TPE corrections}
\label{ssec:general}
Expression (\ref{eq:delta_TPE}) contains infrared (IR) divergences arising from
the elastic intermediate state when the momentum $q_i$ of either photon goes to
$0$. In analyzing the TPE corrections for $ep$ scattering, it is convenient to
separate terms into the ``soft'' parts, which are independent of hadronic
structure, and the ``hard'' parts, which are model-dependent. Soft here implies
that the interaction of one of the virtual photon with the proton occurs with
vanishingly small momentum transfer. As the soft parts are independent of hadron
structure, they are therefore universal, {\em i.e.} the same for protons as they
are for scattering from point-like particles. All of the IR divergences for the
virtual diagrams are contained in the soft parts, and these divergences cancel
in the total amplitude when added to the inelastic bremsstrahlung contributions
involving the emission of a real, soft photon.

Because the soft, IR-divergent part of the TPE amplitude is already included
with other model-independent radiative corrections in experimental analyses, it
has become conventional to consider only the hard part of the TPE amplitude in
discussions of TPE effects. This can be accomplished by an appropriate
subtraction of a conventionally defined soft part from the full TPE amplitude.
Two conventions in common use for the soft, model-independent part of the TPE
amplitude are those of Mo and Tsai~\cite{Tsai:1961zz, Mo:1968cg}, and Maximon
and Tjon~\cite{Maximon:2000hm}. These are discussed extensively in a previous
TPE review by Arrington, Blunden, and Melnitchouk~\cite{Arrington:2011dn}, and
we defer to that paper for details. The explicit expressions are:
\begin{eqnarray}
\delta_{\rm IR}({\rm MoT})
&=& -\frac{2\alpha}{\pi}
   \left[ \log\eta\,
		      \log\left( \frac{2 M \sqrt{E E'}}{\lambda^2} \right)
		- {\rm Li}_2 \left( 1 - \frac{M}{2 E} \right)
       	+ {\rm Li}_2 \left( 1 - \frac{M}{2 E'} \right)
   \right]\,,\label{eq:delMTsai}\\
\delta_{\rm IR}({\rm MTj})
&=& -\frac{2\alpha}{\pi} \log{\eta}\, \log\frac{Q^2}{\lambda^2}\,,
\label{eq:delMTjon}
\end{eqnarray}
with $\eta=E/E'$ the ratio of incident to final electron energies, and ${\rm
Li}_2$ is the dilogarithm function. For historical reasons, the Mo-Tsai
expression is the one generally used in existing experimental computer codes.
However, the Maximon-Tjon expression has also been used in more recent
experimental analyses~\cite{Bernauer:2013tpr}. A meaningful comparison to data
can therefore be made by considering the difference
\be
\delta_{\gamma\gamma}
\equiv\ \delta_{\rm TPE} - \delta_{\rm IR}({\rm MoT})\,,
\label{eq:delta_diff}
\ee
for which the IR divergences cancel, and is therefore independent of $\lambda$.
Here $\delta_{\g2}$ represents TPE effects that are unaccounted for after
applying the standard radiative corrections to data. The measured reduced cross
section $\sigma_R^{\rm meas}$ is therefore related to $\sigma_R$ by
\be
\sigma_R^{\rm meas} = \sigma_R \left(1+\delta_{\g2}\right)\, .
\label{eq:sigmaRmeas}
\ee
Therefore a {\em positive} slope for $\delta_{\g2}$ versus $\eps$ means that the 
form factor $G_E^p$ inferred from $\sigma_R^{\rm meas}$ is {\em larger} than
its actual value, which is what the PT data is telling us.

\subsubsection{One-loop methods}
\label{ssec:elastic}

The box (plus crossed-box) TPE contributions were evaluated by Blunden and
collaborators in a series of papers using one-loop integration techniques.
Intermediate states they considered include elastic ($N$)~\cite{Blunden:2003sp,
Blunden:2005ew} and inelastic $\Delta$ excitations~\cite{Kondratyuk:2005kk}
evaluated within a hadronic framework. The contribution of the most important
heavier spin-\sfrac{1}{2} and spin-\sfrac{3}{2} resonances was also
estimated~\cite{Kondratyuk:2007hc}, using nucleon Compton scattering
calculations at low energies as input parameters. In addition, both
electromagnetic $\gamma\gamma$ and electroweak $\gamma Z$ box contributions to
parity-violating electron scattering have been calculated within the same
framework~\cite{Tjon:2007wx, Tjon:2009hf, Nagata:2008uv, Zhou:2009nf}. The key
findings of these investigations were discussed in the Review by Arrington
\etal~\cite{Arrington:2011dn}. In this section we briefly summarize these
results, and then discuss recent advances, particularly in the contribution of
excited intermediate states, before turning to the approach of using dispersion
relations.

\begin{figure}[tb]
\centering
\begin{minipage}{0.51\textwidth}
\centering
\includegraphics[width=\linewidth]{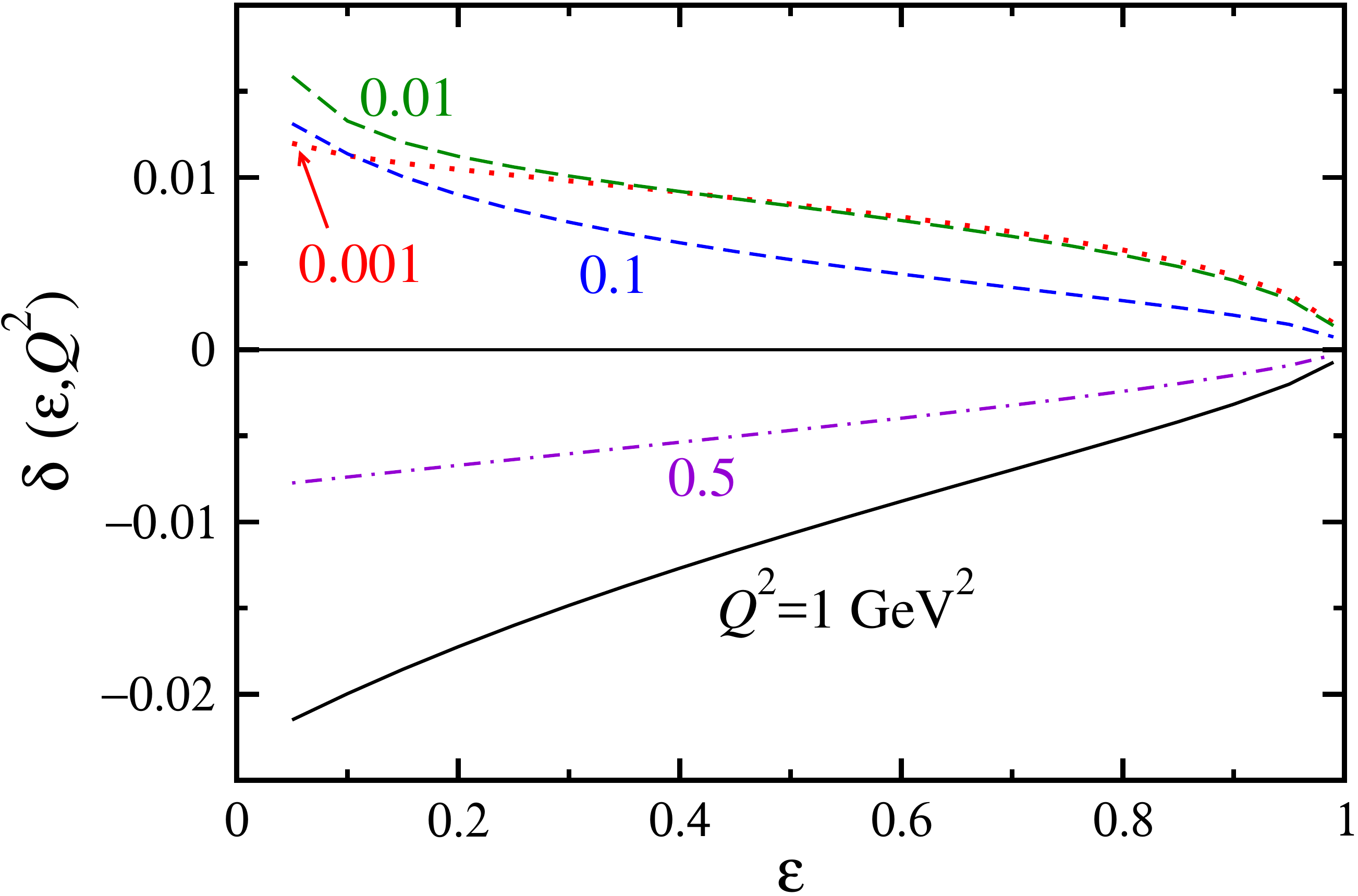}
\end{minipage}%
\begin{minipage}{0.48\textwidth}
\centering
\includegraphics[width=\linewidth]{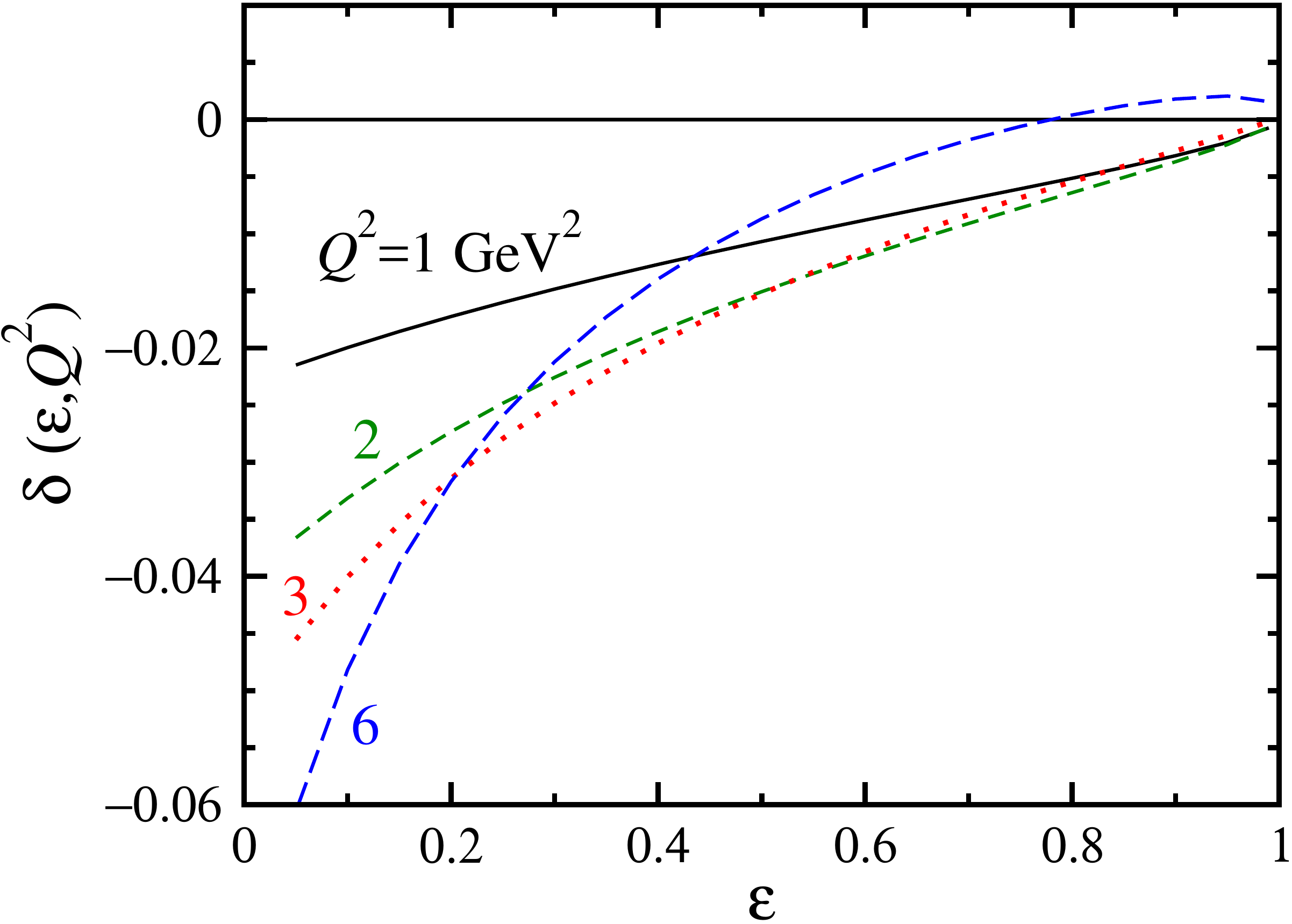}
\end{minipage}%
\caption{TPE correction of Eq.~(\ref{eq:delta_diff}) to
	elastic $ep$ scattering for intermediate nucleon states~\cite{Blunden:2003sp}.
	Curves are for $Q^2=0.001$--1~GeV$^2$ (\textit{left}), and $Q^2=1$--6~GeV$^2$
	(\textit{right}).}
\label{fig:del_gg}
\end{figure}

\paragraph{\bf Elastic contribution}
The elastic contribution, $\delta_N$, dominates the TPE corrections at low to
moderate $Q^2$. Results for $\delta_N$ relative to the Mo-Tsai approximation are
shown in Fig.~\ref{fig:del_gg} as a function of $\eps$ for several values of
$Q^2$, from 0.001 to 1~GeV$^2$ (left panel) and 1 to 6~GeV$^2$ (right panel).
These curves use the form factor parametrization of
Arrington~\etal~\cite{Arrington:2007ux}, although in practice the results are
fairly insensitive to the details of the form factors.

The hadron structure-dependent corrections are most significant at small $\eps$,
where they range from $\sim +1.5$\% at low $Q^2$ to $\sim -6$\% at
$Q^2=6$~GeV$^2$. At high $Q^2$ the magnetic form factor $G_M$ dominates in the
loop integrals. The TPE effect vanishes as $\eps\to 1$, a requirement linked to
unitarity. As noted by Blunden~\etal~\cite{Blunden:2003sp}, the positive slope
in $\eps$ at high $Q^2$ is of the right sign and magnitude to explain the
observed ratio $G_E/G_M$. At lower $Q^2$ values, $\delta_N$ is approximately
linear in $\eps$, but significant deviations from linearity are observed with
increasing $Q^2$, especially at small $\eps$. For $Q^2\lessapprox 0.3$~GeV$^2$,
$\delta_N$ becomes positive, and the electric form factor $G_E$ dominates in the
loop integrals.

At very low $Q^2\lessapprox 0.01$~GeV$^2$, $\delta_N$ approaches the static
limit for a structureless, massive target,
\be
\delta_N \xrightarrow[Q^2\to 0]{} \frac{\alpha \pi}{x+1}\,,
\qquad x=\sqrt\frac{1+\eps}{1-\eps}\,.
\label{eq:delFesh}
\ee
This result was first derived in the second Born approximation by
McKinley and Feshbach~\cite{McKinley:1948zz}, who expressed it in terms
of $\sin{(\theta_e/2)}=1/x$.

\begin{figure}[tb]
\begin{center}
\includegraphics[width=0.5\textwidth]{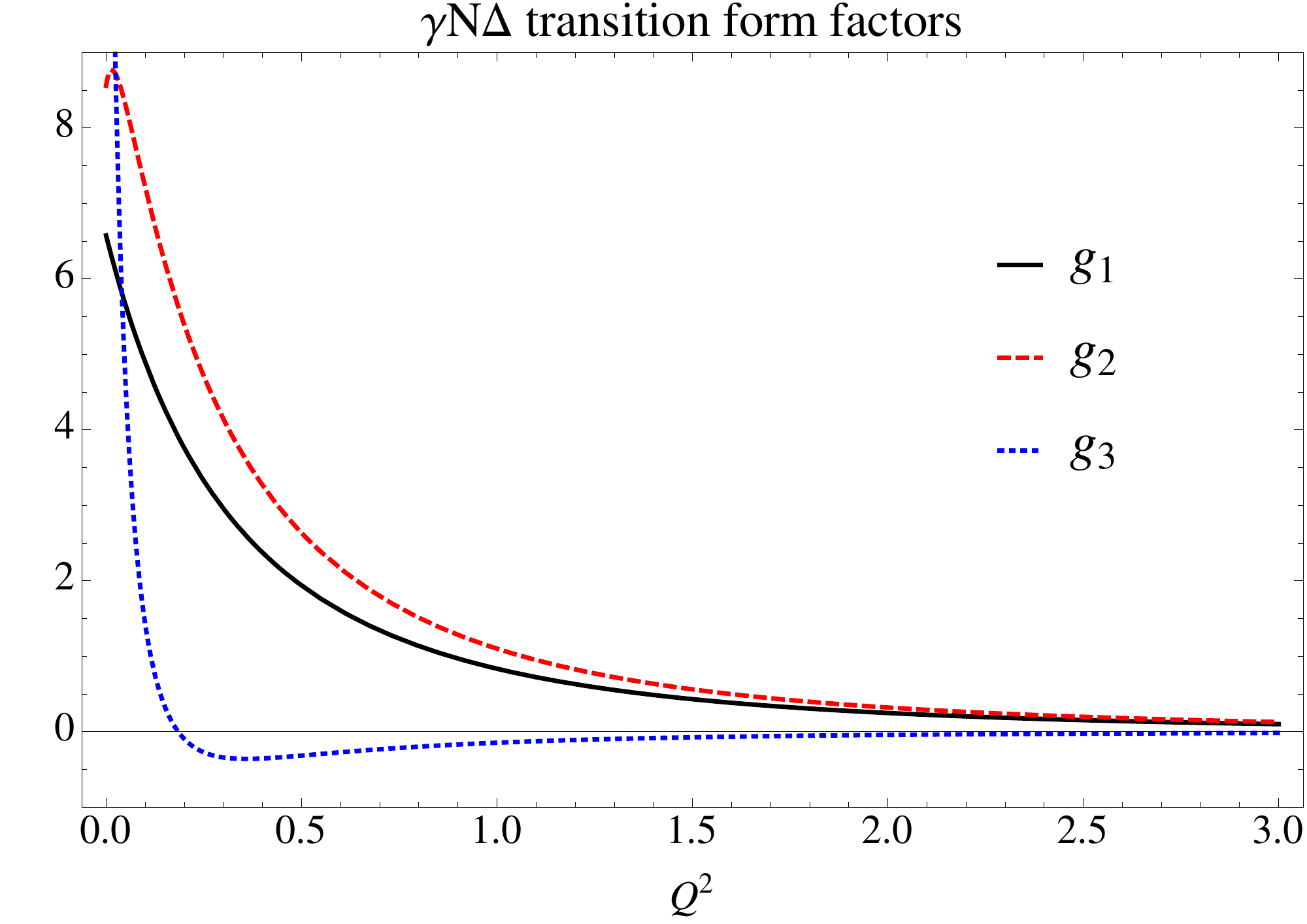}
\caption{The $\gamma N\to \Delta$ transition form factors based on a fit
	of Ref.~\cite{Aznauryan:2011qj}. Over this range of $Q^2$, $g_1(Q^2)$
	is well approximated by a dipole form factor with mass parameter
	$\Lambda_\Delta=0.76$~GeV. }
\label{fig:DeltaFF}
\end{center}
\end{figure}

\paragraph{\bf Inelastic contributions}
In view of the prominent role of the $\Delta(1232)$ resonance in the
electromagnetic excitation spectrum of the nucleon, it is important to evaluate
its contribution to the TPE amplitude. The $\gamma N\to\Delta$ electromagnetic
transition can be expressed in terms of three Jones-Scadron transition form
factors, $G_M^*(Q^2)$, $G_E^*(Q^2)$, and $G_C^*(Q^2)$, corresponding to
magnetic, electric, and Coulomb multipole excitations, respectively. The
magnetic multipole dominates in this transition. Although the $\gamma N\to
\Delta$ cross section is diagonal in these functions, they are cumbersome to
work with in the transition vertex function. For that purpose, an on-shell
equivalent parametrization of the $\gamma N\to\Delta$ vertex
is~\cite{Kondratyuk:2005kk}
\begin{eqnarray}
\Gamma_{\gamma N \to \Delta}^{\alpha\mu}(p_\Delta,q)
&=& \frac{1}{2 M_\Delta^2}\sqrt\frac{2}{3}
\Big\{
  g_1(Q^2)
  \left[ g^{\alpha \mu} \slashed{q}\slashed{p}_\Delta
    - \slashed{q} \gamma^\alpha p_\Delta^\mu 
    - \gamma^\alpha \gamma^\mu q\cdot p_\Delta
    + \slashed{p}_\Delta\,\gamma^\mu q^\alpha 
  \right]				\nn\\
& & 
+\ g_2(Q^2)
  \left[ q^\alpha p_\Delta^\mu - g^{\alpha\mu} q\cdot p_\Delta
  \right]\nn\\
&&+\ \frac{g_3(Q^2)}{M_\Delta}
  \left[ q^2 \left( \gamma^\alpha p_\Delta^\mu 
		  - g^{\alpha\mu} \slashed{p}_\Delta
	     \right)
       + q^\mu \left( q^\alpha \slashed{p}_\Delta
		    - \gamma^\alpha q\cdot p_\Delta
	       \right)
  \right]
\Big\} \gamma_5\,,
\label{eq:gND}
\end{eqnarray}
where $p_\Delta$ and $q$ are the momenta of the {\em outgoing} $\Delta$ and {\em
incoming} photon.  A parametrization of all three $g_i(Q^2)$ transition form
factors, based on recent fits to electro-production data by Aznauryan and
Burkert~\cite{Aznauryan:2011qj}, is shown in Fig.~\ref{fig:DeltaFF}.
Briefly, $g_1$ is determined by the dominant $G_M^*$ magnetic form factor,
$(g_2-g_1)$ is primarily sensitive to the electric form factor $G_E^*$, and
$g_3$ is sensitive to the Coulomb form factor $G_C^*$. Over the range
$0<Q^2<3$~GeV$^2$, the $g_1$ form factor is well approximated by a dipole, with
mass parameter $\Lambda_\Delta=0.76$~GeV.

In Ref.~\cite{Kondratyuk:2005kk} the $g_1$ and $g_2$ form factors were assumed
to have dipole shapes, $g_i(Q^2) = g_i(0)/(1+Q^2/\Lambda_\Delta^2)^2$, with a
dipole mass $\Lambda_\Delta$ ranging between hard ($\Lambda_\Delta=0.84$~GeV)
and soft ($\Lambda_\Delta=0.69$~GeV) to gauge the sensitivity of their shape on
contribution of the $\Delta$ to TPE. The $g_3$ form factor was set to 0. No
attempt was made to make a detailed fit of these form factors to available data.
To summarize the key findings of these early investigations: The TPE $\Delta$
intermediate state contribution is smaller and of opposite sign to the $N$ one,
thereby attenuating the $N$ contribution somewhat; it increases in magnitude as
$Q^2$ increases; and it diverges as $\eps\to 1$, in apparent violation of the
unitarity constraint. This divergence is most apparent for $\eps\gtrapprox 0.9$,
and is sensitive to the transition form factor (soft or hard). These features
are also found in the calculations of Refs.~\cite{Tjon:2009hf, Nagata:2008uv,
Zhou:2009nf}.

\paragraph{\bf Recent work}
More recently, Zhou and Yang~\cite{Zhou:2014xka},
Graczyk~\cite{Graczyk:2013pca}, and Lorenz~\etal~\cite{Lorenz:2014yda} improved
on this early work, primarily through the use of $\gamma N\Delta$ transition
form factors with a closer fit to data.

\begin{figure}[tb]
\begin{center}
\includegraphics[width=0.4\textwidth]{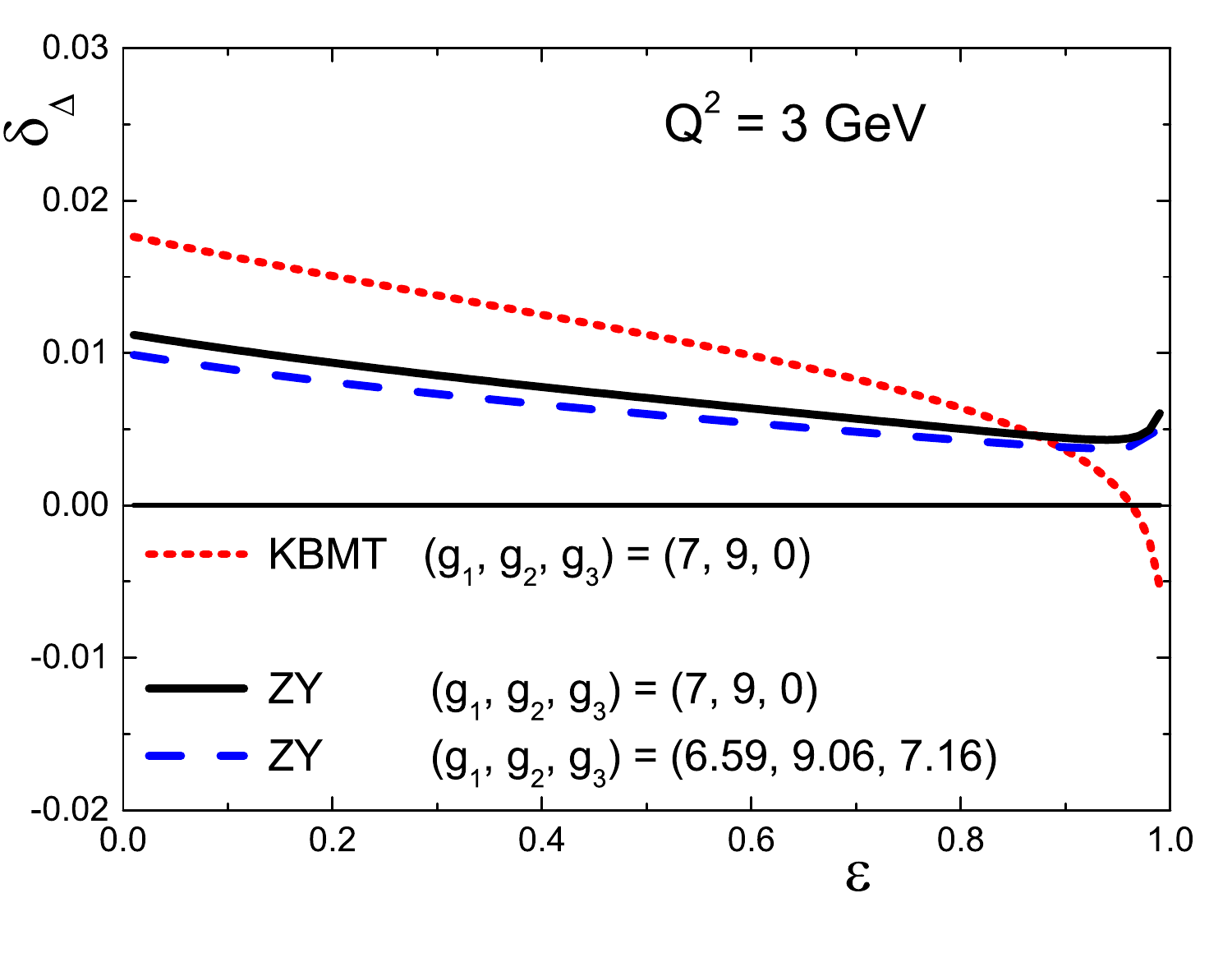}
\caption{Contribution of $\delta_\Delta$ at $Q^2=3$~GeV$^2$, adapted from Zhou and
	Yang~\cite{Zhou:2014xka}. The dotted curve is based on a calculation by
	Kondratyuk~\etal~\cite{Kondratyuk:2005kk}, and uses magnetic and electric
	form factors of dipole form ($\Lambda_\Delta=0.84$~GeV). The solid curve
	uses the same coupling strengths but with a realistic form factor shape fit
	to data~\cite{Zhou:2014xka}, while the long-dashed curve shows the effect of
	using all three transition form factors with coupling strengths and shapes
	fit to data~\cite{Zhou:2014xka}.}
\label{fig:Yang}
\end{center}
\end{figure}

Zhou and Yang~\cite{Zhou:2014xka} improved the treatment of the $\gamma N\Delta$
transition form factors by fitting a sum of monopoles (or dipoles) to existing
electroproduction data. They also included all three form factors: $g_1$, $g_2$,
and $g_3$. Figure~\ref{fig:Yang} shows the effect on $\delta_\Delta$ in
comparison to the simpler parametrization of
Kondratyuk~\etal~\cite{Kondratyuk:2005kk}. The overall contribution with
realistic form factors is somewhat smaller than the hard-dipole fit of
Ref.~\cite{Kondratyuk:2005kk}. The effect of including the Coulomb contribution
(arising from $g_3$) is found to be small.

\begin{figure}[tb]
\centering
\begin{minipage}{0.5\textwidth}
\centering
\includegraphics[width=\linewidth]{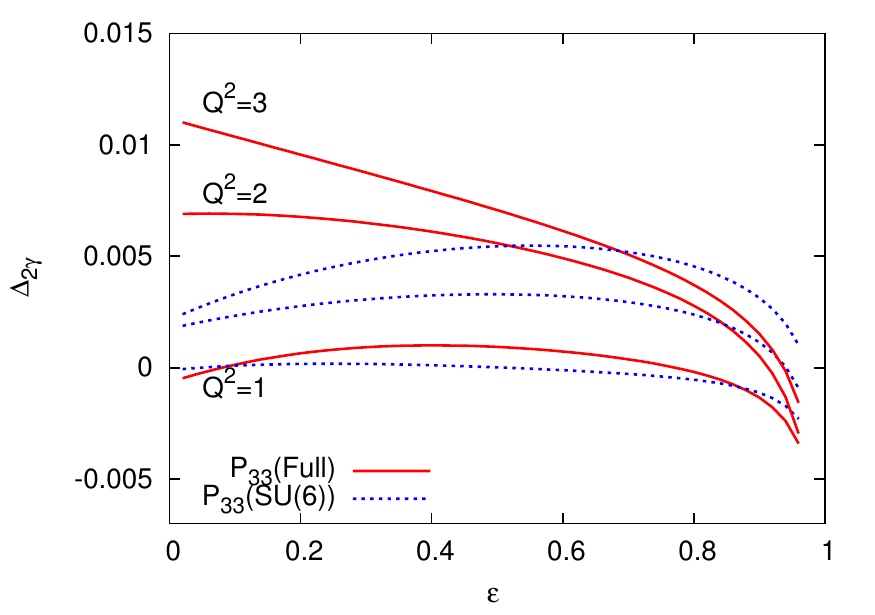}
\end{minipage}%
\begin{minipage}{0.5\textwidth}
\centering
\vspace*{-1ex}
\includegraphics[width=\linewidth]{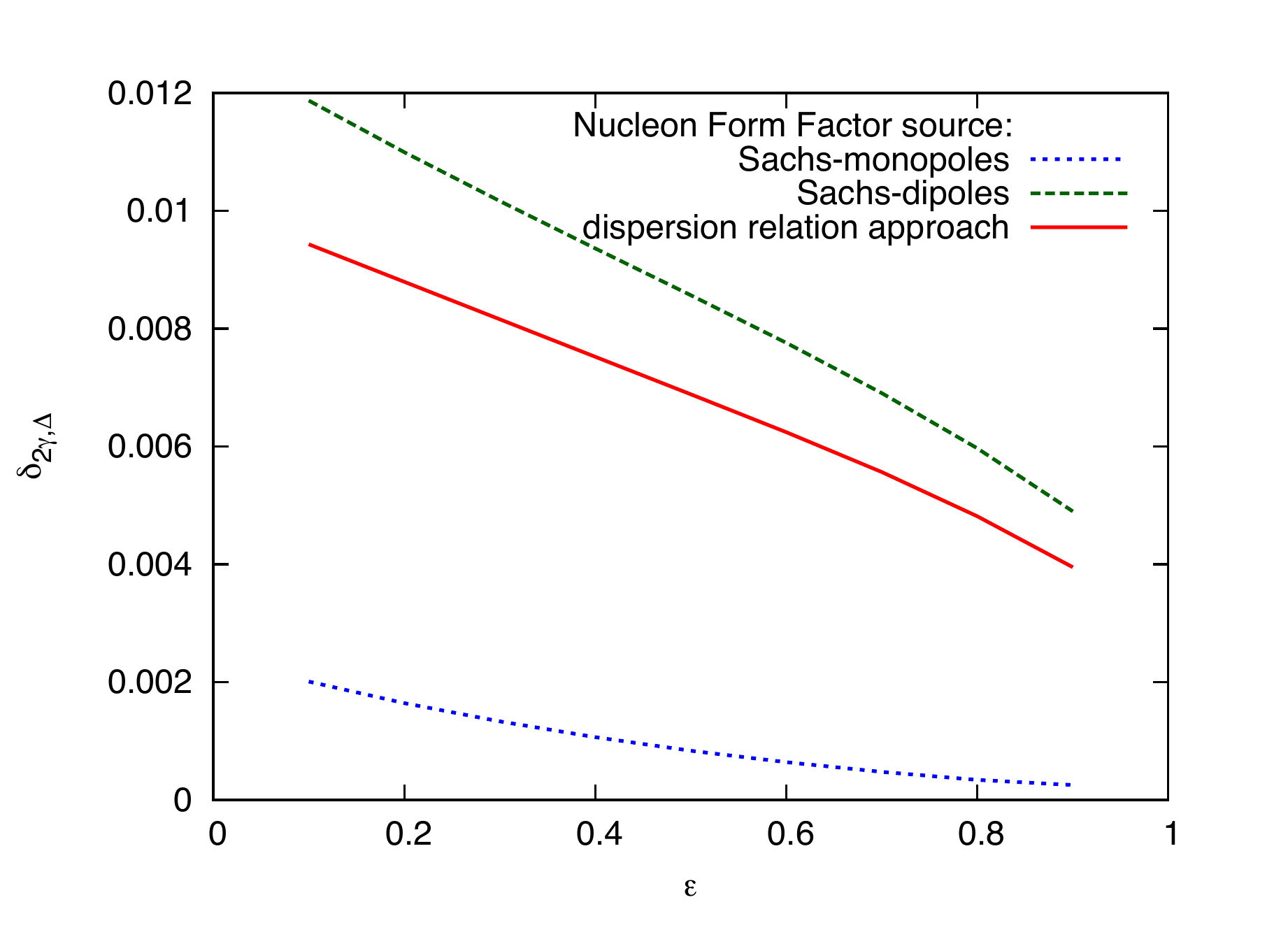}
\end{minipage}
\caption{\textit{Left:} The $\Delta$ TPE contribution $\delta_\Delta$ from
	Graczyk~\cite{Graczyk:2013pca}, showing the effect of different $\gamma N\Delta$
	form factor parametrizations for three values of $Q^2$.
	\textit{Right:} $\delta_\Delta$ at $Q^2=3$~GeV$^2$ from
	Lorenz~\etal~\cite{Lorenz:2014yda}, showing the effect of different
	transition form factor models.}
\label{fig:del_GracLor}
\end{figure}
\begin{figure}[tb]
\centering
\begin{minipage}{0.5\textwidth}
\centering
\hspace*{3em}
\includegraphics[width=0.7\linewidth]{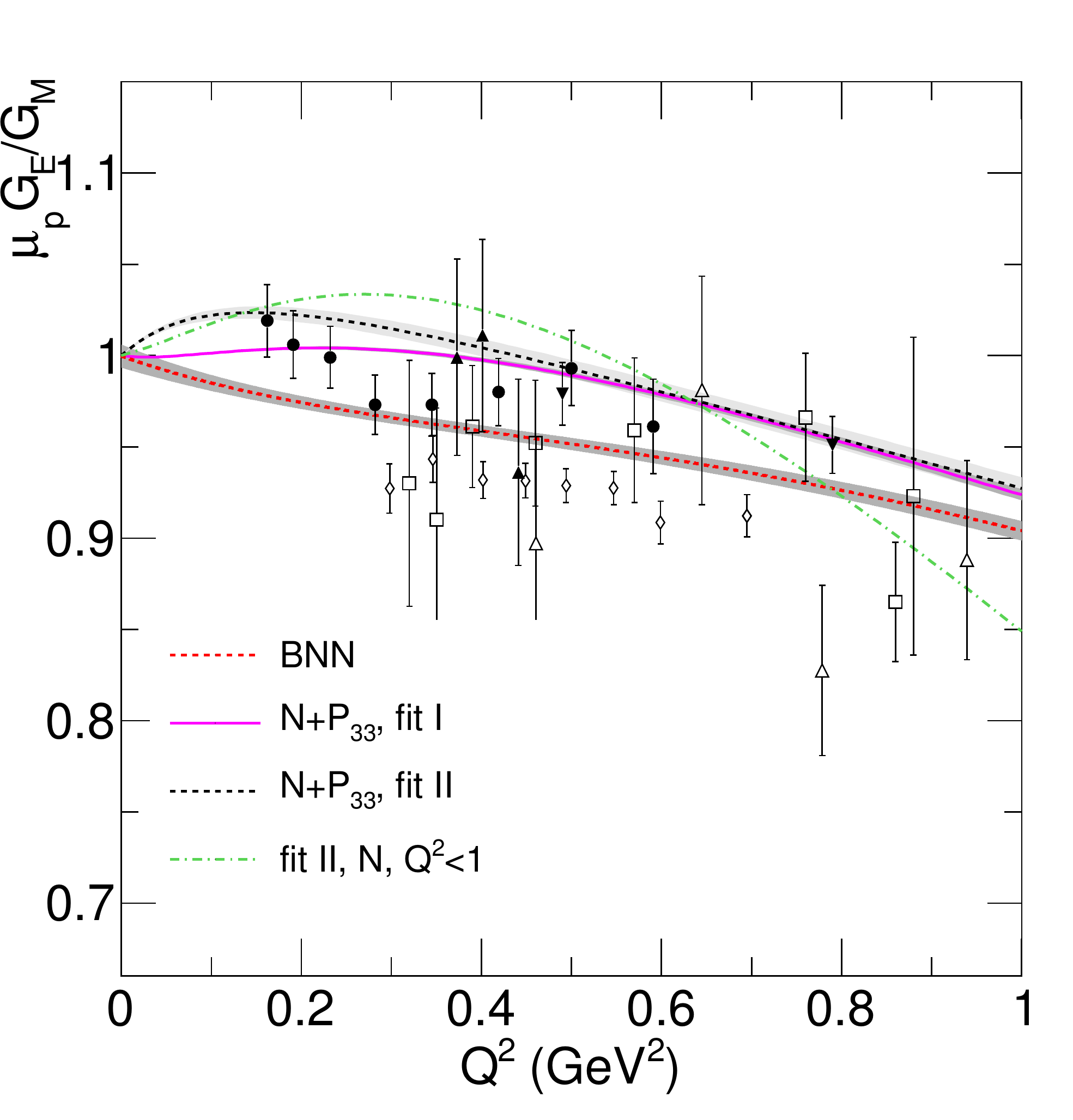}
\end{minipage}%
\begin{minipage}{0.5\textwidth}
\centering
\includegraphics[width=0.7\linewidth]{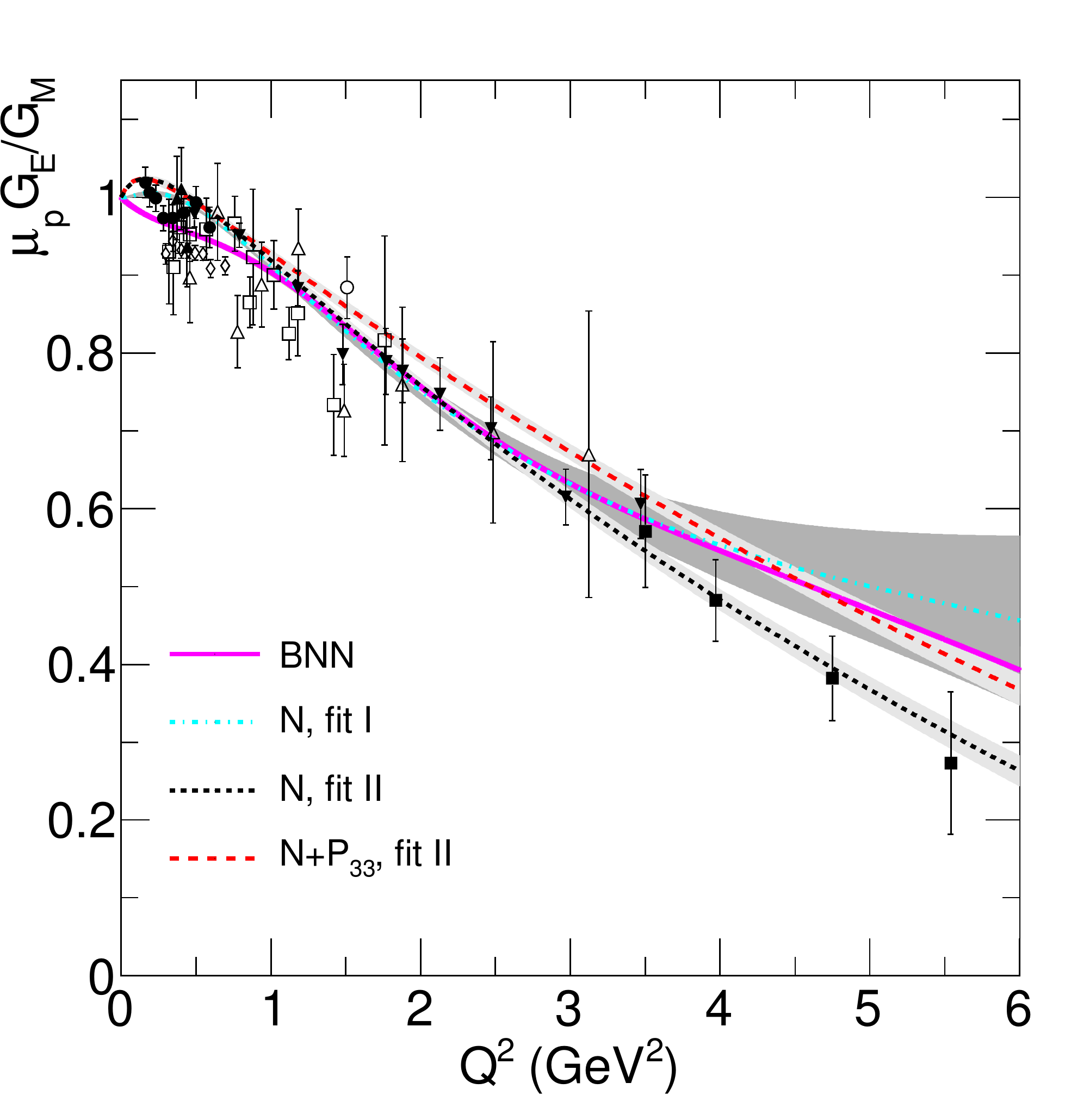}
\end{minipage}%
\caption{The ratio $R=\mu_p G_E/G_M$ calculated by TPE fits
	using intermediate states $N$ only and $N+P_{33}$,
	as well as the BNN fit. The shaded area shows the $1\sigma$ error
	from the fit. The ratio $R$ at low $Q^2$ (\textit{left})
	and at high $Q^2$ (\textit{right}).
	Figures taken from Gracyzk~\cite{Graczyk:2013pca}. }
\label{fig:ratioR}
\end{figure}

Graczyk~\cite{Graczyk:2013pca} obtained predictions for the TPE correction to
the unpolarized $ep$ elastic cross section in two different approaches. The
first is a standard calculation using one-loop box diagrams with $N$ and
$\Delta$ ($P_{33}$ resonance) as hadronic intermediate states. Different form
factor parametrizations of the $\gamma N \Delta$ transition form factors were
taken into consideration. In the second approach the phenomenological TPE
correction was extracted from experimental data by applying the Bayesian neural
network (BNN) statistical framework. The BNN response was constrained by
assuming that the PT data are not sensitive to TPE effects. Predictions of the
two methods agree well in the intermediate $Q^2$ range of 1-3~GeV$^2$, and agree
at the $2\sigma$ level above this range. Below $Q^2=1$~GeV$^2$ the two methods
disagree (see Fig.~\ref{fig:del_GracLor}). The effect on the ratio $G_E/G_M$ of the
combined $N+\Delta$ TPE contributions, as evaluated by Graczyk~\cite{Graczyk:2013pca},
is shown in Fig.~\ref{fig:ratioR}. There is reasonably good agreement with data over
the range of $Q^2$ given.

Lorenz~\etal~\cite{Lorenz:2014yda} also evaluated the $\Delta$ contribution to
the TPE amplitude, using $\gamma N\Delta$ transition vertices matched to
helicity amplitudes from electroproduction of nucleon resonances. They also
considered the effect of nucleon form factors fits constrained by analyticity
and unitarity, denoted as ``dispersion relation'' in Fig.~\ref{fig:del_GracLor}.
Their results are in good agreement with those of Refs.~\cite{Graczyk:2013pca,
Zhou:2014xka}.

Despite the improved fits of the transition form factors in
Refs.~\cite{Graczyk:2013pca, Zhou:2014xka, Lorenz:2014yda} the divergence of the
$\Delta$ contribution as $\eps\to 1$ that was seen in earlier
work~\cite{Kondratyuk:2005kk, Tjon:2009hf, Nagata:2008uv, Zhou:2009nf} is still
apparent.

In general the contributions of all the heavier resonances are much smaller than
those of the nucleon and $\Delta$ ($P_{33}$)~\cite{Kondratyuk:2007hc}. However,
there is an interesting interplay between the contributions of the
spin-\sfrac{1}{2} and spin-\sfrac{3}{2} resonances, which is analogous to the
partial cancellation of the two-photon exchange effects of the nucleon and
$\Delta$ intermediate states, found in Ref.~\cite{Kondratyuk:2005kk}.
Notwithstanding the smallness of the resonance contributions, their inclusion in
the TPE diagrams leads to a better agreement between the Rosenbluth and
polarization transfer data analyses, especially at higher values of the
momentum-transfer squared $Q^2$.

Inclusion of contributions of intermediate states with masses larger than $\sim
2$~GeV becomes impractical within a hadronic approach when one moves beyond the
resonance region.  Here it becomes more efficient to use partonic degrees of
freedom~\cite{Chen:2004tw, Afanasev:2005mp, Borisyuk:2008db, Kivel:2009eg}.
These were discussed in the Review by Arrington~\etal~\cite{Arrington:2011dn}.
We give a brief summary of these contributions in Sec.~\ref{ssec:highQ}.

\subsubsection{Dispersive methods}
\label{ssec:dispersive}

As noted previously, ${\cal M}_{\gamma\gamma}$ of Eq.~(\ref{eq:Mggbox}) has both real
and imaginary parts. The imaginary parts come from the box diagram, and are
completely determined by terms in the amplitude where the electron and hadron
intermediate states are on-shell. In the loop integral method, the imaginary
parts are completely contained in the Passarino-Veltman functions. The real and
imaginary parts are related by dispersion relations~\cite{ Gorchtein:2006mq,
Borisyuk:2008es}, which forms the basis of the dispersive method discussed in
this section.

Following the formalism introduced by Guichon and
Vanderhaeghen~\cite{Guichon:2003qm}, ${\cal M}_{\gamma\gamma}$ can be mapped
onto three generalized form factors $\tilde{F}_i(Q^2,\nu)$:
\begin{eqnarray}
{\cal M}_{\gamma\gamma}&\longrightarrow& -\frac{e^2}{q^2}\,
\bar{u}_e(k') \gamma_\mu u_e(k)\,\times\nn\\
&&\bar{u}_N(p') \left[ \tilde{F}_1(Q^2,\nu)\,\gamma^\mu\,
+\, \tilde{F}_2(Q^2,\nu)\, \frac{i \sigma^{\mu\nu} q_\nu}{2 M}\,+\,
\tilde{F}_3(Q^2,\nu)\,\frac{\slashed{K} P^\mu}{M^2} \right] u_N(p)\, ,
\end{eqnarray}
with $K=(k'+k)/2$ and $P=(p'+p)/2$.
Elastic electron scattering observables (including polarization transfer) can be
expressed in terms of these generalized form factors~\cite{Guichon:2003qm}. To
calculate their imaginary parts explicitly, we can put the intermediate electron
and hadron on-shell in Eq.~(\ref{eq:Mggbox}), thus reducing the four-dimensional loop
integral to a two-dimensional integral over $Q_1^2$ and $Q_2^2$ -- the
four-momentum squared of each virtual photon.

The generalized form factors satisfy fixed-$t$ dispersion
relations~\cite{Gorchtein:2006mq}
\begin{subequations}
\label{eq:genFF}
\begin{eqnarray}
\Re \tilde{F}_1(Q^2,\nu) &=& \frac{2}{\pi} {\cal P} \int_{\nu_{\rm th}}^\infty d\nu'\ 
\frac{\nu}{\nu'^2-\nu^2}\, \Im \tilde{F}_1(Q^2,\nu')\, ,\\
\Re \tilde{F}_2(Q^2,\nu) &=& \frac{2}{\pi} {\cal P} \int_{\nu_{\rm th}}^\infty d\nu'\ 
\frac{\nu}{\nu'^2-\nu^2}\, \Im \tilde{F}_2(Q^2,\nu')\, ,\\
\Re \tilde{F}_3(Q^2,\nu) &=& \frac{2}{\pi} {\cal P} \int_{\nu_{\rm th}}^\infty d\nu'\ 
\frac{\nu'}{\nu'^2-\nu^2}\, \Im \tilde{F}_3(Q^2,\nu')\, ,
\end{eqnarray}
\end{subequations}
where ${\cal P}$ denotes the Cauchy principal value integral, and $\nu_{\rm
th}=-\tau$. This extends the integral into the unphysical region
$\cos{\theta_e}<-1$, which requires knowledge of the transition form factors in
the timelike region. For the interaction of point particles, such as in $e\mu$
scattering, the real parts generated in this way agree completely with those
obtained directly from the four-dimensional loop integrals of
Eq.~(\ref{eq:Mggbox})~\cite{Tomalak:2014dja}.

\begin{figure}[tb]
\centering
\begin{minipage}{0.42\textwidth}
\centering
\includegraphics[width=.9\linewidth]{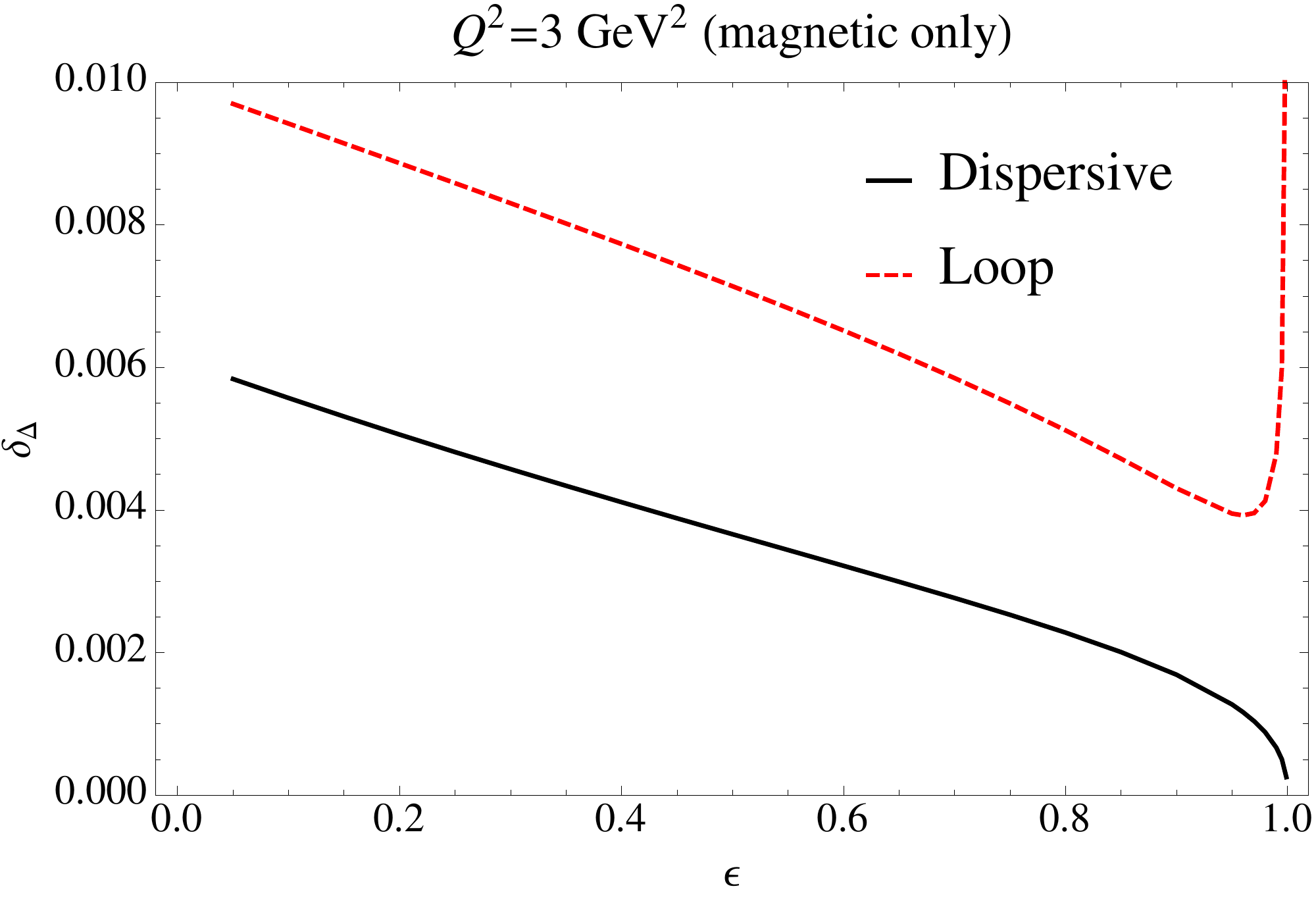}
\end{minipage}%
\begin{minipage}{0.42\textwidth}
\centering
\includegraphics[width=.9\linewidth]{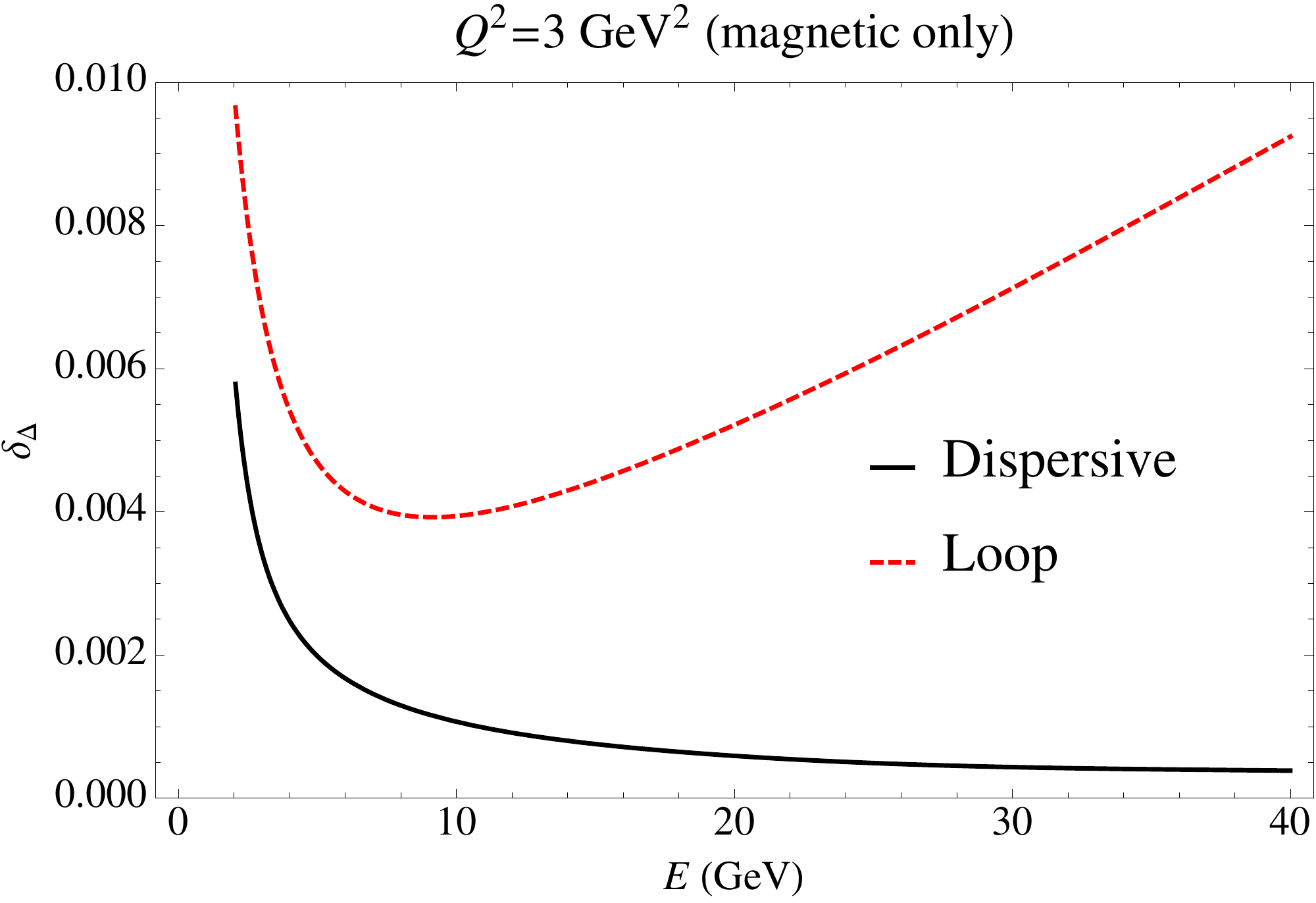}
\end{minipage}
\caption{A comparison of $\delta_\Delta$ calculated using the loop integration
	method with that calculated in the dispersive method. Plotted is the result at
	$Q^2=3$~GeV$^2$ for magnetic $\gamma N\Delta$ transition only, using a realistic
	transition form factor. The left figure shows $\delta_\Delta$ plotted against
	$\eps$. The right figure shows same result plotted against electron energy $E$,
	where the linear divergence with $E$ becomes apparent.
	Figure adapted from Blunden and Melnitchouk~\cite{Blunden:2016}.}
\label{fig:disp_loop}
\end{figure}

However, this equality no longer holds when the transition current operator
depends on the momentum of the hadron, as it does for the $\gamma N\Delta$
vertex of Eq.~(\ref{eq:gND})~\cite{Blunden:2016}. As a specific example, the
left plot in Fig.~\ref{fig:disp_loop} compares the TPE correction
$\delta_\Delta$ calculated using the dispersive method (solid curve) with the
correction calculated using loop integration (dashed curve), plotted as a
function of $\eps$ at $Q^2=3$~GeV$^2$~\cite{Blunden:2016}. Only the dominant
magnetic transition is considered here (by setting $g_1=g_2$ and $g_3=0$ in
Eq.~(\ref{eq:gND})), and the realistic form factor parametrization shown in
Fig.~\ref{fig:DeltaFF}~\cite{Aznauryan:2011qj} is used for $g_1(Q^2)$. The
imaginary parts of $\delta_\Delta$ (not shown) are identical, but the real parts
(as plotted) differ significantly. The dispersive result vanishes as $\eps\to
1$, and it is smaller in magnitude over the whole range of $\eps$. The
high-$\eps$ behaviour can be seen more easily in the right plot, where
$\delta_\Delta$ is plotted against electron energy $E$. The limit $\eps\to 1$
corresponds to $E\to \infty$ at fixed $Q^2$. The divergence in the loop integral
method is linear in $E$ due to the unconstrained linear factors of hadron
momentum $p$ at each $\gamma N\Delta$ vertex. By contrast, the dispersive curve
falls off like $1/E$. A linear growth in $E$ exceeds the Froissart
bound~\cite{Froissart:1961ux}, and signals a violation of unitarity due
to the unphysical off-shell behaviour of the transition current operators.
Including the electric and Coulomb contributions does not alter these
behaviours~\cite{Blunden:2016}.
Thus while the loop integration method gives qualitatively the right behaviour in
$\delta_\Delta$ for $\eps\lessapprox 0.9$, it is likely unreliable for precision
comparisons with experimental data.

\begin{figure}[tb]
\centering
\begin{minipage}{0.48\textwidth}
\centering
\includegraphics[width=\linewidth]{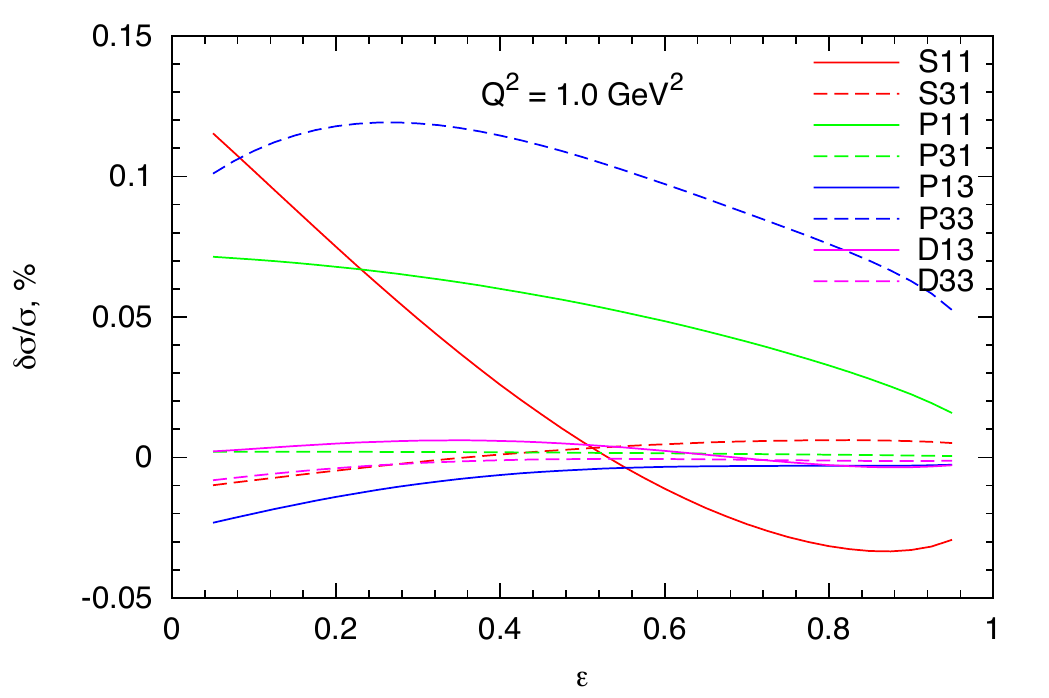}
\end{minipage}%
\begin{minipage}{0.48\textwidth}
\centering
\includegraphics[width=\linewidth]{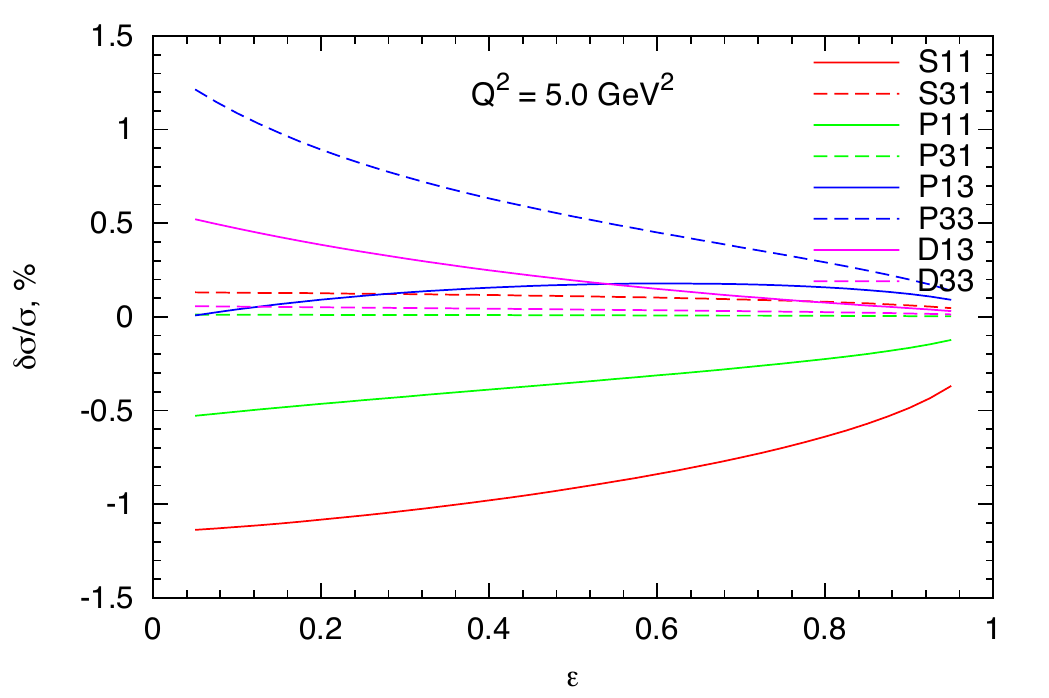}
\end{minipage}
\caption{TPE contributions to $\delta_{\g2}$ from different $\pi N$ channels at two
	different values of $Q^2$. Note the different scales.
	Figure taken from Borisyuk and Kobushkin~\cite{Borisyuk:2015xma}.}
\label{fig:BK2015}
\end{figure}

\paragraph{\bf Recent work}
Borisyuk and Kobushkin~\cite{Borisyuk:2008es, Borisyuk:2012he} evaluated both
$N$ and $\Delta$ contributions using the dispersive method. Empirical $\gamma
N\Delta$ transition form factors were fit to a sum of monopoles parametrization to
allow for analytic evaluation. More recently, they have extended their model to
include the effect of $\pi N$ intermediate hadronic states, concentrating on the
$P_{33}$~\cite{Borisyuk:2013hja}, and seven other spin-\sfrac{1}{2} and
spin-\sfrac{3}{2} channels~\cite{Borisyuk:2015xma}. Their calculation includes a
resonance width and shape, as well as background contributions. A sample of
their calculations of the TPE contributions to $\delta_{\g2}$ from different $\pi N$
channels at $Q^2=1$~GeV$^2$ and 5~GeV$^2$ is shown in Fig.~\ref{fig:BK2015}. As
expected, the $P_{33}$ channel dominates, but substantial contributions from
$P_{11}$ and $S_{11}$ channels are also present. The inelastic contributions are
small at low $Q^2$, and all vanish in the limit $\eps\to 1$.

Tomalak and Vanderhaeghen~\cite{Tomalak:2014sva} also used the dispersive method
to evaluate TPE effects for $N$ intermediate states. They used a subtracted
dispersion relation to improve the fits to observables. This paper introduces a
contour integration method to evaluate the two-dimensional loop integrals needed
for the imaginary part of the generalized form factors in the unphysical region
($-1<\eps<0$).

\begin{figure}[tb]
\centering
\begin{minipage}{0.33\textwidth}
\centering
\includegraphics[width=\linewidth]{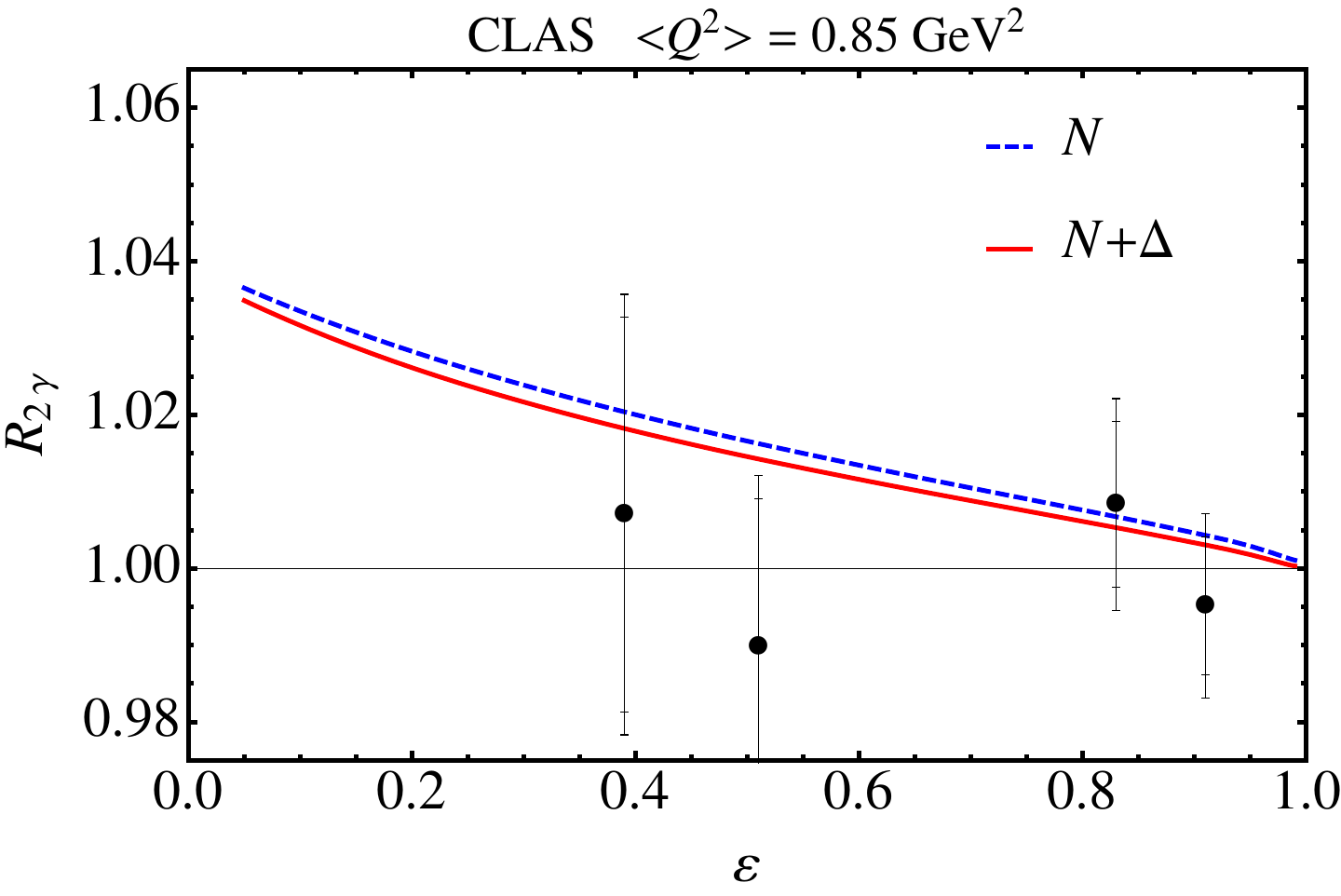}
\end{minipage}%
\begin{minipage}{0.33\textwidth}
\centering
\includegraphics[width=\linewidth]{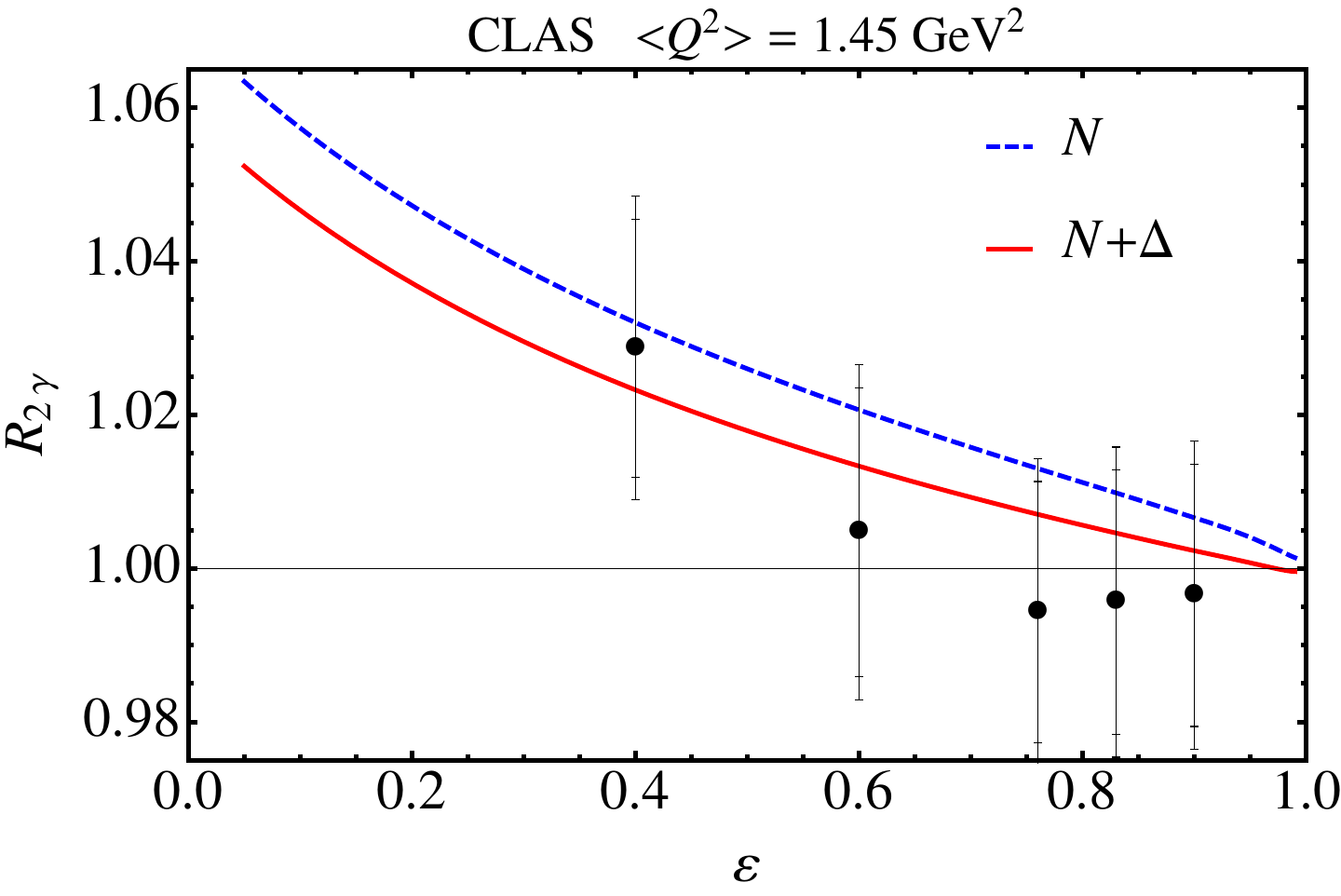}
\end{minipage}%
\begin{minipage}{0.33\textwidth}
\centering
\includegraphics[width=\linewidth]{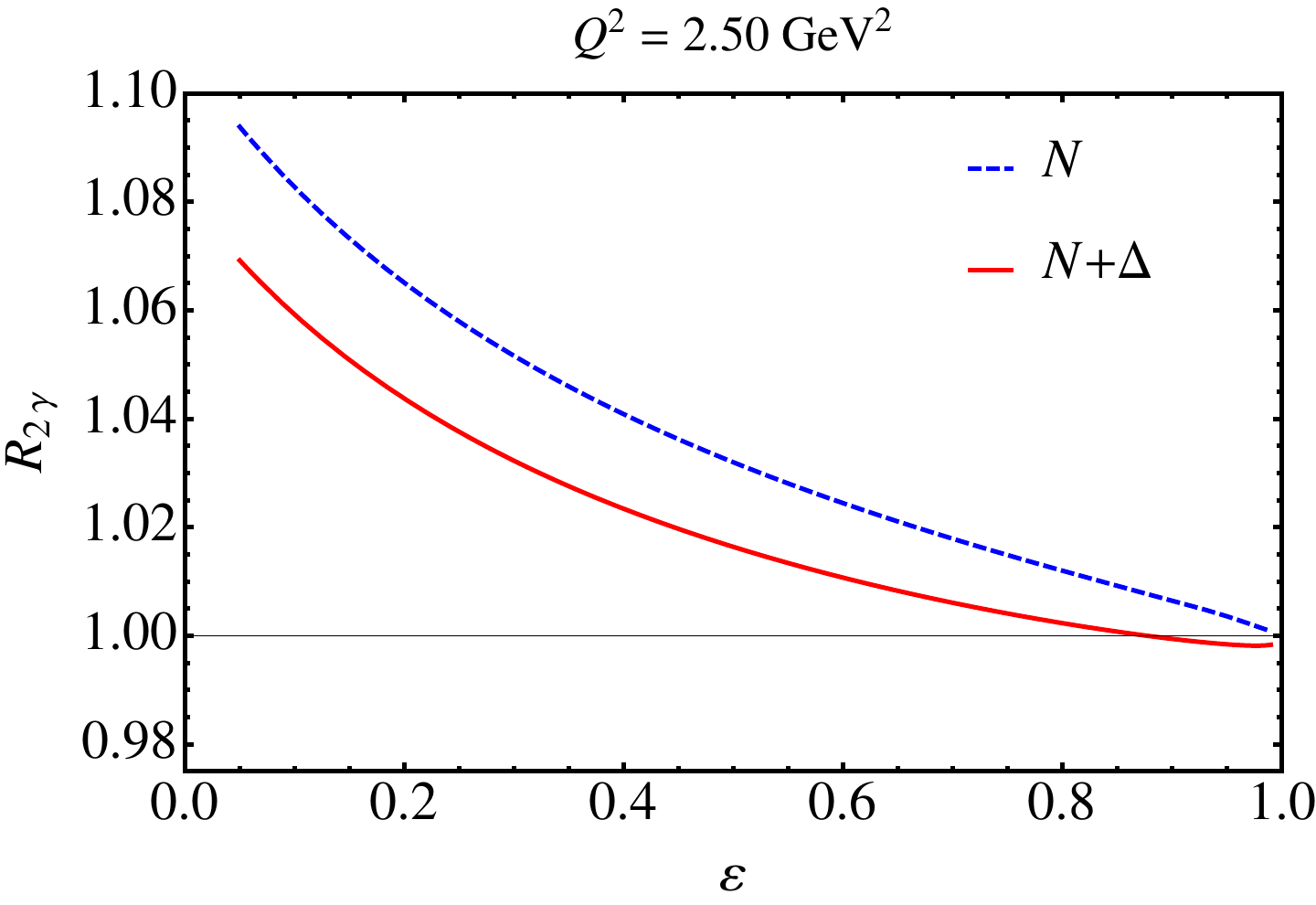}
\end{minipage}
\caption{Ratio of $e^+p/e^-p$ elastic cross sections at various values of $Q^2$,
	after accounting for conventional radiative corrections.
	The data are from Ref.~\cite{Rimal:2016toz}.
	Figure taken from Ref.~\cite{Blunden:2016}.}
\label{fig:R2gBlunden}
\end{figure}

Blunden and Melnitchouk~\cite{Blunden:2016} evaluated both $N$ and $\Delta$
contributions using the dispersive method, with realistic $\gamma NN$ and
$\gamma N\Delta$ transition form factors taken from electro-production data. By
using numerical contour integration in the unphysical region, based on the
approach of Tomalak and Vanderhaeghen~\cite{Tomalak:2014sva}, transition form
factors of arbitrary shape can be used. In particular, no reparametrization of
the form factors in terms of a sum of monopoles is needed. Their calculation of
the ratio of $e^+p$ to $e^-p$ cross sections over the range
$Q^2=0.85-2.50$~GeV$^2$ is shown in Fig.~\ref{fig:R2gBlunden}. The increasing
importance of the $\Delta$ contribution as $Q^2$ increases is apparent from
these plots.

The ratio of $e^+p$ to $e^-p$ cross sections, denoted by
$R_{2\gamma}\approx 1-2\delta_{\g2}$, is a direct measure of TPE effects. This is
discussed more fully in Section~\ref{sec:expt}. Gorchtein~\cite{Gorchtein:2006mq,
Gorchtein:2014hla} has recently made a model-independent analysis of corrections
to $R_{2\gamma}$ in forward kinematics (forward angles, low $Q^2$) using the dispersive
method. In forward kinematics the imaginary part of the TPE amplitude is related
to the total photoabsorption cross section, a result first derived by
Brown~\cite{Brown:1970te}.

Tomalak and Vanderhaeghen~\cite{Tomalak:2015aoa} also considered the $e^+p/e^-p$
ratio at forward angles and small $Q^2$. The inelastic TPE contribution was
expressed as a dispersion integral over the unpolarized proton structure functions
$F_1^\gamma$ and $F_2^\gamma$. Their calculation goes beyond the known leading terms
in the $Q^2$ expansion by keeping the full momentum-dependence of the proton structure
functions. In the range of small $Q^2$, their result is in good agreement with the
empirical TPE fits to existing data.

\subsubsection{QCD-based approaches at high $Q^2$}
\label{ssec:highQ}

\begin{figure}[tb]
\centering
\begin{minipage}{0.4\textwidth}
\centering
\includegraphics[width=0.6\linewidth]{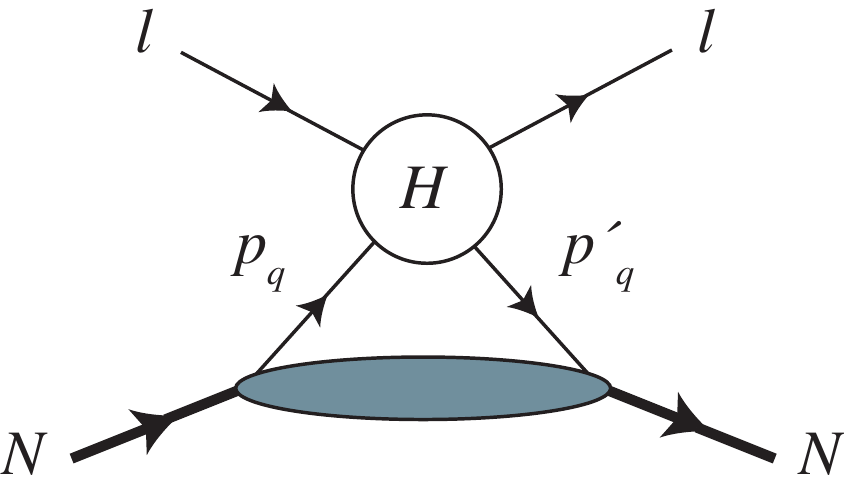}
\end{minipage}%
\begin{minipage}{0.6\textwidth}
\centering
\includegraphics[width=0.4\linewidth]{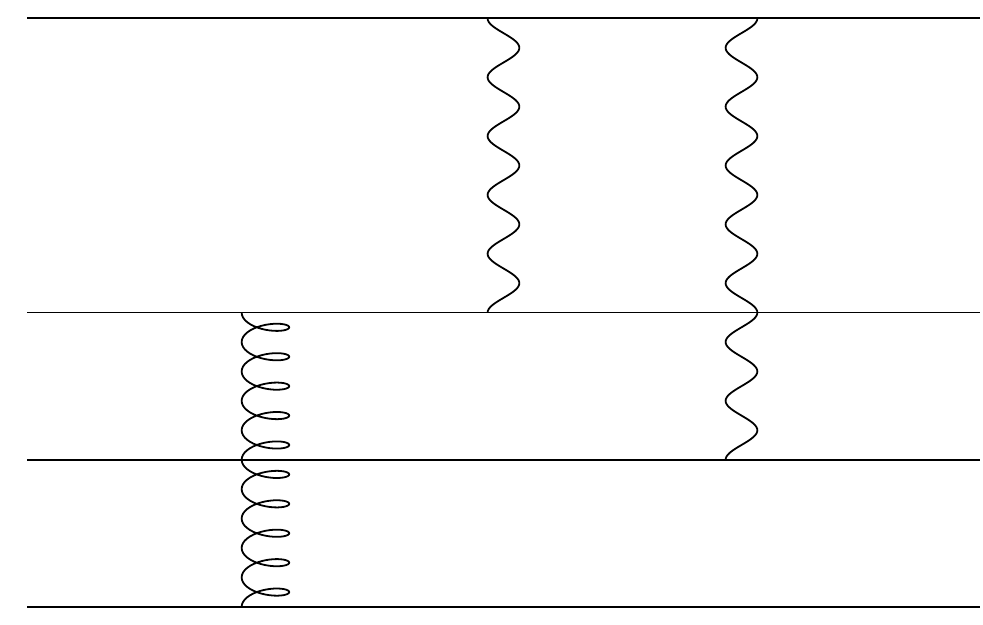}\hfil%
\includegraphics[width=0.4\linewidth]{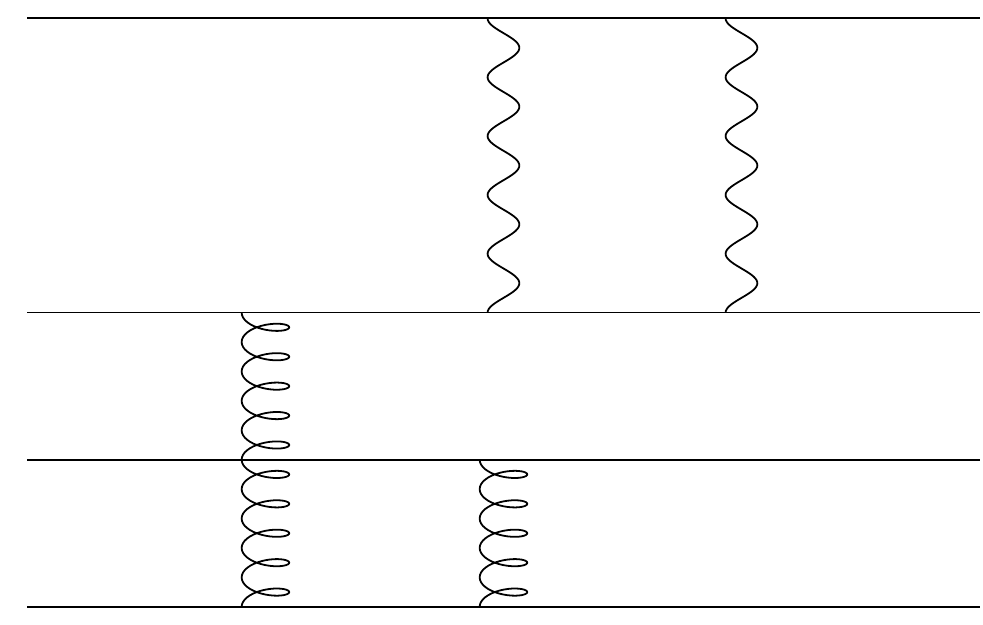}
\end{minipage}
\caption{\textit{Left:} TPE diagram in the GPD-based approach to $eN$
	scattering at high $Q^2$~\cite{Chen:2004tw, Afanasev:2005mp}. Both
	photons interact with the same quark, while the others are spectators.
	\textit{Right:} Sample TPE diagrams in the QCD factorization approach.
	For the leading order term the photons interact with different quarks,
	with a single gluon exchange. The interaction of two photons with the
	same quark is of subleading order in this approach, as it involves two gluons.
	Figures taken from Ref.~\cite{Borisyuk:2008db}.}
\label{fig:partonic}
\end{figure}

The partonic approach for TPE calculations was developed in
Refs.~\cite{Chen:2004tw, Afanasev:2005mp}. It is based on applying a formalism
of Generalized Parton Distributions (GPD) to the amplitude of wide-angle nucleon
Compton scattering~\cite{Radyushkin:1998rt}. This approach assumes that a
handbag mechanism is dominant (Fig.~\ref{fig:partonic}), with the electron-parton
scattering proceeding on an individual parton. parametrizations of GPDs are
obtained by fitting the global data on nucleon form
factors~\cite{Guidal:2004nd}.

Calculations of TPE using quark degrees of freedom for large transferred momenta
include the QCD factorization approach~\cite{Borisyuk:2008db, Kivel:2009eg}, and
parametrizations backed by Soft-Collinear Effective Theory
(SCET)~\cite{Kivel:2012vs}. Borisyuk and Kobushkin~\cite{Borisyuk:2008db}
pointed out that within a pQCD framework, the TPE diagram where the photons both
interact with the same quark is subleading in comparison to the diagram where
the photons interact with different quarks. As illustrated in
Fig.~\ref{fig:partonic}, the second diagram requires two gluons, whereas the first
diagram requires only one. Similar conclusions were reached by Kivel and
Vanderhaeghen~\cite{Kivel:2009eg}.
\begin{figure}[tb]
\centering
\begin{minipage}{0.5\textwidth}
\centering
\includegraphics[width=0.8\linewidth]{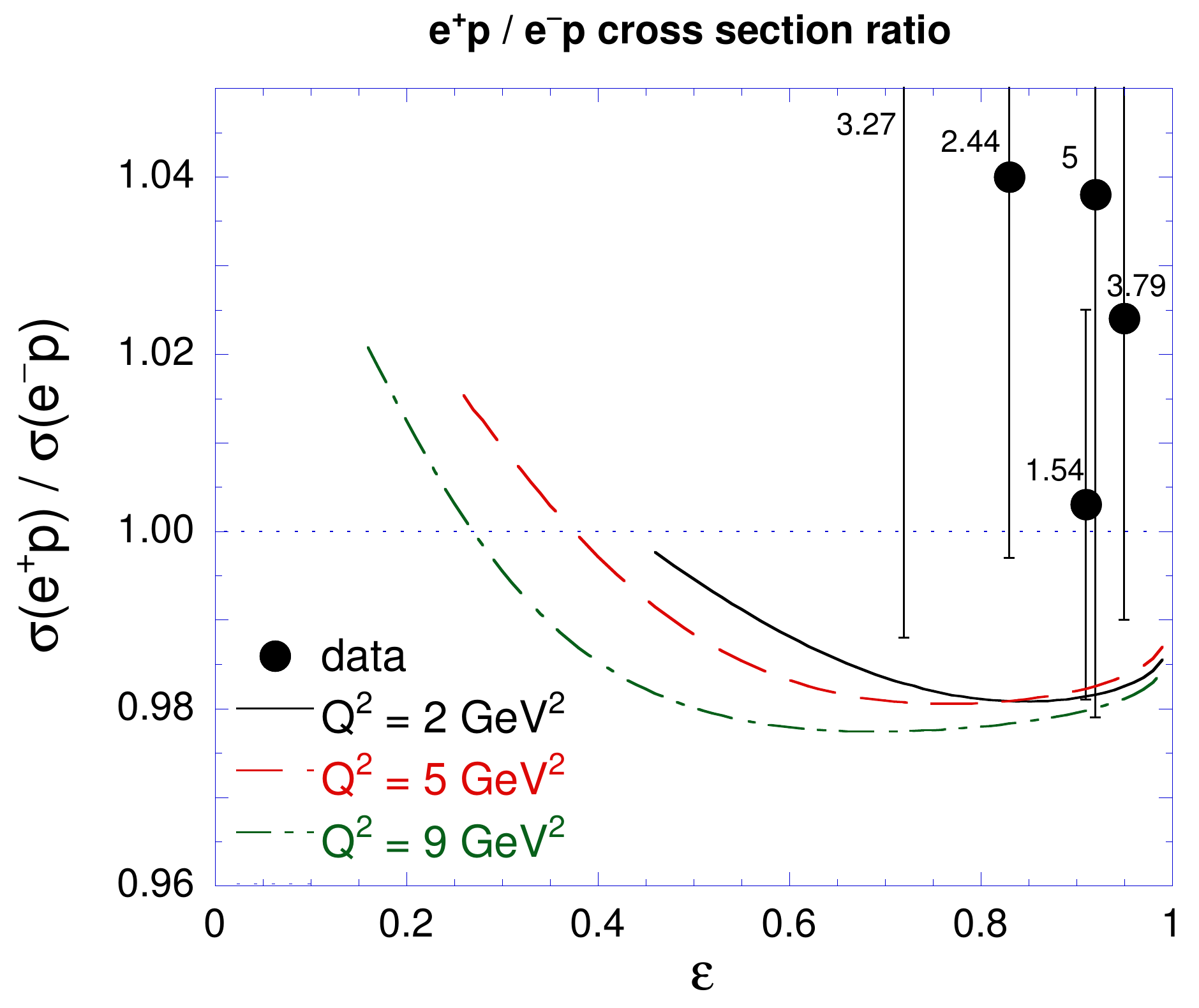}
\end{minipage}%
\begin{minipage}{0.5\textwidth}
\centering
\vspace*{3ex}
\includegraphics[width=\linewidth]{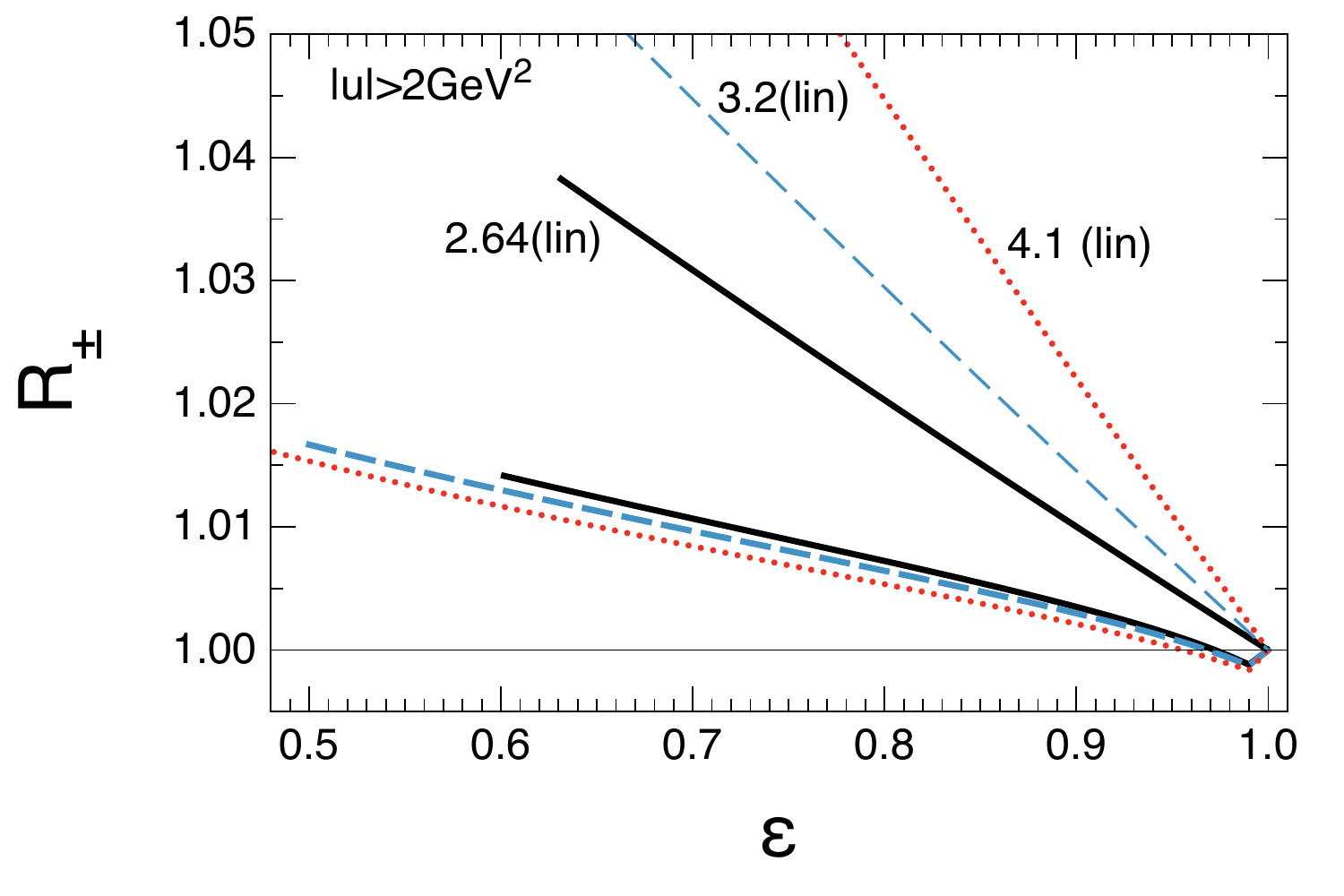}
\end{minipage}
\caption{\textit{Left:} Ratio of $e^+p/e^-p$ elastic cross sections, taken from
	Ref.~\cite{Afanasev:2005mp}.
	The GPD calculations for the TPE correction are for three fixed $Q^2$
	values of 2, 5, and 9 GeV$^2$, for the kinematical range where $-u$ is above
	$M^2$. Also shown are early SLAC data~\cite{Mar:1968qd}, with $Q^2$ above
	1.5 GeV$^2$.  The numbers near the data give $Q^2$ for that point in GeV$^2$.
	\textit{Right:} Ratio of $e^+p/e^-p$ at high $Q^2$ calculated in the QCD
	factorization approach~\cite{Kivel:2012vs}. Also shown for comparison are the
	results (labelled {\em lin}) from the from the phenomenological fits of
	Ref.~\cite{Guttmann:2010au}.
	Figure taken from Ref.~\cite{Kivel:2012vs}.}
\label{fig:positron}
\end{figure}

Kivel and Vanderhaeghen~\cite{Kivel:2012vs} calculated the TPE corrections in
the QCD factorization approach formulated in the framework of soft-collinear
effective theory. This technique allows them to develop a description for the
soft-spectator scattering contribution, which is found to be important in the
region of moderately large scales. Predictions of the GPD and QCD factorization
models for the ratio of $e^+p/e^-p$ elastic cross sections at high $Q^2$ are
shown in Fig.~\ref{fig:positron}. Also shown for comparison are the predictions from
a phenomenological fit to polarization and cross section
data~\cite{Guttmann:2010au}. The predictions from phenomenological fits are
considerable larger than the QCD factorization predictions.

\subsection{Two-photon exchange for spin polarization effects}
\label{ssec:spin}

Two-photon exchange also has an effect on the polarization observables of
electron-nucleon scattering. For double-polarization correlations, with both
polarized electrons and a polarized recoil nucleon (or the nucleon target),
TPE alters the dependence of the target asymmetry or recoil polarization, given
in Born approximation by the expressions of Eqs.~(\ref{eq:sigL}) and
(\ref{eq:sigT}). This is caused by the additional spin dependence of the real
part of the TPE amplitude beyond soft-photon contributions,
Eqs.~(\ref{eq:delMTsai}) and (\ref{eq:delMTjon}). The ratio of the recoil proton
polarizations in the scattering plane can be expressed in terms of the real part
of the generalized TPE form factors of
Eq.~(\ref{eq:genFF})~\cite{Guichon:2003qm, Afanasev:2005mp}.

\begin{figure}[tb]
\begin{center}
\includegraphics[width=.5\textwidth, trim = 14mm 42mm 15mm 23mm, clip]{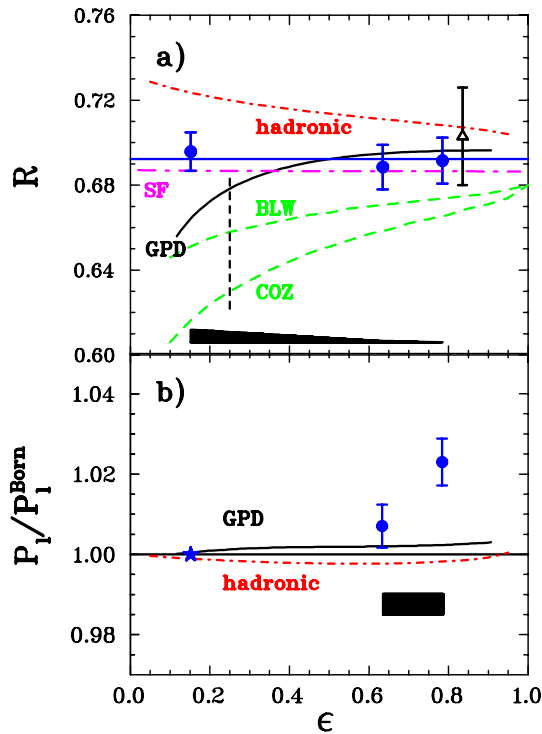}
\end{center}
\caption{
a) Experimental results for $R$ as a function of $\epsilon$ with
statistical uncertainties, from Ref.~\cite{Meziane:2010xc}. The theoretical
predictions are from: \cite{Blunden:2009dm} (hadronic), \cite{Afanasev:2005mp}
(GPD), \cite{Kivel:2009eg} (COZ and BLW).
b) $P_{\ell}/P^{Born}_{\ell}$ as a function of $\epsilon$. The point-to-point
systematic uncertainties, shown with a band in both panels, are relative to the
largest $\epsilon$ kinematic in a) and relative to the smallest $\epsilon$
kinematic in b). The star indicates the $\epsilon$ value at which the analyzing
power is determined.}
\label{ratiopl}
\end{figure}

Experimental tests were conducted at Jefferson Lab~\cite{Meziane:2010xc}, and
the results are shown in Fig.~\ref{ratiopl}. While the polarization ratio is
consistent with the Born approximation within experimental errors, the longitudinal
polarization transfer $P_L$ revealed a noticeable deviation at forward scattering
angles.

In the plane-wave Born approximation, single-spin asymmetries (SSA) are zero due to
time-reversal invariance and parity conservation in the electromagnetic
interaction. But TPE is capable of generating SSA via the imaginary part of the
amplitude. The corresponding spin-momentum correlation can be defined by a
parity-even product $\bm \xi\cdot (\bm k \times \bm k')$, where $\bm{k}\,
(\bm k')$ is the initial (final) electron momentum, and the vector $\bm \xi$ defines
the polarization vector of either the electron or the nucleon. This form implies
that the scattering asymmetry arises due to the polarization component oriented
normal to the scattering plane.
   
A similar asymmetry in a pure QED process of electron-muon scattering due to
two-photon exchange was calculated by Barut and Fronsdal~\cite{Barut:1960zz}.
Later calculations~\cite{Dixon:2004qg} for Moller scattering $e^-+e^-\to
e^-+e^-$ also included radiative corrections to the asymmetry. SLAC experiment
E158 confirmed the theoretical predictions within the statistical
uncertainty~\cite{Anthony:2005pm}: $A_n({\rm exp})=7.04\pm 0.25$~(stat)~ppm
versus $A_n({\rm theory})=6.91\pm 0.04$~ppm.

Early calculations of SSA in elastic $ep$-scattering were done by De Rujula~\etal
~\cite{DeRujula:1972te, DeRujula:1973pr} for the case of a transversely
polarized proton target. They pointed out that a nonzero SSA is due to the absorptive
part of the non-forward Compton amplitude for off-shell photons scattering from
nucleons.  Pasquini and Vanderhaeghen~\cite{Pasquini:2004pv} modelled the virtual
Compton amplitude by single-pion intermediate states, in the approach that is
well justified by $S$-matrix unitarity for energies below the threshold of
two-pion production. They made predictions for both beam and target SSA, and
obtained good agreement with MAMI data~\cite{Maas:2004pd}.

In the formalism of Refs.~\cite{Chen:2004tw, Afanasev:2005mp}, the SSA, denoted
as $A_n$, can be expressed in terms of the imaginary part of the generalized TPE
form factors of Eq.~(\ref{eq:genFF})~\cite{Guichon:2003qm, Afanasev:2005mp}:
\begin{eqnarray}
A_n &=& \sqrt{2 \, \eps \, (1+\eps )\tau} \,
	\frac{1}{\sigma_R^{\rm meas}} 
	\left\{ - \, G_M \, \Im 
	\left(\tilde F_1-\tau\tilde F_2 + \nu \tilde F_3 \right)
\right.\nonumber \\
	&+ &\, \left. G_E \, \Im \left(\tilde F_1 + \tilde F_2
	+ \left( \frac{2 \eps}{1 + \eps} \right) 
	\nu \tilde F_3 \right) \right\} 
	       \, .
\label{eq:tnsa}
\end{eqnarray}

\begin{figure}[tb]
\begin{center}
\includegraphics[width=0.5\linewidth]{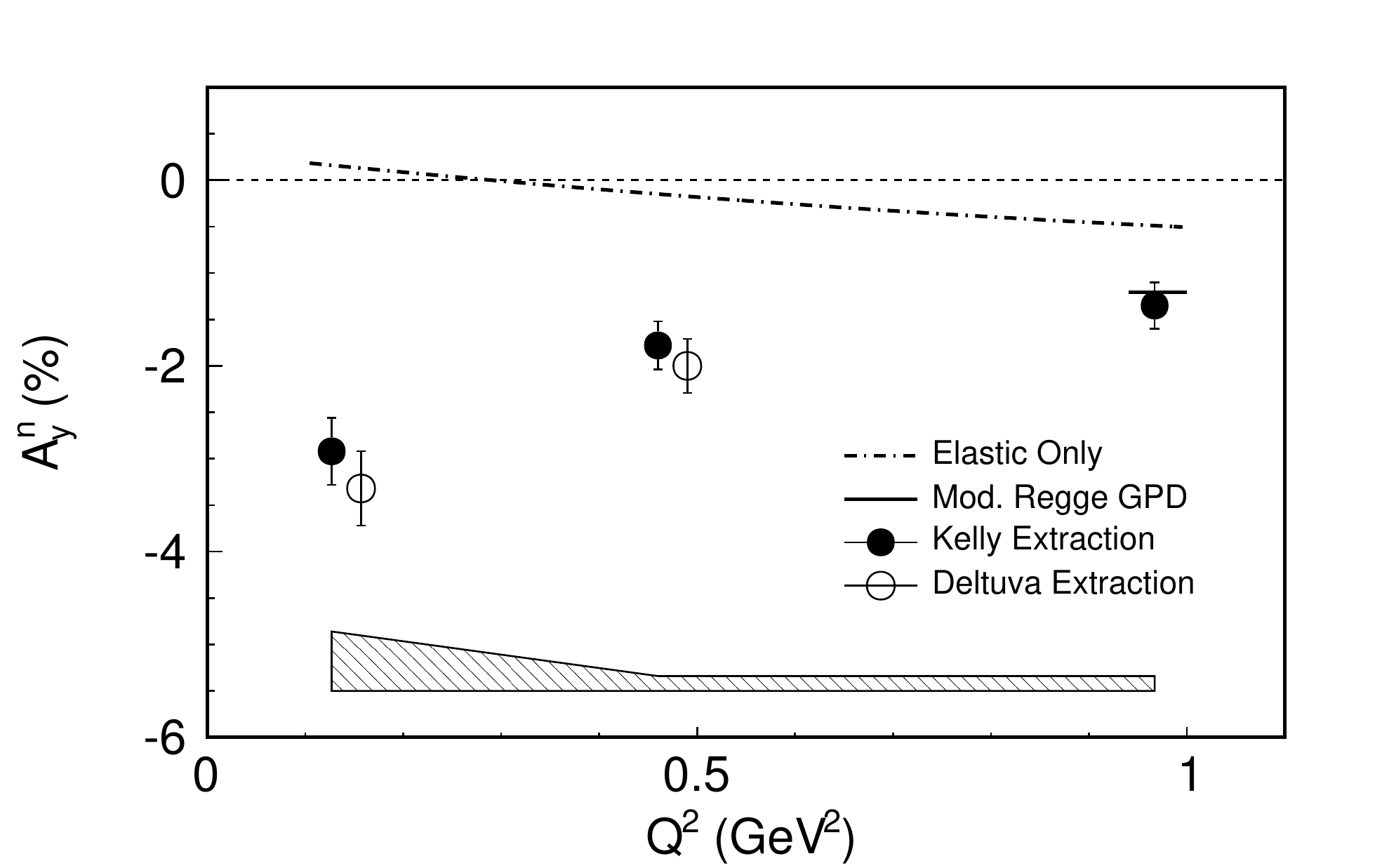}
\caption{
Results for the neutron asymmetries~\cite{Zhang:2015kna} $A\mathrm{_n ^n}$, as
a function of $Q^2$.  Uncertainties shown on the data points are statistical,
while the systematic uncertainties are shown by the band at the bottom.  The
elastic contribution to the intermediate state is shown by the dot-dash
line~\cite{Afanasev:2002gr}, and at $Q^2=0.97$~GeV$^2$, the GPD calculation of
Refs.~\cite{Chen:2004tw, Afanasev:2005mp} is shown by the short solid line.}
\label{fig:normal-ABC}
\end{center}
\end{figure}

The asymmetry $A_n$ was measured in Jefferson Lab experiment
E05-015~\cite{Zhang:2015kna} on a polarized $^3$He target, designed to obtain
the asymmetry on a single neutron. The results are compared with the theoretical
predictions~\cite{Chen:2004tw, Afanasev:2005mp} in Fig.~\ref{fig:normal-ABC} at
$Q^2\approx 1$~GeV$^2$, where the GPD approach is believed to be applicable.

If elastic scattering occurs on an unpolarized target, but the electron beam
has a polarization component normal to the scattering plane, the corresponding
asymmetry is also nonzero due to TPE. Transverse beam SSA for scattering in the
Coulomb field of a nucleus was calculated a long time ago by Mott~\cite{Mott29}.
Examples of more recent work for the scattering on large-$Z$ nuclei may be found
in~\cite{Cooper:2005sk, Jakubassa-Amundsen:2014oaa}, where the Dirac equation is
solved for the electron wave function in Coulomb field of nuclei. For low-$Z$
targets both TPE and Distorted-Wave Born Approximation calculations yield the
same results for the beam SSA to order {\cal O}($\alpha$) that has the following
behavior at small scattering angles $\theta_e <<1$:
\begin{eqnarray}
\label{eq:forward-coulomb}
A_n^e \propto \frac{\alpha m_e \theta_e^3}{E_e}\, .
\end{eqnarray}

For electron scattering at GeV energies this formula predicts rather small
asymmetries, around $10^{-10}$ for $\theta <10^\circ$ at Jefferson Lab
conditions~\cite{Cooper:2005sk}. However, when we consider inelastic excitations
of a nucleon, the predictions for the asymmetry are dramatically different.
Theoretical calculations for a nucleon target~\cite{Afanasev:2004hp,
Afanasev:2004pu, Gorchtein:2005za} predicted that above the nucleon resonance
region the beam SSA (a) does not decrease with beam energy, and (b) is enhanced
by large logarithms due to exchange of hard collinear virtual photons. The
expression for the asymmetry is the simplest in the diffractive regime and small
scattering angles, where the virtual Compton amplitude can be related via the
optical theorem to the total cross section of photoproduction on a nucleon by
real photons, $\sigma_{\gamma p}$, that varies only slowly with energy. It allows
for an exact calculation of the loop integral for the imaginary part of TPE
amplitude, resulting in the following expression for the beam asymmetry in the
high-energy diffractive regime:
\be
\label{eq:asymp-tsna}
A_n^e=-\sigma_{\gamma p} \frac{m_e \sqrt{Q^2}}{8 \pi^2} \frac{G_E}{F_1^2+\tau
F_2^2} \left(\log{\left(\frac{Q^2}{m_e^2}\right)}-2\right)
\exp{(-b Q^2)}\, ,
\ee
where the parameter $b$ describes the slope of a non-forward Compton amplitude
for the nucleon target, and $F_{1(2)}$ are Dirac (Pauli) form factors of a
nucleon. If we compare the magnitude of the beam SSA predicted by
Eqs.~(\ref{eq:forward-coulomb}) and (\ref{eq:asymp-tsna}), one can see that for
small angles and high energies the diffractive mechanism may exceed the Coulomb
one by several orders of magnitude. We can also apply this approach in the
nucleon resonance region, where the photoabsorption cross section, $\sigma_{\gamma p}$,
strongly varies with photon energy. In this case the corresponding factor in the
asymmetry is an integral~\cite{Gorchtein:2008dy} over the lab photon
energy $\omega=(W^2-M^2)/(2M)$, with $W$ being an invariant mass of the
intermediate excited nucleon state:
\be
\Im A_1^{\rm inel}=\frac{1}{4\pi^2}
\frac{M}{E_{e}}
\int_0^{E_{e}} \! d\omega\ \omega\,\sigma_{\gamma N}(\omega)\,
\log{\left(\frac{Q^2}{m^2}\left(\frac{E_{e}}{\omega}-1\right)^2\right)}\, .
\label{eq:ima1_inelast}
\ee

\begin{figure}[tb]
\begin{center}
\includegraphics[width=0.6\textwidth]{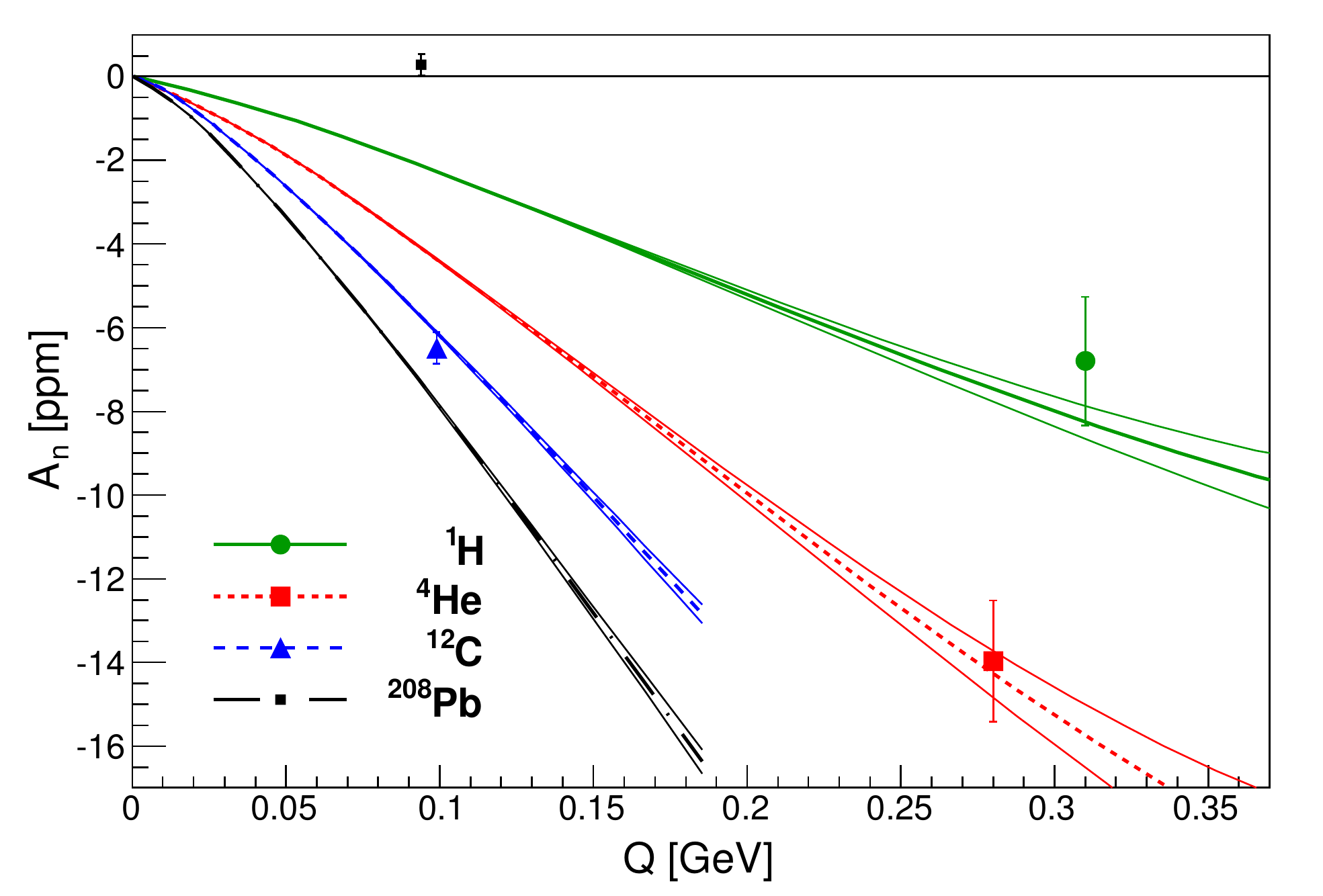}
\caption{Extracted physics asymmetries $A_n$ versus $Q$ from
Ref.~\cite{Abrahamyan:2012cg}. Each curve, specific to a particular nucleus as
indicated, is a theoretical calculation from Ref.~\cite{Gorchtein:2008dy}.}
\label{fig:HaPrex}
\end{center}
\end{figure}

As long as the electron scattering angle is small, so that the nucleon Compton
amplitude can be obtained by extrapolation from the forward limit, this
unitarity-based approach gives a good description of experimental data that
accompany the measurements on parity-violating electron scattering for a broad
range of electron energies. Extension to the case of nuclei is straightforward,
with the cross section replaced by the corresponding photonuclear cross
section~\cite{Gorchtein:2008dy}. In particular, there is good agreement with
data from HAPPEX and PREX experiments at Jefferson Lab~\cite{Abrahamyan:2012cg}
obtained on a proton and light nuclei, $^4$He and $^{12}$C
Fig.~\ref{fig:HaPrex}. However, a significant disagreement with theory was
observed for a high-$Z$ target $^{208}$Pb both in sign and magnitude:
$A_n=0.28\pm 0.25$~ppm (experiment) against $\approx-8$~ppm (theory).  A
possible reason for the disagreement is an effect of Coulomb distortion that
grows linearly with a charge of a nucleus and may become significant for this
case. A theoretical approach that combines Coulomb distortion and
intermediate-state inelastic excitations is required for this case, while
experiments with intermediate-mass nuclei, such as
$^{48}$Ca~\cite{Horowitz:2013wha}, could provide valuable information on
transition between different dynamical mechanisms for the asymmetry generation.
In the meantime, new preliminary results from the QWeak collaboration at Jefferson
Lab~\cite{Waidyawansa:2016znm} appear to be in good agreement with theory. In
summary, SSA on the nucleon and nuclei provide valuable new information on the
nucleon Compton amplitude and multi-photon effects in scattering on nuclei.

\section{Experimental observation of TPE}
\label{sec:expt}

\subsection{Nonlinearity of Rosenbluth data}
\label{ssec:indirect}

One technique employed as an indirect determination of TPE is to search for
nonlinearity in the Rosenbluth data as a function of $\eps$. In the OPE
approximation, the reduced cross section $\sigma_R$ of \cref{eq:sigmaR} depends
linearly on $\eps$. From \cref{fig:del_gg} it is clear that $\delta_{\g2}$
itself varies {\em at least} linearly with $\eps$.  Therefore the measured
reduced cross section $\sigma_R^{\rm meas}$ of \cref{eq:sigmaRmeas} will
have a quadratic or higher order dependence on $\eps$. Assuming a parametrization
of the form $\delta_{\g2}=A(1-\eps)$, consistent with the condition that
$\delta_{\g2}\to 0$ as $\eps\to 1$, one finds
\be
\label{eq:nonline}
 \sigma_R^{\rm meas}\approx (1+A)\tau G_M^2+ \left[\left(1+A\right)
 G_E^2 - \tau G_M^2\right]\eps - A G_E^2 \eps^2\, .
\ee
A similar quadratic dependence would appear in elastic electron-nucleus or
inelastic electron scattering data used in $\sigma_L/\sigma_T$ separations.

\begin{figure}[tb]
\begin{center}    
\includegraphics[width=0.49\textwidth]{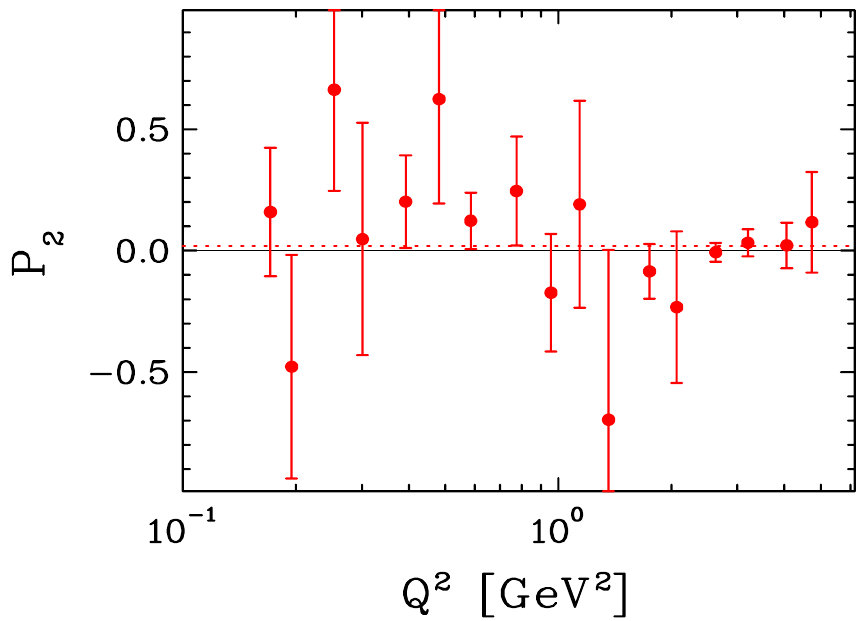}
\includegraphics[width=0.49\textwidth]{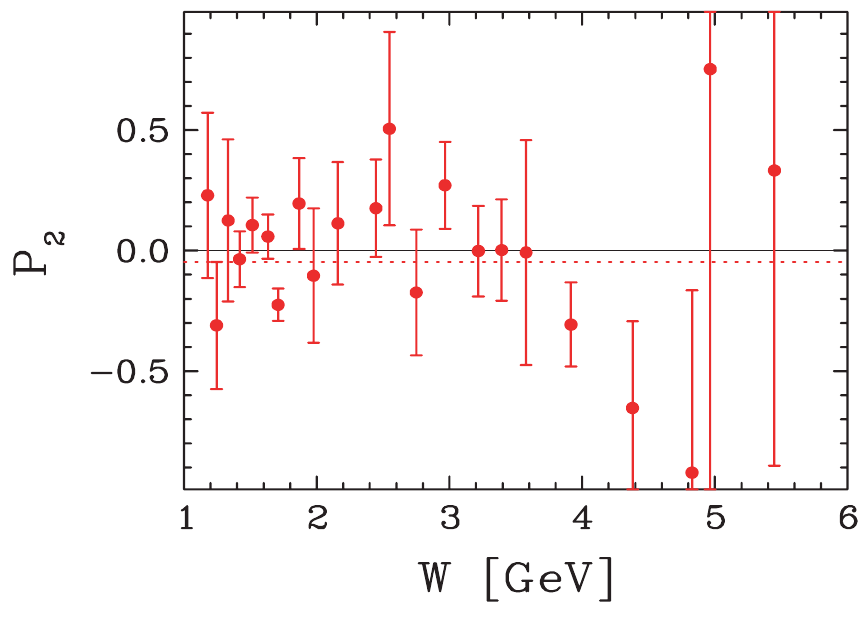}
\caption{
	Values of $P_2$ for elastic (\textit{left}) and inelastic data
	(\textit{right}) as functions of $Q^2$ and $W$, respectively.  The red
	dotted lines correspond to $\langle P_2\rangle=0.019\pm0.027$ for the
	elastic data and $-0.048\pm0.036$ for the inelastic data. Figure adapted
	from Ref.~\cite{Tvaskis:2005ex}.}
\label{fig:nonline}
\end{center}    
\end{figure}

At low values of $Q^2$, where $G_E$ dominates, any indication of a quadratic
behavior of $\sigma_R^{\rm meas}$ would be strong empirical evidence for TPE
effects, as seen by the last term in \cref{eq:nonline}.  An analysis of both
elastic and inelastic electron scattering data was performed by Tvaskis {\it et
al.}~\cite{Tvaskis:2005ex} in which they fit the reduced cross section with a
parametrization
\be
\label{eq:nonline2}
 \sigma_R=P_0\left[1+P_1\left(\eps-0.5\right)+P_2\left(
 \eps-0.5\right)^2\right]\, .
\ee
The form was chosen so that the nonlinear term, $P_2$, is a fractional
contribution relative to the average cross section. It does not require  
$\delta_{\g2}\to 0$ as $\eps\to 1$. The resulting values of
$P_2$ are shown in Fig.~\ref{fig:nonline}.  The average values, $\langle
P_2\rangle=0.019\pm 0.027$ for elastic data and $-0.048\pm 0.036$ for the
inelastic data, are both consistent with no nonlinear effects.  One also sees
that the vast majority of the points in both plots lack the sensitivity to
discern nonlinear effects.  The three most precise values of $P_2$ come from the
Jefferson Lab ``Super Rosenbluth'' experiment~\cite{Qattan:2004ht}. 
Individually, these points are entirely consistent with $P_2=0$. A calculation
of $P_2$~\cite{Chen:2004tw, Afanasev:2005mp} using partonic
calculations~\cite{Afanasev:2005ex, Chen:2004tw} yielded values consistent with
what was observed in Ref.~\cite{Tvaskis:2005ex}. Similar analyses on
$ep$~\cite{TomasiGustafsson:2004ms} and $ed$~\cite{Rekalo:1999mt} elastic
scattering found similar negligibly small nonlinearities.

\begin{figure}[tb]
\begin{center}    
\includegraphics[width=0.5\textwidth]{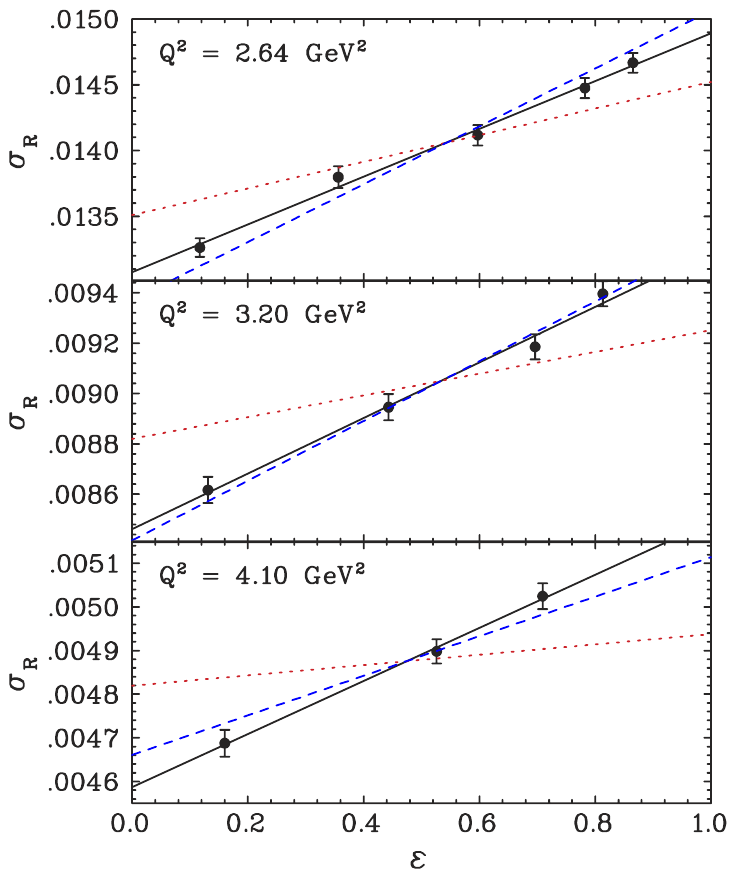}
\caption{
	Reduced cross sections with linear fit (solid line) used to extract the
	proton form factors. The dotted line indicates the slope predicted by the
	polarization transfer experiments and the dashed line shows the slope
	expected from scaling ($\mu_p G_E/G_M$).  Figure adapted from
	Ref.~\cite{Qattan:2004ht}.}
\label{fig:RosenQattan}
\end{center}    
\end{figure}

Equation~(\ref{eq:nonline}) reveals the challenge of finding evidence for TPE
effects. The nonlinear term requires one to look at low $Q^2$ where $G_E$ is
largest, but this is precisely where one expects TPE effects to be small. At
higher values of $Q^2$, where the $G_E$ term is suppressed,
Eq.~(\ref{eq:nonline}) suggests the $\eps$ dependence would come mostly from the
$G_M$ part of the second term.  However, a Rosenbluth separation by itself would
not reveal any TPE effect, but rather only lead to an incorrect extraction of
the form factors.  Since one expects a positive slope for $\delta_{\g2}$ as a
function of $\eps$, {\it i.e.} $B>0$, this would lead to an increase in the
$\eps$-dependent slope in the of the measured reduced cross section. 
Figure~\ref{fig:RosenQattan} shows the Rosenbluth separation results from
Ref.~\cite{Qattan:2004ht} that also includes lines with slopes inferred from
polarization recoil data, which are, indeed, less than the fitted slopes.

\subsection{Direct measurements of TPE corrections to $\sigma_R$}
\label{ssec:direct}

One way to directly determine the size of the TPE effect is to measure the ratio
of $e^-p$ and $e^+p$ elastic scattering cross sections~\cite{Arrington:2003ck,
Arrington:2009qd}. The TPE correction, $\delta_{\g2}$, arising from the
interference between one- and two-photon exchange amplitudes, has opposite signs
for electrons and positrons, whereas most of the other radiative corrections
cancel to first order in the ratio. (From \cref{eq:delta_TPE}, ${\cal M}_\gamma$
changes sign for positrons, whereas ${\cal M}_{\g2}$ does not.) The largest of
the corrections that don't cancel in the ratio is the interference between
lepton and proton bremsstrahlung radiation, $\delta_{{\rm b},ep}$, which is of
comparable size to $\delta_{\g2}$. Taken together, $\delta_{\rm odd} \equiv
\delta_{\g2}+\delta_{{\rm b},ep}$ constitute the charge-odd radiative
corrections.

The charge-even radiative corrections, denoted $\delta_{\rm even}$, include
logarithmically enhanced terms $\sim \log{(Q^2/m_e^2)}$ arising from electron
bremsstrahlung, vacuum polarization, and vertex corrections, and are therefore
not small. No such enhancements occur for $\delta_{\rm odd}$, so that typically
$|\delta_{\rm odd}|\ll |\delta_{\rm even}|$.

As detailed in Refs.~\cite{Rimal:2016toz, Moteabbed:2013isu}, the ratio of the
measured $e^\pm p$ elastic scattering cross sections can therefore be written as
\begin{eqnarray}
R_{\rm meas}^{\pm}=  \frac{\sigma(e^+p)}{\sigma(e^-p)} \approx
   \frac{1+\delta_{\rm even}-\delta_{\rm odd}}{1+\delta_{\rm even}+\delta_{\rm odd}}
   \approx 1 - \frac{2 \delta_{\rm odd}}{1+\delta_{\rm even}}\, .
\end{eqnarray}
After correcting the experimental ratio $R_{\rm meas}^{\pm}$ for the calculated
$\delta_{{\rm b},ep}$ and $\delta_{\rm even}$ corrections, $\delta_{\g2}$ can be
extracted from the ratio
\be
\label{eq:R2g}
	R_{2\gamma} \approx 1 - 2 \delta_{\g2}\, .
\ee

There have been several attempts to measure TPE corrections from the ratio of
positron scattering to electron scattering cross sections. These include several
experiments in the 1960's and 1970's~\cite{yount62, Browman:1965zz,
Anderson:1966zzf, Bartel:1967dsa, Cassiday67, Anderson:1968zzc, Bouquet:1968yqa,
Mar:1968qd, Hartwig:1975px}. These experiments showed no significant TPE effect,
and will be discussed in Section~\ref{ssec:oldexpt}. With the rise of the
Rosenbluth polarization transfer discrepancy there was a renewed effort at a
direct measurement of the TPE effect. Three recent experiments have released
results:
\begin{itemize}
\item The VEPP-3~\cite{VEPP-3, Rachek:2014fam} experiment used monoenergetic
beams from a storage ring incident on an internal gas target.  The VEPP-3
experiment used beams of 1.0 and 1.6~GeV with non-magnetic spectrometers at
angles from 15{\degree} to 105{\degree} and is discussed in
Section~\ref{ssec:vepp3}.
\item The CLAS TPE experiment~\cite{Adikaram:2014ykv, Rimal:2016toz} ran at the
Thomas Jefferson National Accelerator Facility (Jefferson Lab) and used a mixed
beam of positrons and electrons with beam energies from 0.8 to 3.3~GeV allowing
for simultaneous detection of electron and positron scattering events in the
CEBAF Large Acceptance Spectrometer (CLAS), while also covering a wide range in
$\eps$ and $Q^2$.  This experiment will be discussed in
Section~\ref{ssec:clas}.
\item The OLYMPUS~\cite{Henderson:2016dea}, like the VEPP-3 experiment, used a
monoenergetic beam from a storage ring incident on an internal gas target. The
beam energy was 2.01~GeV and the scattered particles were detected with the MIT
Bates BLAST detector and covered a continuous range of scattering angles from
25{\degree} to 75{\degree}.  OLYMPUS is discussed in Section~\ref{ssec:olympus}.
\end{itemize}

\subsection{Early measurements of TPE}
\label{ssec:oldexpt}

Although even early on the size of the TPE effect was expected to be small, there
were several attempts during the 1960s and 1970s to do a direct measurement of
it by measuring the ratio $R_{2\gamma}$. The earliest of these was Yount {\it et
al.}~\cite{yount62} in 1962 using the Stanford Mark III electron accelerator. 
The electrons were either sent directly to the target or used to produce a
positron beam. The leptons passed through a reversible-field momentum-analyzing
magnet on their way to a liquid hydrogen target with beam currents determined by
a Faraday cup and, independently, by ion chambers in the beamline. Scattered
leptons were detected in a magnetic spectrometer. Four data points were
measured, one at 205 MeV and a scattering angle of 30{\degree} and three at 307
MeV and scattering angles of 30{\degree}, 45{\degree}, and 130{\degree}. All
four data points were consistent with $R_{2\gamma}=1$.
\begin{figure}[tb]
\begin{center}
\includegraphics[width=0.7\textwidth]{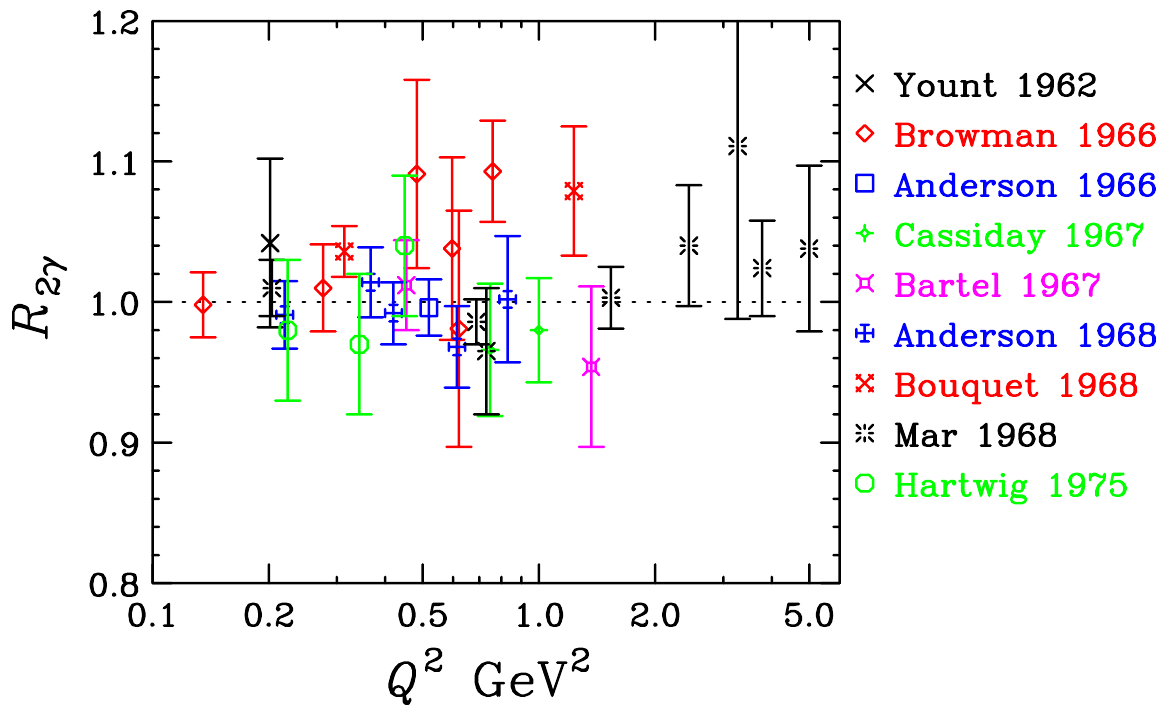}
\caption{
	$R_{2\gamma}$ data from the 1960's and 70's. Data are plotted with a
	logarithmic scale in $Q^2$. The key for the points are shown for Yount 1966
	\cite{yount62}, Browman 1966 \cite{Browman:1965zz}, Anderson 1966
	\cite{Anderson:1966zzf}, Cassiday 1967 \cite{Cassiday67}, Bartel 1967
	\cite{Bartel:1967dsa}, Anderson 1968 \cite{Anderson:1968zzc}, Bouquet 1968
	\cite{Bouquet:1968yqa}, Mar 1968 \cite{Mar:1968qd}, and Hartwig 1975
	\cite{Hartwig:1975px}.}
\label{fig:OldData}
\end{center}
\end{figure}

Following Yount~\etal~\cite{yount62}, a series of eight more electron-positron
experiments were conducted over the next 13 years using much the same technique.
 These included experiments at the Stanford Mark III
accelerator~\cite{Browman:1965zz}, the Cornell
Synchrotron~\cite{Anderson:1966zzf, Anderson:1968zzc, Cassiday67},
DESY~\cite{Bartel:1967dsa, Hartwig:1975px}, the Orsay Linear
Accelerator~\cite{Bouquet:1968yqa}, and SLAC~\cite{Mar:1968qd}.  The results of
these experiments are shown in Fig.~\ref{fig:OldData} as a function of $Q^2$
though the data vary in their value of $\eps$.  It is clear from this
plot that these measurements are statistically consistent with $R_{2\gamma}=1$. 
If, as expected, the TPE effect is only a few percent, the uncertainties of
these results are simply too large to discern the effect.  It must be said, that
these measurements are extremely difficult, even with today's technology, and
maintaining precision control of systematic uncertainties is daunting as is
obtaining the statistical accuracy needed for a few percent measurement.

\begin{figure}[tb]
\begin{center}
\includegraphics[width=0.7\textwidth]{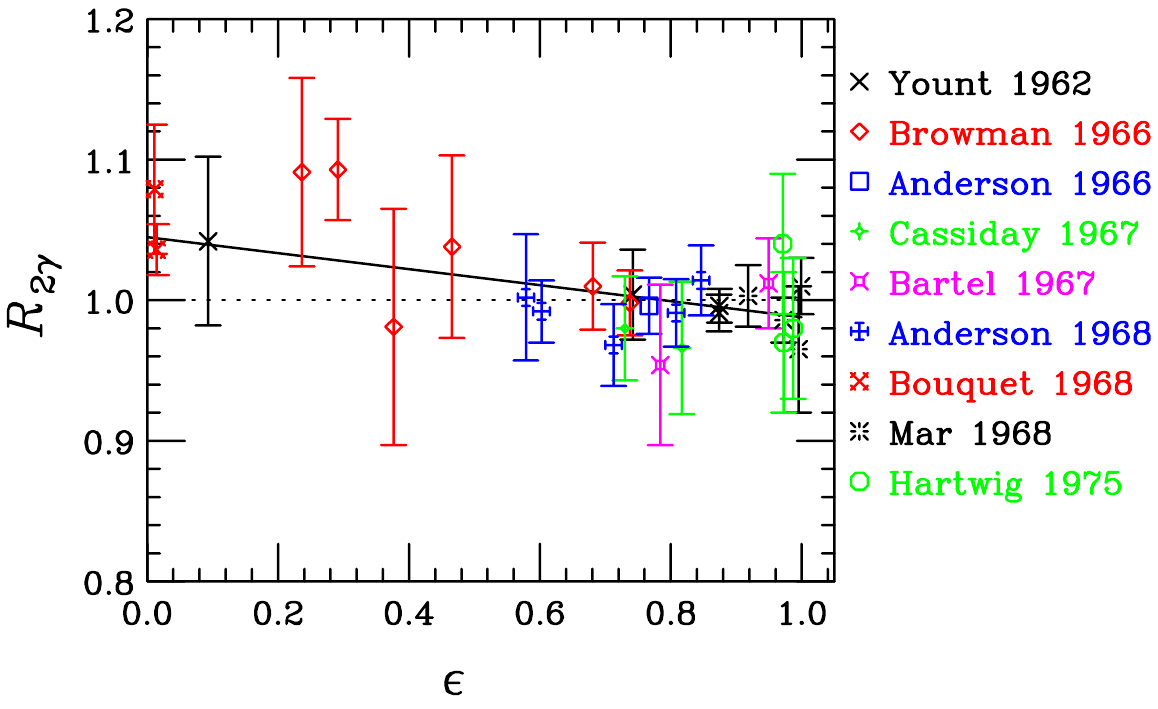}
\caption{
	$R_{2\gamma}$ data with $Q^2< 2.0$~GeV$^2$ from the 1960's and 70's. The key
	for the points are the same as in Fig.~\ref{fig:OldData}. The line is a
	reproduction of the fit from Ref.~\cite{Arrington:2003ck}. }
\label{fig:OldDataEps}
\end{center}
\end{figure}

Soon after it became apparent that there was a discrepancy between Rosenbluth
and polarization-transfer measurement of the proton form factors, these data
were reexamined~\cite{Arrington:2003ck}.  While no discernible $Q^2$ dependence
was found, a linear fit as a function of $\eps$ for $Q^2<2$~GeV$^2$
resulted in a 5.7\% increase of $R_{2\gamma}$ with decreasing $\eps$. 
Though the uncertainty on the fit was quoted as $\pm 1.8$\%, the fit is largely
driven by just few low $\eps$ data points, most of which have $Q^2< 0.5$,
which is well below where there is any significant discrepancy in the form
factor data. Given the lack of detail in most of the papers from which these
data came, specifically regarding the radiative corrections, it is unclear if
the data have been consistently corrected. A plot of these data as a function of
$\eps$ are shown in Fig.~\ref{fig:OldDataEps} along with a reproduction
of the fit from Ref.~\cite{Arrington:2003ck}.

\subsection{TPE in experiments with spin}
\label{ssec:exptwspin}

As discussed in \cref{ssec:TPE}, TPE can have a significant effect on the
Rosenbluth or longitudinal-transverse (LT) separation technique's determination
of $\mu_p G_E/G_M$ while having a lesser effect on the results obtained using
polarization transfer techniques (PT).  Nevertheless, as outlined in
\cref{ssec:spin}, TPE must also be taken into account in spin-dependent
electron-nucleon scattering as well as elsewhere.  A number of experiments
measuring polarization observables have investigated the role of TPE.  Some of
these experiments will be briefly reviewed in this section.

The GEp2$\gamma$ Collaboration~\citep{Meziane:2010xc} investigated effects
beyond one-photon exchange (OPE) or the Born approximation by measuring
polarization transfer in the reaction H$(\vec{e},e^\prime\vec{p})$.  The
experiment ran in Hall~C at Jefferson Lab using a longitudinally polarized
electron beam with 82\%--86\% polarization incident on a 20~cm long liquid
hydrogen target. Scattered electrons were detected in the BigCal lead glass
calorimeter.  The coincident protons were detected in the High Momentum
Spectrometer and two, 55~cm thick, CH$_2$ analyzer blocks in the spectrometer's
focal plane determined the proton polarization. Three electron beam energies
were used: 1.87, 2.84, and 3.63~GeV and the detector angles were adjusted to
obtain results at a common $Q^2=2.49$~GeV$^2$.  The data were analyzed to
determine the proton form factor ratio, $\mu_p G_E/G_M$, that sparked the
current interest in TPE. The results have have been shown already in
\cref{ratiopl}.

The ratio is consistent with the Born approximation over the measured range
$0.152<\eps<0.785$ at the 1.5\% level suggesting little or no effect at this
$Q^2$ or possibly that TPE effects cancel in the ratio. They also measured the
longitudinal polarization transfer $P_L$, which showed a significant ($\sim2$\%)
increase with $\eps$ compared to the Born expectation.  This implies a similar,
cancelling behavior for $P_T$ to explain the result for $R$ and that TPE may not
be negligible in polarization transfer observables at forward angles.

The SAMPLE experiment~\citep{Wells:2000rx} ran at MIT-Bates.  The experiment
employed polarized, 200~MeV electrons incident on a 40~cm long liquid hydrogen
target.  The polarization was $\sim36\%$. The electrons, scattered at back
angles $130\degree<\theta_e<170\degree$, were detected in a large acceptance
($\sim1.5$~sr) air \v{C}erenkov detector.  Initially, SAMPLE ran with
longitudinally polarized electrons and measured the parity violating asymmetry,
$A_{PV}$, to constrain the strange weak magnetic form
factor~\citep{Spayde:1999qg}. Later, transversely polarized electrons were
employed.  Single spin asymmetries (SSA) as already pointed out in
\cref{ssec:spin}, vanish in the Born approximation because of time-reversal
invariance and parity conservation in purely electromagnetic interactions. 
However, TPE can produce SSA when either the beam or target is transversely
polarized.  The SAMPLE result was $A_n=-15.4\pm5.4$~ppm (parts per million) at
the average scattering angle of $146.1\degree$ corresponding to
$Q^2=0.1$~GeV$^2$.  The result is almost two standard deviations below the
expected asymmetry from a calculation that assumed only a proton propagator for
the intermediate proton state in TPE. 

Similarly, the A4 experiment~\citep{Maas:2004pd} at the MAMI accelerator in
Mainz measured $A_n$ at a similar $Q^2$ but with higher electron beam energies
of 569.3 and 855.2~MeV.  A 10~cm liquid hydrogen target was used and an
azimuthally symmetric array of PbF$_2$ crystals covered the angular range
$30\degree<\theta_e<40\degree$. The average squared momentum transfers were
0.106 and 0.230~GeV$^2$.  The A4 results are:
\begin{subequations}
\begin{eqnarray}
  A_n\left(Q^2=0.106~{\rm GeV}^2\right)&=&-8.59\pm0.89\,({\rm stat})\pm0.75\,
  ({\rm syst})\,({\rm ppm})\, ,\\
  A_n\left(Q^2=0.230~{\rm GeV}^2\right)&=&-8.52\pm2.31\,({\rm stat})\pm0.87\,
  ({\rm syst})\,({\rm ppm})\, .
\end{eqnarray}
\end{subequations}
Comparison with theoretical calculations at the time showed no agreement for the
simple proton intermediate state though calculations including $\pi N$
intermediate states were much closer but still needed some tuning.

The E158 experiment~\citep{Anthony:2005pm} at SLAC was designed to make a
precise measurement of the weak mixing angle, $\sin{\theta_W}^2$, in M{\o}ller
scattering.  The experiment scattered longitudinally polarized electrons at
50~GeV from a liquid hydrogen target and detected them 60~m downstream in an
azimuthally symmetric ring of calorimeters that were segmented both radially and
azimuthally.  The parity violating asymmetry, $A_{PV}$, observed by reversing
the helicity of the incident electron beam was measured to determine the weak
mixing angle. However, in addition to measuring the longitudinal asymmetry, E158
made a series of runs with transverse electron polarization.  The SSA, $A_n$,
was measured at E158 for both M{\o}ller and $ep$ scattering and the results
reported for Run~2 in a Technical note~\citep{Laviolette:2009aa}.  The results
are in good agreement with theoretical calculations by Dixon and
Schreiber~\citep{Dixon:2004qg}:
\begin{subequations}
\begin{eqnarray}
  A_n^{ee}({\rm expt})&=&-7.04\pm0.25\,({\rm stat})\pm0.37\,({\rm syst})\,({\rm ppm})\, ,\\
  A_n^{ee}({\rm theory})&=&-6.91\pm0.04\,({\rm ppm})\, ,\\
  A_n^{ep}({\rm expt})&=&+2.89\pm0.36\,({\rm stat})\pm0.17\,({\rm syst})\,({\rm ppm})\, ,\\
  A_n^{ep}({\rm theory})&=&+3.16\pm0.32\,({\rm ppm})\, .
\end{eqnarray}
\end{subequations}

Similarly, the G0 experiment~\citep{Armstrong:2005hs,Androic:2011rha} was
designed to measure the parity violating asymmetry, $A_{PV}$ in $ep$ elastic
scattering and in quasi-elastic scattering $ed$.  The experiment ran in Hall~C
at Jefferson Lab.  Longitudinally polarized electrons at 3.031~GeV were directed
onto a 20~cm long liquid hydrogen target. A large, super-conducting, eight
sector, toroidal spectrometer was used to momentum-analyze the reactions.  This
spectrometer was utilized in two modes: a forward angle mode detecting the
recoil protons and a backward angle mode detecting the scattered electrons. By
combining information from the electromagnetic nucleon form factors with form
factors for the neutral weak currents the contributions of the lightest quark
flavors can be extracted.

However, the G0 experiment also ran with transverse electron polarization and
extracted information on TPE in forward-angle mode~\citep{Armstrong:2007vm} and
also in backward-angle mode~\citep{Androic:2011rh}.  With the spectrometer in
forward angle mode, measurements were made at $Q^2=0.15$ and $0.25$~GeV$^2$
detecting the recoil proton, and found the SSAs
\begin{subequations}
\begin{eqnarray}
  A_n^{ep}\left(Q^2=0.15~{\rm GeV}^2\right)&=&-4.06\pm0.99\,({\rm stat})\,({\rm ppm})\, ,\\
  A_n^{ep}\left(Q^2=0.25~{\rm GeV}^2\right)&=&-4.82\pm1.87\,({\rm stat})\,({\rm ppm})\, .
\end{eqnarray}
\end{subequations}
Again, calculations using simply the proton intermediate state did not approach
the measured results while calculations including the $\pi N$ intermediate state
were better but still not in agreement.  Another calculation based on the
optical theorem and total photo-production cross section over-shot the data.
Running in backward angle mode, detecting the scattered electron in $ep$ elastic
scattering, yielded
\begin{subequations}
\begin{eqnarray}
  A_n^{ep}\left(Q^2=0.362~{\rm GeV}^2\right)&=&-176.5\pm9.4\,({\rm stat})\,({\rm ppm})\, ,\\
  A_n^{ep}\left(Q^2=0.687~{\rm GeV}^2\right)&=&-21.0\pm24.0\,({\rm stat})\,({\rm ppm})\, .
\end{eqnarray}
\end{subequations}
These results, together with results from SAMPLE and A4, were in good agreement
with the theoretical calculations of Pasquini and
Vanderhaeghen~\citep{Pasquini:2004pv} that interpreted the imaginary part of TPE
in terms of doubly virtual Compton scattering using a phenomenological analysis
of pion electroproduction observables.

The G0 Collaboration also measured $A_n$ for $ed$ scattering:
\begin{subequations}
\begin{eqnarray}
  A_n^{ed}\left(Q^2=0.362~{\rm GeV}^2\right)&=&-108.6\pm7.2\,({\rm stat})\,({\rm ppm})\, ,\\
  A_n^{ed}\left(Q^2=0.687~{\rm GeV}^2\right)&=&-55.7\pm78.0\,({\rm stat})\,({\rm ppm})\, .
\end{eqnarray}
\end{subequations}
The deuteron SSA measurements can be unfolded to estimate the contributions from
the proton, $A_n^p$, and neutron, $A_n^n$, separately:
\begin{subequations}
\begin{eqnarray}
  A_n^{p}\left(Q^2=0.362~{\rm GeV}^2\right)&=&-176.5\pm49.4\,({\rm stat})\,({\rm ppm})\, ,\\
  A_n^{p}\left(Q^2=0.687~{\rm GeV}^2\right)&=&-21.0\pm24.0\,({\rm stat})\,({\rm ppm})\, ,\\
  A_n^{n}\left(Q^2=0.362~{\rm GeV}^2\right)&=&+86.6\pm41.0\,({\rm stat})\,({\rm ppm})\, ,\\
  A_n^{n}\left(Q^2=0.687~{\rm GeV}^2\right)&=&-138.0\pm268.0\,({\rm stat})\,({\rm ppm})\, .
\end{eqnarray}
\end{subequations}

The HAPPEX and PREX collaborations~\citep{Abrahamyan:2012cg} measured the
transverse beam asymmetry for elastic electron scattering from a number of
different targets: H, $^4$He, $^{12}$C, and $^{208}$Pb.  The results have
already been shown (\cref{fig:HaPrex}) and discussed.

The QWeak Collaboration~\citep{Allison:2014tpu} is also measuring $A_n$
for a number of different targets.  For $ep$ elastic scattering at 1.155~GeV and
$Q^2=0.025$~GeV$^2$, QWeak~\citep{Waidyawansa:2016znm} found
\be
A_n^{p}\left(Q^2=0.025~{\rm GeV}^2\right)=-5.350\pm0.07\,({\rm stat})\pm0.15\,({\rm syst})\,({\rm ppm})\, .
\ee

In summary, the single spin asymmetry measurements, while small, potentially can
constrain the imaginary parts of the generalized form factors in a
spin-dependent calculation of two-photon exchange.  It is also clear that
theoretical calculations must include resonance states in the intermediate
propagator.

\subsection{The VEPP-3 experiment}
\label{ssec:vepp3}

\begin{figure}[tb]
\begin{center}    
	\includegraphics[width=0.8\textwidth]{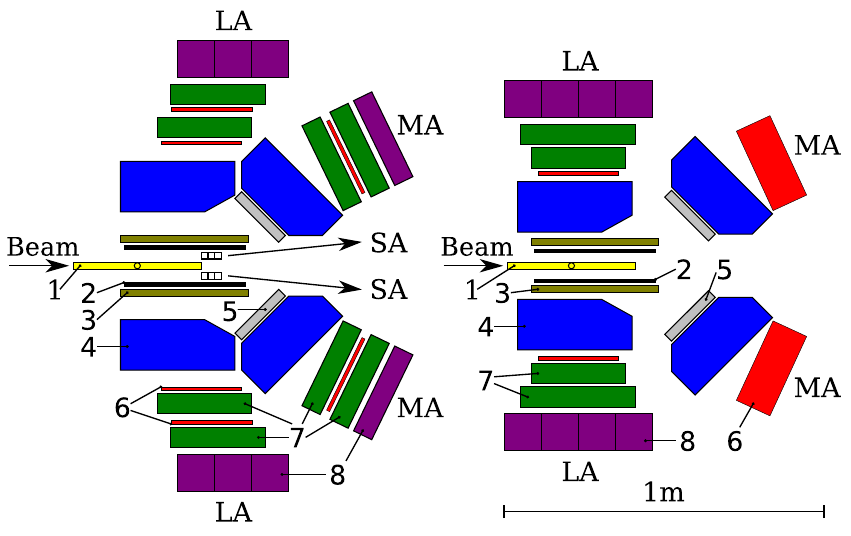}
	\caption{
	VEPP-3 TPE experiment detector configurations for the 1.6~GeV run (left) and
	the 1.0~GeV run (right). The labels refer to 1--target cell,; 2--beryllium
	sheet; 3--multiwire proportional chamber; 4--drift chamber; 5--acrylic
	glass; 6--plastic scintillator; 7--CsI crystals; 8--NaI crystals.  LA
	corresponds to the large angle spectrometer, MA to the medium angle
	spectrometer, and SA to the small angle calorimeter. Figure adapted from
	Ref.~\cite{Rachek:2014fam}.}
\label{fig:VEPPexpt}
\end{center}    
\end{figure}

The first of the new era of direct TPE experiments to take data was done using
the VEPP-3 storage ring in Novosibirisk, Russia \cite{Gramolin:2011tr,
Rachek:2014fam}. It used an internal hydrogen gas target and took data with beam
energies of 1.6 and 1.0~GeV during two separate running periods. The experiment
used non-magnetic spectrometers (shown in Fig.~\ref{fig:VEPPexpt}), which
guarantees identical acceptances for $e^+p$ and $e^-p$ events. This constituted
a relative advantage for the VEPP-3 experiment compared to the CLAS and OLYMPUS
experiments, which both used magnetic spectrometers and can lead to acceptance
differences for the two types of events. There were two pairs of left-right
symmetric spectrometers -- large angle (LA) spectrometers and medium angle (MA)
spectrometers -- along with a pair of left-right symmetric small angle (SA)
calorimeters. The LA and MA spectrometers were placed at different angles for
the two runs resulting in average lepton angles of
$\langle\theta_e\rangle=66.2\degree$ and $20.8\degree$ for the 1.6~GeV run and
$75.4\degree$ and $21.4\degree$ for the 1.0~GeV run, and were used to detect
both leptons and protons.  The components of each of the spectrometers is
described in the caption of Fig.~\ref{fig:VEPPexpt}. Scattering angles were
primarily determined by the multiwire proportional chambers and the drift
chambers, while the other components were used to measure particle energy and to
determine particle identification.

The experiment alternated between running with positron and electron beams but
did not determine an absolute positron/electron normalizations. Instead,
luminosity normalization points were take at small angles where hard TPE effects
are expected to be small~\cite{Arrington:2002cr, Arrington:2011dn} and
$R_{2\gamma}=1$. The SA arms detected only scattered leptons and were used only
during the 1.6~GeV run to determine the luminosity normalization, while MA
events were used for luminosity normalization during the 1.0~GeV run.  For the
1.6~GeV run a total of luminosity of 320~pb$^{-1}$ was recorded, while for the
1.0~GeV run the luminosity was 600~pb$^{-1}$.

Elastic events were determined by a co-planarity cut and a series of cuts on the
scattering angles and energies of the coincident lepton-proton pair. Remaining
background was removed by simulations that included the most likely background
events: $ep\rightarrow e'n\pi^+$, $ep\rightarrow e'p\pi^0$,
$\gamma^*p\rightarrow n\pi^+$, and $\gamma^*p\rightarrow p\pi^+\pi^-$. The
background was found to be less than 4\% in the LA angle range and negligible
for the other angle ranges.

The data were corrected for first-order radiative effects using a simulation
with an event generator~\cite{Gramolin:2014pva} that used a dipole
parametrization of the for the proton form factors, a QED calculation beyond
the soft-photon approximation for first-order bremsstrahlung, a vacuum
polarization correction that included the hadronic contribution, and the soft
TPE terms according to Mo and Tsai~\cite{Mo:1968cg}. The generated events were
fed into a simulation of the experiment and was run for both positrons and
electrons, with radiative effects turned on and also assuming the exchange of a
single virtual photon. The resulting ratios, $N^+_{\rm sim}/N^0_{\rm sim}$ and
$N^-_{\rm sim}/N^0_{\rm sim}$ were then applied to the measured ratio of $R$ to give
\be
\label{eq:VEPPRad}
R_{2\gamma}= R_{\rm meas}^{\pm} \frac{ N^-_{\rm sim}/N^0_{\rm sim}}{N^+_{\rm sim}/N^0_{\rm sim}}\, .
\ee

The total systematic uncertainty was very low for this experiment and varied
between 0.08\% to 0.32\%. The sources of systematic uncertainties include beam
effects, kinematic cuts, background subtraction, and radiative corrections. For
three of the four VEPP-3 data points the statistical uncertainty is a factor of
3-4 times larger than the systematic, while at the other data point the
systematic uncertainty is about twice as large as the statistical uncertainty.

\begin{figure}[tb]
\begin{center}    
\includegraphics[width=1\textwidth]{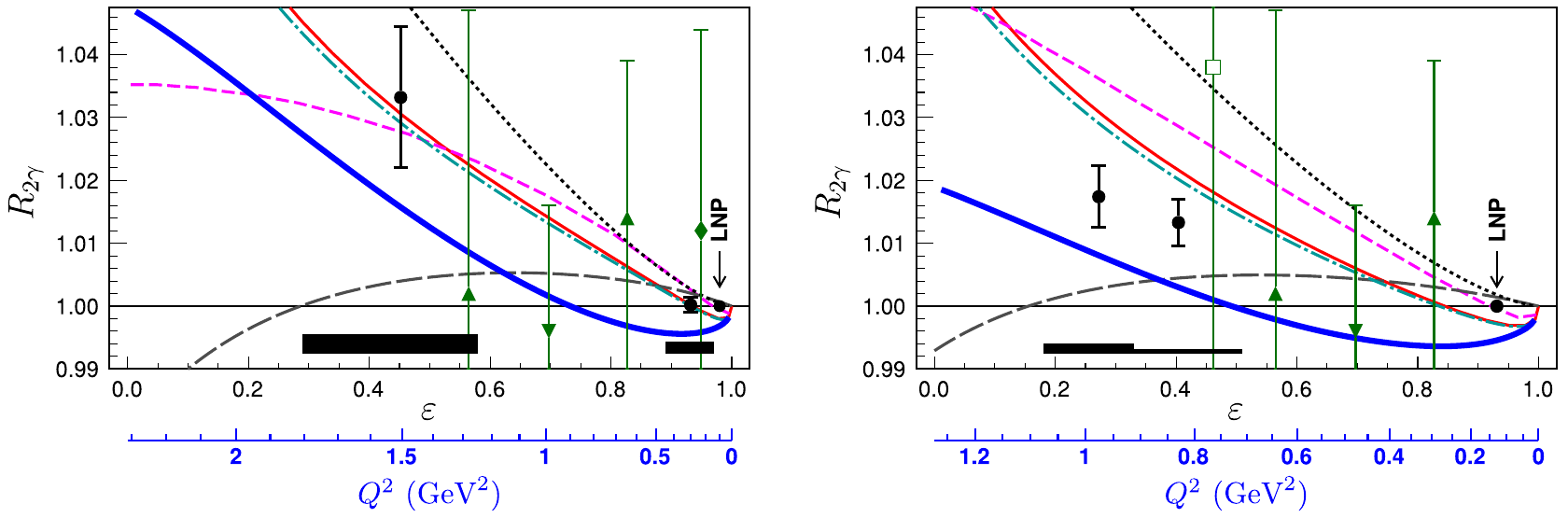}
\caption{
	$R_{2\gamma}$ as a function of $\eps$ and $Q^2$. The black circles
	are the VEPP-3 data points, while the other data points are from the early
	experiments~\cite{Browman:1965zz,
	Anderson:1966zzf, Bartel:1967dsa, Anderson:1968zzc}. The calculation curves
	are cyan dash-dotted~\cite{Borisyuk:2008es}, red solid
	line~\cite{Blunden:2005ew}, blue solid~\cite{Bernauer:2013tpr}, gray
	long-dash~\cite{TomasiGustafsson:2009pw} magenta
	short-dash~\cite{Arrington:2004is}, and black dots~\cite{Qattan:2011ke}.
	Figure adapted from Ref.~\cite{Rachek:2014fam}.}
\label{fig:VEPP3Results}
\end{center}    
\end{figure}

The results of the VEPP-3 experiment are shown in Fig.~\ref{fig:VEPP3Results}
along with older data at similar kinematics~\cite{Browman:1965zz,
Anderson:1966zzf, Bartel:1967dsa, Anderson:1968zzc} and various
predictions~\cite{Borisyuk:2008es, Blunden:2005ew, Bernauer:2013tpr,
TomasiGustafsson:2009pw, Arrington:2004is, Qattan:2011ke}. The results are in
agreement with, but significantly more precise, than the older data.

It is important to note that the results are normalized to the luminosity
normalization points taken at small scattering angles and assumed to have
$R_{2\gamma}=1$. So a direct comparison inferred from the plot does not does not
reveal how well the data and predictions match. In principle, the data should be
shifted so that the normalization point is at the value of $R_{2\gamma}$ of the
curve to which the data are being compared. The authors of
Ref.~\cite{Rachek:2014fam} performed this exercise and found reasonable
agreement with predictions of~\cite{Borisyuk:2008es, Blunden:2005ew,
Bernauer:2013tpr} with $\chi^2_\nu$ from 1.0 to 4.19, while predictions
of~\cite{TomasiGustafsson:2009pw, Arrington:2004is, Qattan:2011ke} had
$\chi^2_\nu>5$. The ``no hard TPE effect'' had $\chi^2_\nu=7.97$, which
corresponds to ruling out the ``no TPE'' hypothesis at better than a 99\% confidence
level.

%
\subsection{The CLAS experiment}
\label{ssec:clas}

\begin{figure}[tb]
\begin{center}
\includegraphics[width=0.95\textwidth]{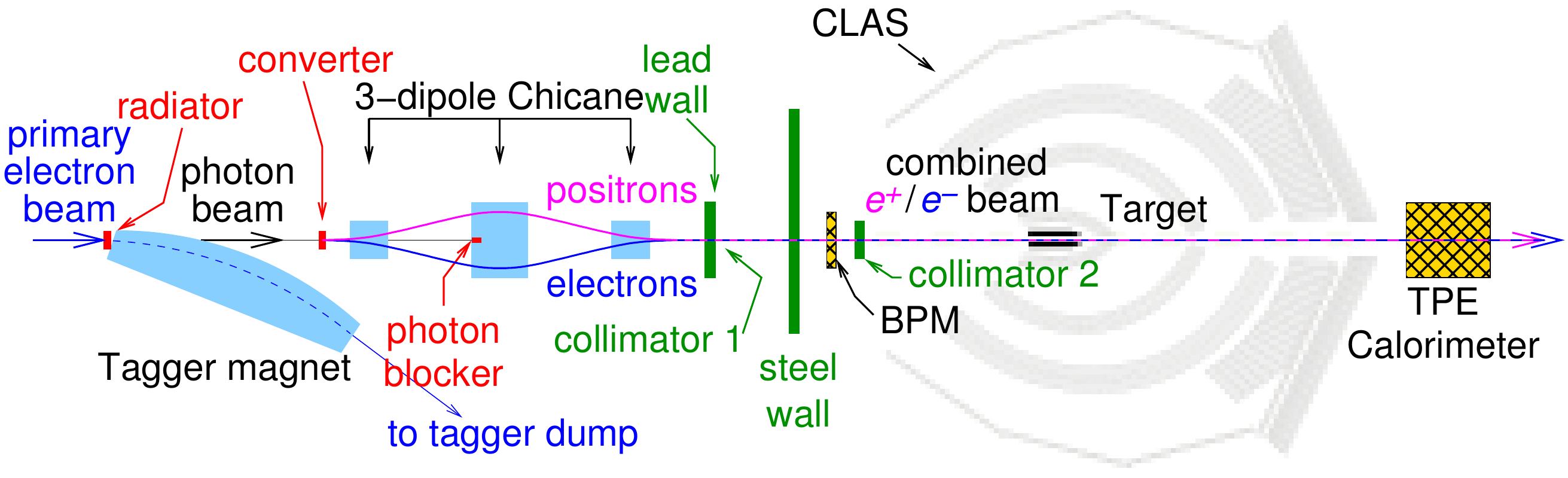}
\caption{
Beamline sketch for the CLAS TPE experiment. The chicane bends the electron and
positron trajectories in the horizontal plane, rather than the vertical plane as
shown in the figure.  The electron and positron directions are selected by the
chicane polarity. The cutaway view of CLAS shows the three regions of drift
chambers (DC), the time-of-flight scintillation counts (TOF), the \v{C}erenkov
Counters (CC), and the Electromagnetic Calorimeters (EC). The TPE Calorimeter was
removable and only placed in the beam for special calibration runs. Drawing is
not to scale.}
\label{fig:CLASExpt}
\end{center}
\end{figure}

The CLAS TPE experiment~\cite{Adikaram:2014ykv, Rimal:2016toz} ran at the Thomas
Jefferson National Accelerator Facility (Jefferson Lab). It had large coverage
of kinematics with $0.85\leq Q^2\leq 1.45$~GeV$^2$ and nearly the entire interval of
$\eps$ from 0.2 to 0.9. The experiment utilized a mixed beam of electrons
and positrons with an effective beam energy of between 0.8 and 3.5~GeV. This
mixed beam was produced by first bombarding a gold radiator (see
Fig.~\ref{fig:CLASExpt}) with a 110-140-nA, 5.6~GeV electron beam. The electrons
were diverted by the CLAS tagger magnet into a beam dump while the collimated
photon beam struck a gold converter to produce electron-positron pairs.  The
leptons then passed through a three-dipole chicane, the first of which separated
the leptons horizontally from the photon beam, which was prevented from reaching
the target by a tungsten block within the chicane. The second and third dipoles
recombined the lepton beams that were collimated en route to a liquid hydrogen
target.  The scattered leptons and the protons were detected in the CEBAF Large
Acceptance Spectrometer (CLAS)~\cite{Mecking:2003zu}.

CLAS is a nearly 4$\pi$ acceptance spectrometer divided into six sectors by
superconducting coils that produce a toroidal magnetic field in the azimuthal
direction. These magnets bent charged particles away or toward the beamline. The
polarity of the CLAS magnetic field was flipped periodically during the
experiment, thus flipping the tracks of electrons and positrons. Drift chambers
(DC), including a set within the magnetic field, measured charged-particle
trajectories enabling momentum and scattering-angle determination. Scintillation
counters (TOF) measured the charged-particle time of flight, which when combined
with momentum measurements provided particle identification. Electromagnetic
Calorimeters (EC), which were generally used in typical CLAS experiments to
measure total energy for charged and neutral particles, were used in the event
read-out trigger.

Using a mixed electron-positron beam allowed the simultaneous detection of
$e^-p$ and $e^+p$ events, thus ensuring an identical experimental configuration
for both. However, a number of unique challenges resulted. In order to produce
sufficient luminosity, the primary electron beam ran at a much higher current
(between 110 and 140 nA) than previous CLAS experiments leading to a large
radiation background from the radiator and the beam dump.  The process of
converting the photon beam to $e^+/e^-$ pairs also produced a large radiation
background.  Extensive shielding was required around the beam dump, the chicane,
and in front of CLAS to prevent the detectors from becoming overwhelmed with
background.

The symmetry of the spatial and energy spectrum of the two lepton beams was
determined by a scintillating fiber beam monitor (BM) and a removable
lead/scintillator calorimeter.  The BM was located at the upstream entrance to
CLAS and was used to adjust the chicane field settings to produce maximally
overlapping beams.  The calorimeter was inserted into the beam every time a the
chicane polarity was reversed (thus flipping which side of the chicane the
electrons and positrons passed through) to determine relative energy
distributions.  Individual electron or positron energy distributions were
measured by blocking one side of the chicane.

The experiment did not measure absolute cross sections because there was no way
to measure the absolute flux of leptons on the target. Instead, the experiment
relied upon the fact that both types of leptons are produced in equal numbers at
the relevant energies and that differences in various particle acceptances will
cancel by forming multiple ratios. For a given torus polarity, $t=\pm$, and
chicane polarity, $c=\pm$, the ratio of detected elastically-scattered
positrons, $N^+_{tc}$, and electrons, $N^-_{tc}$ was measured:
\be
\label{eq:R1a}
	R_{tc}=\frac{N^+_{tc}}{N^-_{tc}}\, .
\ee
This cancels out any proton acceptance and detector efficiency factors for the
two different lepton events. The yield $N^\pm_{tc}$ is proportional to the
elastic-scattering cross section, $\sigma^\pm$ (here $\pm$ refers to the lepton
charge), the lepton-charge-related detector efficiency and acceptance function,
$f^\pm_t$, as well as the luminosity for a given chicane polarity, $L^\pm_c$, 
so that
\be
\label{eq:R1b}
	R_{tc}=\frac{\sigma^+ f^+_t L^+_c}{\sigma^-f^-_t L^-_c}\, .
\ee
The square root of the product of measurements done with both torus polarities
but a fixed chicane polarity leads to
\be
	R_c= \sqrt{R_{+c}R_{-c}}= \sqrt{\frac{N^+_{+c}}{N^-_{+c}}
	\frac{N^+_{-c}}{N^-_{-c}}}
	= \sqrt{\frac{\sigma^+ f^+_+ L^+_c}{\sigma^- f^-_+ L^-_c}
	\frac{\sigma^+ f^+_- L^+_c}{\sigma^- f^-_- L^-_c}}
	=\frac{\sigma^+}{\sigma^-}  \frac{L^+_c}{L^-_c}\, ,  \label{eq:R2}
\ee
where it was assumed that $f^+_+=f^-_-$ and $f^+_-=f^-_+$.  That is, the
unknown detector efficiency and acceptance functions for positrons cancel those
for electrons when the torus polarity is switched and are expected to cancel out
in this double ratio.

By reversing the chicane current the spatial positions of the oppositely charged
lepton beams is swapped so that $L^+_+=L^-_-$ and $L^+_-=L^-_+$. Then taking the
square root of the product of the double ratios defined in Eq.~(\ref{eq:R2}) leads
to
\be
	R= \sqrt{R_{++}R_{-+}R_{+-}R_{--}} =\sqrt{\frac{N^+_{++}}{N^-_{++}}
	\frac{N^+_{-+}}{N^-_{-+}} \frac{N^+_{+-}}{N^-_{+-}}
	\frac{N^+_{--}}{N^-_{--}}}
	=	\sqrt{\frac{\sigma^+ L^+_+}{\sigma^- L^-_+} \frac{\sigma^+
	L^+_-}{\sigma^- L^-_-}} =\frac{\sigma^+}{\sigma^-}\, , \label{eq:Rmeas}
\ee
thus eliminating any flux-dependent differences between the two lepton beams. 

Though this process in principle eliminates any acceptance differences between
$e^+p$ to $e^-p$ events, further corrections were necessary to account for
detection inefficiencies and the fact that the experiment ran with a
``minitorus''.  The minitorus was a fixed polarity magnet system that bent
low-energy M{\o}ller electrons in the forward direction and out of the detector.
 These remaining acceptance differences were accounted for by an algorithm that
kept events only if the oppositely-charged lepton event also would have been
within the detector acceptance and also by a Monte Carlo simulation that
included the minitorus field and produced a residual acceptance correction
factor.

The identification of elastic events for this experiment relied upon the fact
that elastic scattering kinematics are overdetermined when the momenta and
scattering angles of both the scattered leptons and protons are experimentally
measured. The analysis utilized a co-planarity cut ($\delta\phi$) and a series
of three other cuts related to the kinematics of the events. The incident lepton
beam energy was unknown but was reconstructed from the measured kinematic
variables.

Though background was very small (less than 1\% of the signal) for
forward-scattered lepton events, at large angles a few percent background was
present in the final data set. The background was modelled by taking events from
the sideband of one of the energy distributions and were projected onto the
$\delta\phi$ distribution. It was found to be Gaussian in shape so a Gaussian
background model was used to removed background from all data bins.

\begin{figure}[tb]
\begin{center}    
\includegraphics[width=0.49\textwidth]{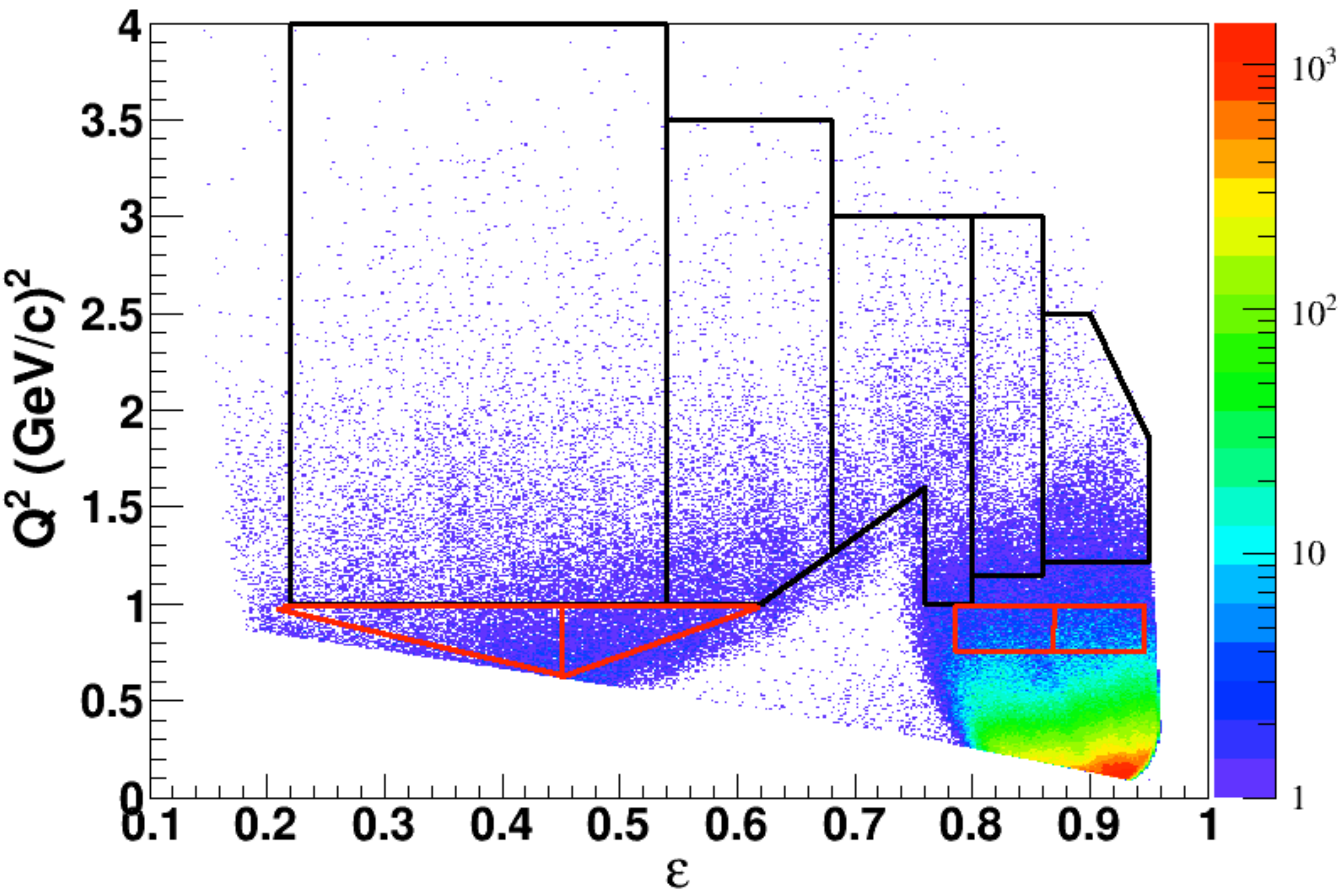}
\includegraphics[width=0.49\textwidth]{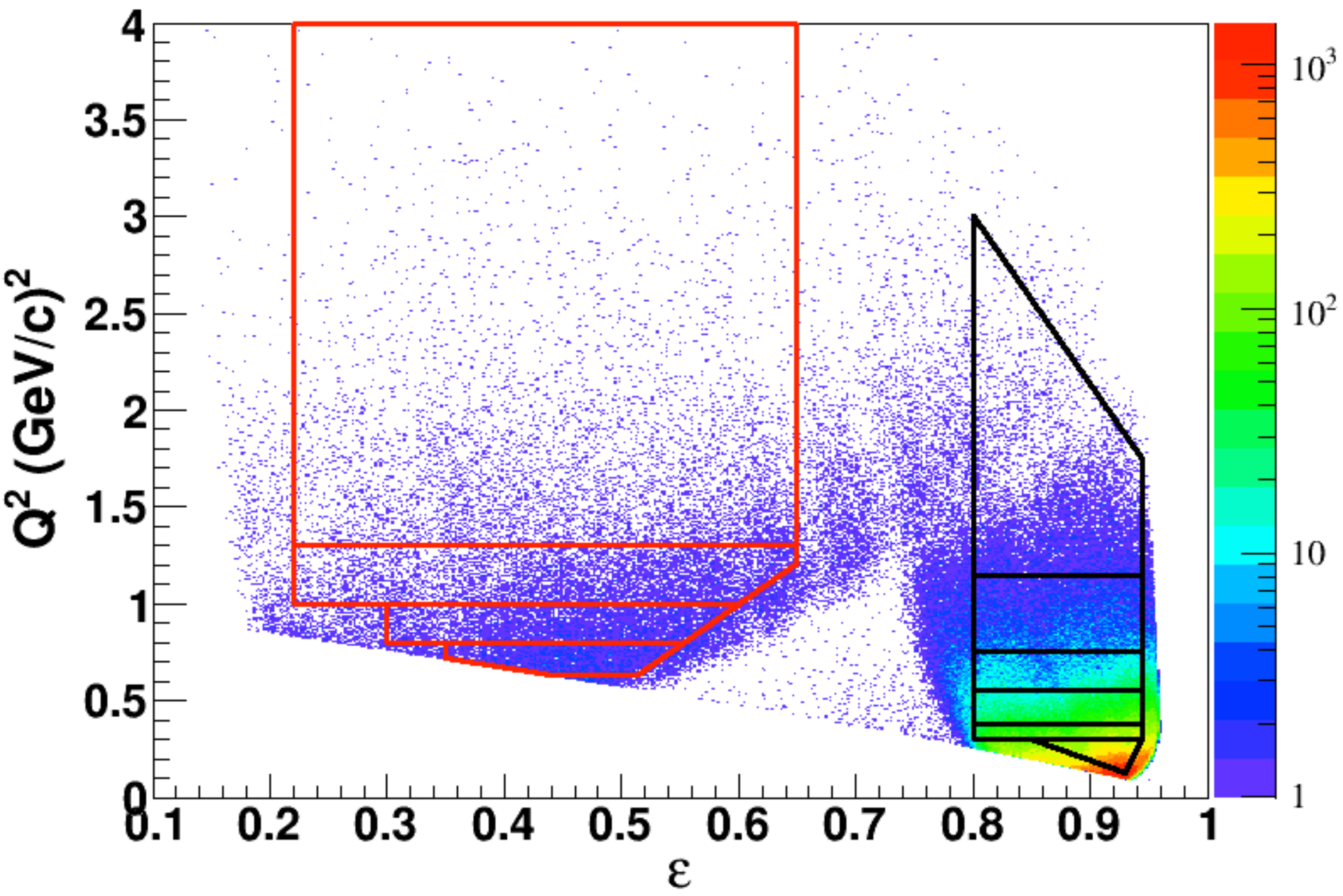}
\caption{
	Data binning in $Q^2$ and $\eps$ overlaid on positive torus $e^{+}p$
	events. The left plot shows the two sets of bins for the $\eps$
	dependence (red and black boxes for $\langle Q^2\rangle=0.85$ and 1.45
	GeV$^2$, respectively), while the right plot shows the two binning choices
	for the $Q^2$ dependence (red and black boxes for
	$\langle\eps\rangle=0.45$ and 0.85, respectively.)}
\label{fig:bins}
\end{center}    
\end{figure}

The kinematic coverage of $Q^2$ versus $\eps$ for the CLAS experiment is
shown in Fig.~\ref{fig:bins} . The hole in the distribution at
$\eps\approx 0.7$ and lower values of $Q^2$ is due to the trigger used in
the experiment, which required one particle track hitting the forward TOF and
the EC.  Events where neither particle had a lab-frame scattering angle of less
than about 45$^\circ$ did not trigger the CLAS readout.  The trigger hole is
largest for $e^+p$, positive torus events, which ultimately limited the
kinematic coverage. Data near the edges of the distributions, where the
acceptance for in-bending and out-bending particles vary rapidly, were not
included in the analysis. This binning choice led to some overlap in the data
bins so not all of the published results are independent.

The measured $e^+p$ to $e^-p$ ratio as defined in Eq.~(\ref{eq:Rmeas}) and
corrected for additional acceptance effects required correction for radiative
effects. The largest contribution is from charge-even terms, which are the same
for positrons and electrons and acts as a dilution factor in the measured ratio.
The charge-odd terms include both the TPE contribution and the interference
between real photon emission from the proton and from the leptons
(bremsstrahlung interference).  These corrections were determined by simulating
radiative effects as described in Ref.~\cite{Ent:2001hm}, using the ``extended
peaking approximation.''  Simulations were run for electron-proton scattering
and again for positron-proton scattering. The average of the two simulation
yields give the charge even correction $\delta_{\rm even}$ and the ratio of
these yields give the charge asymmetry that corresponds to the no-TPE limit of
Eq.~(\ref{eq:nonline}). The difference between uncorrected and corrected results
varied from 0.003 at high $\eps$ and low $Q^2$ to 0.034 at low $\eps$ and high
$Q^2$. The quoted uncertainties in the radiative corrections were a scale
uncertainty of roughly 0.3\% and a point-to-point uncertainty of 15\% of the
correction and were generally small compared to the statistical and other
systematic uncertainties.

The total instrumental systematic uncertainties for the ratio $R_{2\gamma}$ for
this experiment varied between 0.0042 and 0.0187. These are typically dominated
by effects of the kinematic cuts and variations of the measured ratio from
sector to sector, indicative of acceptance and efficiency variations. For
comparison, statistical uncertainties varied between 0.0067 and 0.0125.

\begin{figure}[tb]
\begin{center}    
\includegraphics[width=1\textwidth]{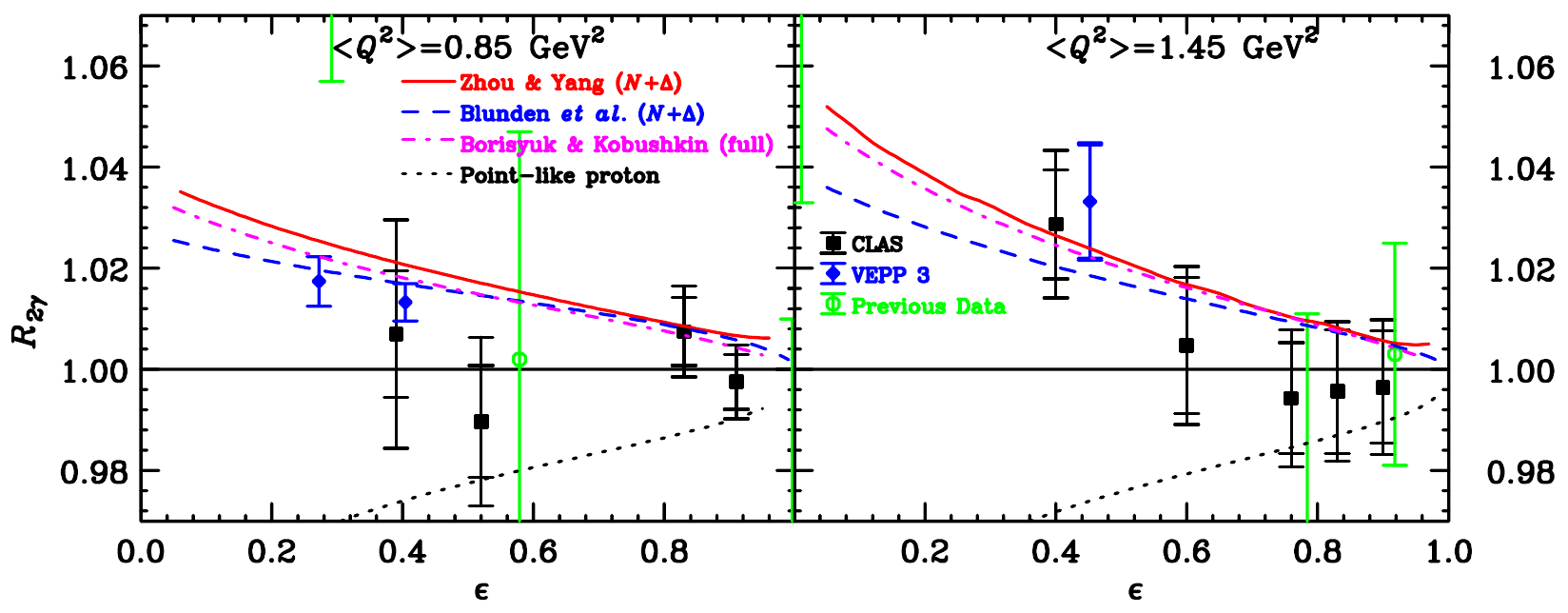}
\caption{
	$R_{2\gamma}$ as a function of $\eps$ at $Q^2 \approx 0.85$~GeV$^2$
	(\textit{left}) and 1.45~GeV$^2$ (\textit{right}) extracted from the
	measured ratio of $e^+p$/$e^-p$ cross sections corrected for both
	$\delta_{{\rm b},ep}$ and $\delta_{\rm even}$. The filled black squares show
	the results of the CLAS experiment~\cite{Rimal:2016toz}, while the filled
	blue diamonds are from VEPP-3~\cite{Rachek:2014fam} at similar kinematics.
	The open green circles show the previous world data at $0.7\leq Q^2 \leq
	1.0$~GeV$^2$ and $1.2\leq Q^2\leq 1.53$~GeV$^2$ in the left and right plots,
	respectively~\cite{Browman:1965zz, Bartel:1967dsa, Anderson:1968zzc, Mar:1968qd}.
	The error bars reflect statistical and point-to-point systematic
	uncertainties combined. The line at $R_{2\gamma}=1$ is the limit of no TPE.
	The solid red curve shows the calculation by Zhou and
	Yang~\cite{Zhou:2014xka} including $N+\Delta$ intermediate states.  The
	dashed blue curve shows the calculation by
	Blunden~\etal~\cite{Blunden:2005ew, Blunden:2016} including $N+\Delta$
	intermediate states (full). The dot-dashed purple curve shows the calculation
	by Borisyuk and Kobushkin~\cite{Borisyuk:2015xma}. The black dot-dashed line
	shows the calculation of TPE effects on a structureless point
	proton~\cite{Arrington:2011dn}.}
\label{fig:RatioEpsDep}
\end{center}    
\end{figure}

The results of the experiment as presented in Ref.~\cite{Rimal:2016toz} show an
$\eps$ dependence at $\langle Q^2 \rangle\approx 0.85$ and 1.45~GeV$^2$.
These results, along with results from the VEPP-3 experiment at similar $Q^2$
values are shown in Fig.~\ref{fig:RatioEpsDep} as well as predictions from
Refs.~\cite{Blunden:2016, Borisyuk:2015xma}, along with the no TPE limit and a
structureless proton model~\cite{Arrington:2011dn}.  At the time the CLAS
results were submitted for publication the OLYMPUS results were not available.
The conclusions drawn were that that the CLAS and VEPP-3 results at
$Q^2=1.45$~GeV$^2$ showed a ``moderate'' increase in $R_{2\gamma}$ with
decreasing $\eps$, while at $Q^2=0.85$ no clear change with $\eps$ is apparent.

\begin{figure}[tb]
\begin{center}
\includegraphics[width=1\textwidth]{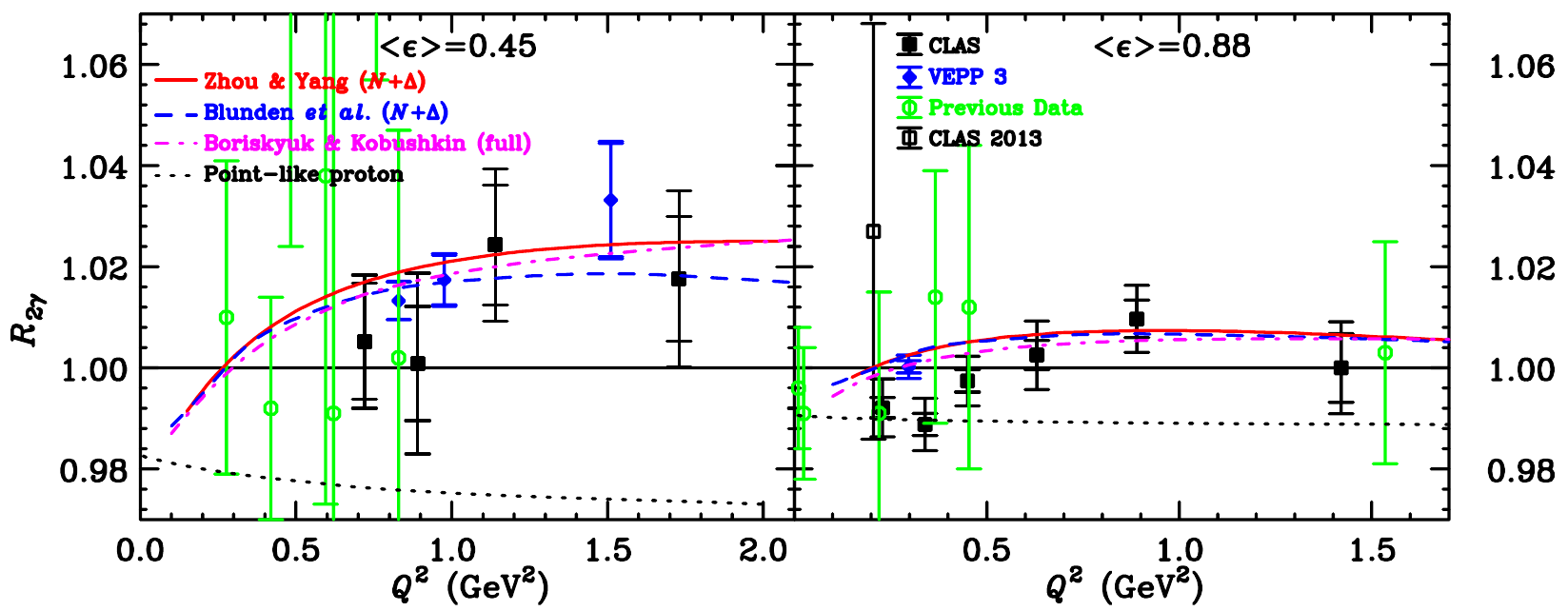}
\caption{
	$R_{2\gamma}$ as a function of $Q^2$ at $\eps \approx 0.45$ (\textit{left})
	and 0.88 (\textit{right}). The data and curves are the same as in
	Fig.~\ref{fig:RatioEpsDep} with an additional point from the CLAS TPE
	test-run experiment~\cite{Moteabbed:2013isu} (black open square). The open
	green circles show the previous world data
	\cite{yount62, Browman:1965zz, Bartel:1967dsa, Anderson:1968zzc, Mar:1968qd,
	Hartwig:1975px} at $0.2\leq\eps\leq 0.7$ and
	$0.7\leq\eps\leq 0.95$ in the left and right panels, respectively.}
\label{fig:RatioQ2Dep}
\end{center}    
\end{figure}

Figure~\ref{fig:RatioQ2Dep} shows the $Q^2$ dependence of the CLAS data, again
with the VEPP-3 results at similar values of $\eps$. Also included  in the
figure for the high $\eps$ data is a single data point from the CLAS TPE test
run~\cite{Moteabbed:2013isu}, which was primarily a proof of principle
experiment that ultimately had a large uncertainty due to data being taken with
only a single chicane setting leading to a large normalization uncertainty. The
seven data points from that experiment were taken at an average $Q^2$ of 0.206
GeV$^2$ and $0.830\leq\eps\leq 0.942$. These seven data points were averaged
together in the single point presented in the figure. The CLAS and VEPP-3 data
showed only a hint of a rise with $Q^2$ at the lower value of $\eps$ and no
indication of a change with $Q^2$ at the higher value of $\eps$.

The CLAS paper provided a global analysis that included both the CLAS data and
the VEPP-3 data but excluded the previous world data due to their large
uncertainties. The global analysis compared 12 independent CLAS data points and
the four non-normalization data points to the hadronic calculations of
Refs.~\cite{Blunden:2005ew, Zhou:2014xka}, the no-TPE assumption, and the
calculation based on a structureless proton~\cite{Arrington:2011dn}. These data
are in good agreement with the hadronic calculations of
Refs.~\cite{Blunden:2005ew, Zhou:2014xka} but of insufficient precision to make
any definitive distinction between them. However, the CLAS and VEPP-3 data
exclude the no-TPE hypothesis at the $5.3\sigma$ level, and rule out the
point-proton result at the $\sim 25\sigma$ level, which is essentially
equivalent to the $Q^2=0$ limit. A summary of the CLAS global analysis is shown
in Table~\ref{tab:Global1}. Again, as we shall discuss later, the inclusion of
the OLYMPUS results leads to a rather different conclusion.

\begin{table}[tb]
\centering
\begin{tabular}{l @{\qquad} S S}
\toprule
TPE calculation	& {$\chi^2_\nu$}	& {Confidence level (\%)} \\
\midrule
Blunden~\etal~($N$)~\cite{Blunden:2005ew}	& 1.23		& 23 \\
Zhou \& Yang~($N$)~\cite{Zhou:2014xka} 		& 1.27		& 20 \\
Zhou \& Yang~($N+\Delta$)~\cite{Zhou:2014xka} 	& 1.19	& 27 \\
$\delta_{\g2}=0$~(No TPE)					& 2.32		& 0.2 \\
Point-proton calculation					& 7.38		& 3.e-15 \\
\bottomrule
\end{tabular}
\captionsetup{width=.65\textwidth}
\caption{
Comparison of the 16 CLAS and VEPP-3 data points to various TPE calculations
showing the reduced $\chi^2$ value and the confidence level.}
\label{tab:Global1}
\end{table}

The CLAS results also included a corrected Rosenbluth separation based upon a
linear fit of all of the data at $Q^2\approx 1.45$~GeV$^2$ shown in
Fig.~\ref{fig:RatioEpsDep}. This fit constrained the line to go to
$R_{2\gamma}=1$ at $\eps=1$.  From this fit, $\delta_{\g2}$ was determined as a
function of $\eps$ and applied as a correction to the reduced cross-section data
of Andivahis~\etal~\cite{Andivahis:1994rq} according to
Eq.~(\ref{eq:sigmaRmeas}). This TPE correction changed the proton form factor
ratio, $R=\mu_p G_E/G_M$, from the original value of $0.910\pm 0.060$ to
$0.820\pm 0.044$.  This brings it into good agreement with the polarization
transfer result of $0.789\pm 0.042$ at $Q^2=1.77$~GeV$^2$~\cite{Punjabi:2005wq}.

The conclusion of this paper is that the CLAS and VEPP-3 data combined indicate
a non-zero TPE effect that is consistent with models that provide a TPE
correction that generally account for the Rosenbluth and polarization transfer
discrepancy.

\subsection{The OLYMPUS experiment}
\label{ssec:olympus}
  
The OLYMPUS experiment was designed to measure the ratio between positron-proton
and electron-proton elastic scattering over a broad angular range,
$25\degree<\theta<75\degree$, with a precision of around~1\%.  Only a brief
description of the experiment will be given here.  A full description can be
found in~\citep{Milner:2013daa}.

OLYMPUS ran on the DORIS positron/electron storage ring at the DESY laboratory,
Hamburg, Germany.  Data were collected in two periods for approximately three
months in total during 2012 before DORIS was shutdown.  The lepton beam energy
was nominally 2.01~GeV with currents up to 70~mA.  The lepton beam species
(electrons or positrons) was changed daily. A total integrated luminosity of
$\sim4.5$~fb$^{-1}$ was collected.

The stored lepton beam passed through a windowless, unpolarized, hydrogen gas
target~\citep{Bernauer:2014pva} internal to the DORIS storage ring.  The typical
areal target density was $\sim 3\times 10^{15}$~atoms/cm$^2$.

\begin{figure}[tb]
\centering
\includegraphics[width=0.6\textwidth]{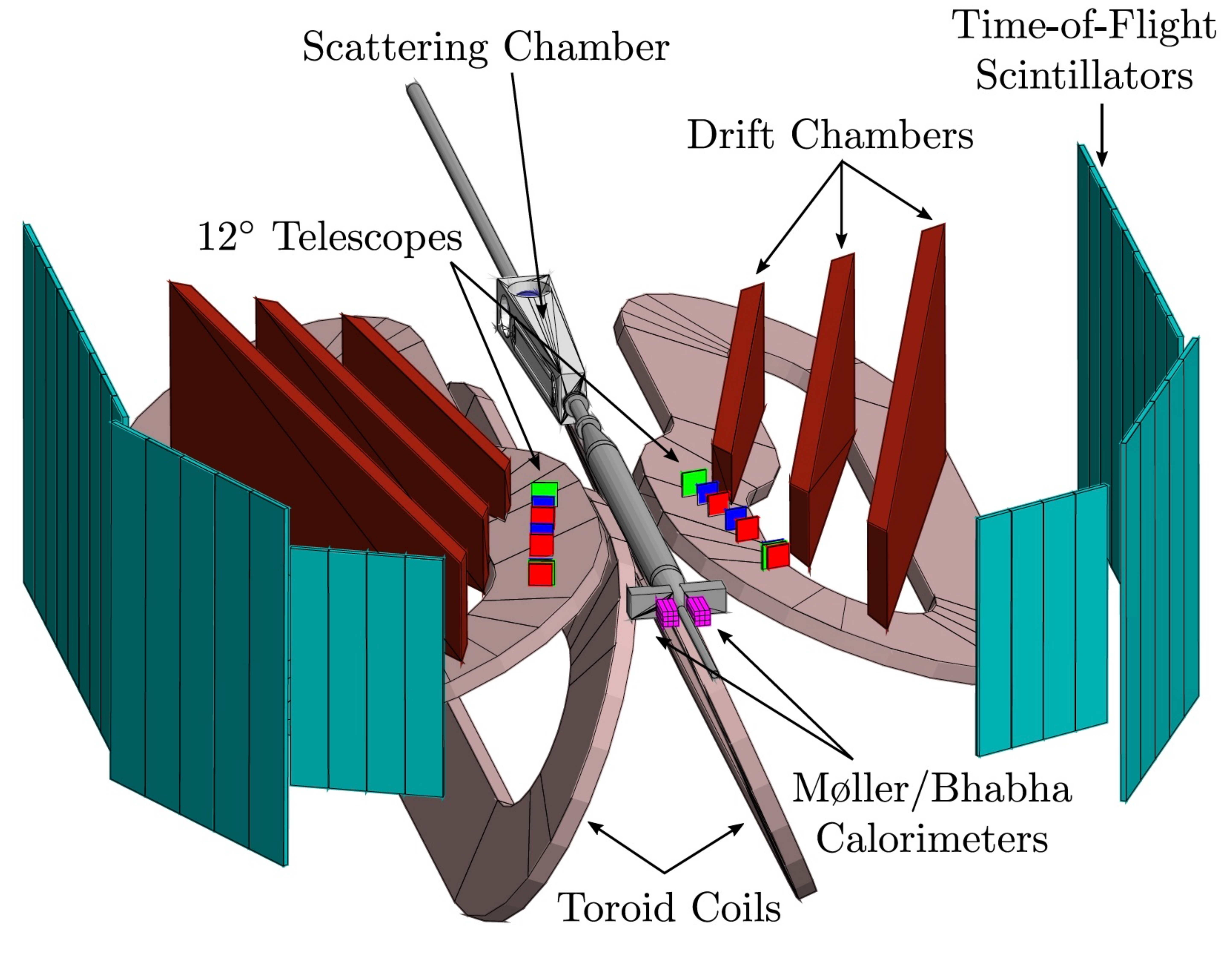}
\caption{
	Schematic representation of the OLYMPUS detector with the top four toroid
	coils removed to reveal the two horizontal, instrumented sectors. Note the
	drift chambers are shown as three separate chambers in each sector but in
	actuality were enclosed in a single gas volume.}
\label{fig:OLYMPUSdetector}
\end{figure}
The OLYMPUS detector (see~\cref{fig:OLYMPUSdetector}) was based on the former
MIT-Bates BLAST detector~\citep{Hasell:2009zza}.  This consisted of an eight
sector, toroidal, magnetic spectrometer with the two horizontal sections
instrumented with large acceptance ($20\degree<\theta<80\degree$,
$-15\degree<\phi<15\degree$) drift chambers (DC) for particle tracking and walls
of time-of-flight (ToF) scintillator bars to trigger the data acquisition system
and for particle identification. To a good approximation the detector system was
left-right symmetric, and this redundancy was used as a cross check by analyzing
and comparing the result determined when the lepton scattered into the left
sector with the result when the lepton scattered into the right sector.

Two new detector systems were designed and built to monitor the luminosity. 
Symmetric M{\o}ller / Bhabha calorimeters~\citep{Benito:2016cmp} (SYMB)
consisting of $3\times3$ arrays of PbF$_2$ crystals behind lead collimators were
situated at $1.29\degree$ left and right of the beam axis and approximately 3~m
downstream from the target.  There were also two detector telescopes of three
triple GEM detectors interleaved with three MWPC detectors mounted to the left
and right drift chambers at $12\degree$.  The $12\degree$ telescopes also had
plastic scintillators, front and back, with SiPM readout used in coincidence to
trigger the readout of the GEM and MWPC tracking detectors.

The first level trigger system used left-right coincidences between ToF bars
loosely corresponding to $e^{\pm}p$ elastic scattering angles. The second level
trigger required at least one hit in the middle and outer drift chambers of each
sector to indicate a potential track.  This helped reduce noise events and
allowed a higher rate of useful events to be collected.

In 2013, immediately after the experimental data runs, cosmic ray data were
collected for one month.  Then a complete optical survey of the detector
positions was made and the magnetic field was mapped throughout the tracking
volume~\citep{Bernauer:2016cc} and the volume between the scattering chamber and
the SYMB.

\subsubsection{Luminosity}

The integrated luminosity for each beam species was monitored by four
independent methods using: the slow control information, the $12\degree$
telescopes, the SYMB, and a multi-interaction event (MIE) method that also used
the SYMB calorimeters.

The slow control system monitored and recorded the beam current, beam position
and slope, and the flow of gas into the target cell in addition to numerous
other parameters.  A detailed molecular flow simulation converted the gas flow
rate into the target areal density.  Taking the product of the beam current and
target density yielded a 5\% absolute luminosity measurement and a 2\% relative
measure between beam species that was available online for quick analyses.

The $12\degree$ MWPC detectors tracked leptons elastically scattered over a
small range of angles around $12\degree$ in both the left and right sectors in
coincidence with the recoil proton tracked in the DC and ToF around $72\degree$.
 Combined with the MC simulation of $e^{\pm}p$ elastic scattering and assuming a
small contribution from two-photon effects, a luminosity determination was
possible at the level of~1\% every 20~minutes and a statistical accuracy on the
order of 0.01\% over the whole experiment.  Including systematic uncertainties,
the $12\degree$ system provided a 2.4\% absolute and a better than 0.5\%
relative luminosity determination.  The GEM detector readout was not used in
tracking or luminosity measurements at $12\degree$ though it was useful in
calibrating and aligning the MWPC.

The symmetric M{\o}ller/Bhabha calorimeters should have provided a fast and
high precision measurement of the luminosity.  Unfortunately the steep and
differing slopes of the M{\o}ller and Bhabha cross sections made it extremely
sensitive to the exact geometry and alignment of beam position, beam slope, and
the collimators in front of the calorimeters.  Ultimately the relative
luminosity was limited to an uncertainty of $\pm$3\%.  In addition,
time-dependent readout issues were encountered which were problematic.
 
However, the systematic and electronics issues with the planned M{\o}ller /
Bhabha measurements could be overcome by comparing the relative rates for the
lepton-lepton coincidences with the rate for detecting a $\sim2$~GeV lepton from
lepton-proton elastic scattering in one of the calorimeters in coincidence with
the lepton-lepton coincidence~\citep{Schmidt:2016wt}.  This multi-interaction
event (MIE) method produced a 0.3\% relative uncertainty in luminosity between
beam species.

The slow control, $12\degree$, and MIE methods for determining the luminosity
were all in excellent agreement lending support for the measurement of the
luminosity.  The MIE was chosen to normalize the analysis because it had the
smallest uncertainty. This had the additional advantage that the $12\degree$
detector system could be used to measure the ratio of
$\sigma_{e^{+}p}/\sigma_{e^{-}p}$ at $12\degree$ providing a measure of
two-photon exchange contributions at low $Q^2$ (high $\eps$) where
two-photon exchange is generally expected to be small.

\subsubsection{\label{sec:radcorr}Radiative corrections}
 
Radiative corrections are an important step in analyzing any electron scattering
experiment and it is important to include all the first order processes in
calculating the radiative corrections.

In the OLYMPUS experiment radiative corrections can not be simply applied to the
measured cross section.  OLYMPUS measured the lepton and proton in coincidence
over a broad kinematic range where the acceptance, efficiency, and energy
resolution vary as a function of $Q^2$.  Therefore it was necessary to build a
radiative generator that could be used in the Monte Carlo simulation of the
experiment.  This was done in parallel with the radiative generator developed by
the VEPP-3 group~\citep{Gramolin:2014pva} and was used to cross-check both
calculations.  The OLYMPUS radiative generator had numerous options to select
the proton form factor (point-like, Kelly, Bernauer, etc.), soft-photon
prescription (Mo-Tsai, Maximon-Tjon, etc.), vacuum polarization calculation, and
whether or not to use exponentiation.  The cut-off energy in OLYMPUS was
typically a few percent, and beyond this radiative corrections were explicitly
calculated.

\begin{figure}[t]
\centering\includegraphics[width=0.5\columnwidth]{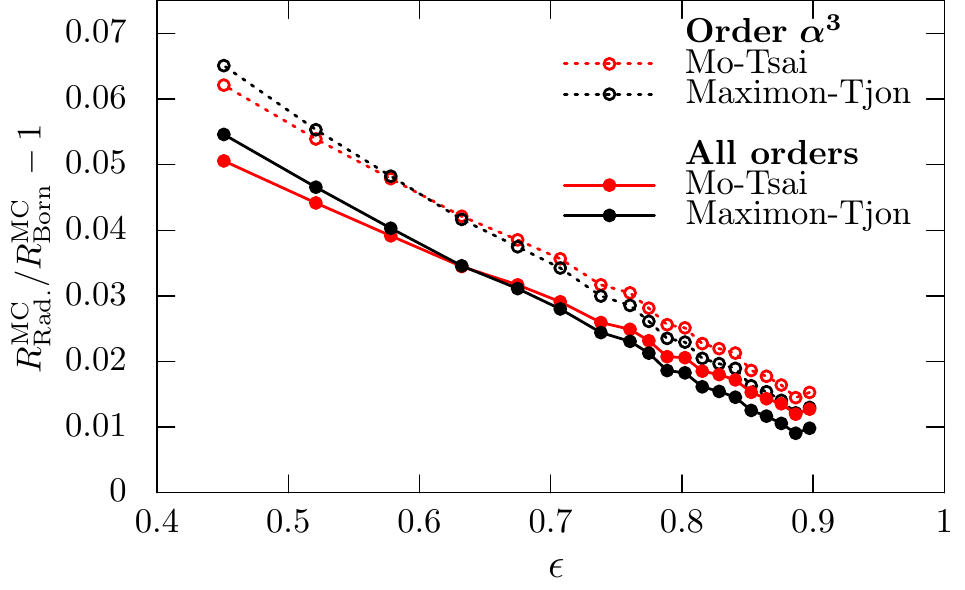}
\caption{
	Effect of radiative corrections relative to the Born result for the two
	prescriptions Mo-Tsai~\citep{Mo:1968cg} and
	Maximon-Tjon~\citep{Maximon:2000hm} and for the $\alpha^3$
	(non-exponentiated) or exponentiated calculation as a function of
	$\eps$.}
\label{fig:rad_corr_compare}
\end{figure}

It is important to note the significance of the soft TPE contributions at these
energies.  Figure~\ref{fig:rad_corr_compare}
compares four calculations of the radiative corrections relative to the Born
result as a function of $\eps$.  The difference between Mo-Tsai and
Maximon-Tjon is not significant.  But as $\eps$ decreases to 0.45 the effect
increases quickly from $\sim1$\% to 5-7\%, and the non-exponentiated
($\alpha^3$) effect is about $\sim1$\% larger than the exponentiated
contribution.

\subsubsection{Analysis}\label{sec:ana}

The analysis presented here is a combination of the results of four independent
analyses (three PhD theses~\citep{Henderson:2016wt, Russell:2016wt,
Schmidt:2016wt} and an analysis performed by Jan Bernauer).  The results are
available at~\citep{Henderson:2016dea}.  Two other theses are nearing
completion~\citep{OConnor:2016aa, Ice:2016aa} and these will be incorporated
into a longer paper with a more thorough description of each analysis.  Each
analysis was developed independently and their results are highly compatible.
For the results shown here and published in~\citep{Henderson:2016dea} we have
averaged the four results without weighting (statistics for each were
comparable) and the spread in the results was included as a point-to-point
systematic uncertainty.

The analyses to date are based on a subset of the total recorded data by
selecting runs with optimal running conditions (without tripped channels, etc).
These correspond to about $3.2$~fb$^{-1}$ of integrated luminosity.

Track reconstruction began with a pattern matching algorithm to identify
potential tracks and to obtain initial estimates for the track parameters. Then
two different tracking algorithms were employed to fit each track candidate to
optimize the track vertex, scattering angles, and momentum.  Both algorithms
produced very similar results.  The final tracked data sets consisted of
candidate track momentum, polar and azimuthal angles, $z$ position in the
target, charge, track path length, time to time-of-flight detector, and energy
deposited in the ToF.

Starting from the same tracked data set each analysis performed an independent
analysis.  For each analysis this included a series fiducial cuts to select good
tracks.  This was followed by applying loose cuts on all combinations of tracks
in an event to select pairs of tracks consistent with elastic $e^{\pm}p$
scattered events and to reduce background events.  Further cuts were then
applied to select the final $e^{\pm}p$ events.  The resulting events were binned
in a common selection of $Q^2$ bins reconstructed from the proton scattering
angle to minimize fake asymmetries (the two lepton charges bend differently and
could have different errors in the reconstructed kinematics). In each bin, the
background was subtracted.  At forward angles (low $Q^2$) the background was
negligible increasing to approximately 20\% at backward angles (high $Q^2$). 
This background fraction was roughly the same for all analyses and also the same
for both electron and positron runs.  The number of events in each $Q^2$ bin
after background subtraction was collected for both electron and positron beams.

A complete Monte Carlo (MC) simulation of the detector and experiment was also
made with full digitization to produce MC data in exactly the same format as the
real data.  This allowed the acceptances, efficiencies, and resolutions of the
DC, ToF, and $12\degree$ detectors to be simulated in the MC and compared with
the real data.

A radiative event generator was developed specifically for OLYMPUS
(see~\cref{sec:radcorr}).  This generated $e^{\pm}p$ events (including inelastic
processes) weighted by the scattering cross sections. Since the radiative
corrections depend on the proton structure and various radiative correction
prescriptions each of these effects were carried as separate cross section
weights.  Carrying the various cross section weights throughout the simulation
allowed their effect to be studied without having to regenerate and re-track the
MC for each.

The generated events were propagated through the detector simulation using
GEANT4.  More MC data were produced than real data to reduce systematic
uncertainties due to the MC simulation.  The MC used slow control information
like the lepton beam energy, position, slope, and instantaneous luminosity to
match the data as closely as possible on a run-by-run basis.  The MC data were
then analyzed using the same code, event selection, and cuts as used on the real
data.

To obtain the ratio between positron-proton and electron-proton elastic
scattering as a function of $Q^2$ (or $\eps$),
$R_{2\gamma}(Q^2)=\sigma_{e^+p}(Q^2)/\sigma_{e^-p}(Q^2)$, we take the luminosity
weighted ratio of the number of events for both data and MC for each $Q^2$ bin:
\be
  R_{2\gamma}= \frac{N_{\rm exp}(e^+)}{N_{\rm exp}(e^-)}\bigg/
\frac{N_{\rm MC}(e^+)}{N_{\rm MC}(e^-)}\, .
\ee
Note that we do not correct the yield from data for efficiencies, acceptances,
or radiative effects.  Rather this is all included and corrected through the
complete Monte Carlo simulation.

In addition to statistical uncertainties there are various uncorrelated
systematic uncertainties that vary from bin to bin and correlated systematic
uncertainties common to all bins.  The systematic uncertainties are given
in~\cref{tab:systematics}.

\begin{table}[tbp]
\centering
\begin{tabular}{l S}
\toprule
Type of contribution & {Uncertainty in $R_{2 \gamma}$}\\
\midrule
{\em Correlated contributions}\\
\qquad Beam energy & 0.04{-0.13}\%\\
\qquad MIE luminosity & 0.36\% \\
\qquad Beam and detector alignment & 0.25\% \\
{\em Uncorrelated contributions}\\
\qquad Tracking efficiency & 0.20\% \\
\qquad Elastic selection and background subtraction & 0.25{-1.17}\%\\
\bottomrule
\end{tabular}
\caption{Contributions to the OLYMPUS systematic
    uncertainty in $R_{2 \gamma}$.}
\label{tab:systematics}
\end{table}

\begin{figure}[tb]
\centering
\includegraphics[width=0.4\columnwidth, viewport=0 10 160 170, clip]{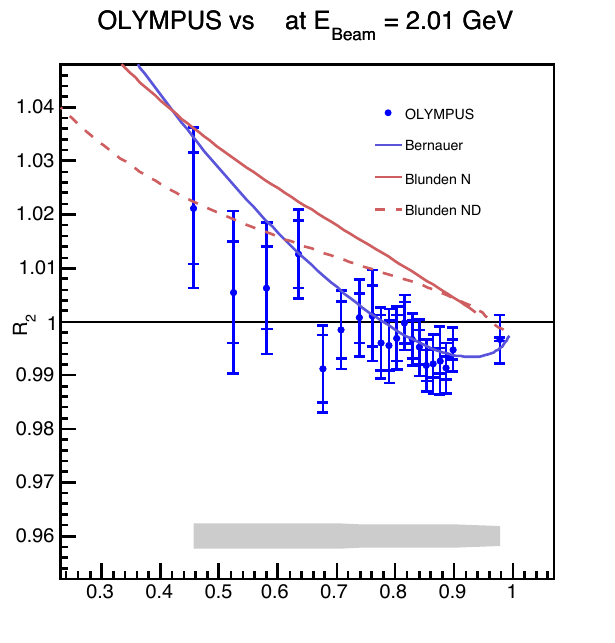}
\caption{
	OLYMPUS results for $R_{2\gamma}$ using the exponentiated Mo-Tsai radiative
	corrections plotted as a function of $\eps$. The results are plotted
	with statistical uncertainties (inner error bars) and uncorrelated
	systematic uncertainties (outer error bars).  The correlated uncertainty is
	represented by the gray bar at the bottom of the figure.  Theoretical
	calculations from Blunden~\citep{Blunden:2016} for $N$
	and $N+\Delta$ and the phenomenological fit to the form factor data
	from Bernauer~\citep{Bernauer:2013tpr} is also shown.}
\label{fig:OLYMPUS_Results}
\end{figure}

The final results for $R_{2\gamma}$ for OLYMPUS are available
in~\citep{Henderson:2016dea} and are shown in~\cref{fig:OLYMPUS_Results}.
The results are plotted with the statistical and systematic uncertainties
together with the theoretical calculations from Blunden for $N$ and $N+\Delta$
~\citep{Blunden:2016} and the predictions from the phenomenological fit to the
existing form factor data by Bernauer~\citep{Bernauer:2013tpr}.  The plotted
results are with the radiative corrections to all orders using the convention of
Mo-Tsai~\citep{Mo:1968cg} for compatibility with the CLAS and VEPP-3 results.

The OLYMPUS results are in general less than unity at high $\eps$ gradually
rising to around 2\% at $\eps=0.456$.  The OLYMPUS results are systematically
lower than the theoretical calculations of Blunden but in reasonable agreement
with the predictions of Bernauer's phenomenological fit. This implies that
perhaps the theoretical calculations that account for the discrepancy in the
form factor ratio at higher $Q^2$ do not extend to this relatively low $Q^2$
region or that other effects need to be taken into account. Bernauer's fit that
includes low $Q^2$ measurements agrees with the data better.

Bernauer's model~\citep{Bernauer:2013tpr} was a fit to all the available form
factor data including the polarization data (1866 data points).  For the
unpolarized data he extracted cross section data and redid the radiative
corrections to standardize the treatment.  The Mainz data, included in the fit,
were the largest and most consistent set of data and extended to the smallest
$Q^2$ values available.  To include the polarization measurements he modelled
the hard two-photon exchange contribution as the Feshbach correction of
Eq.~(\ref{eq:delFesh}) plus $\delta_{\rm hard}$ using the parametrization:
\be
\delta_{\rm hard}=-(1-\eps) a \log{( b\, Q^2 + 1 )}\, ,
\ee
where $a$ and $b$ were included in the fit parameters.  The fit found
$a=0.06894$ and $b=0.3947$~GeV$^{-2}$.  The final fit was the product of a
spline interpolant and the standard dipole form.  The spline used knots at $Q^2$
values of: 0.0, 0.25, 0.5, 0.75, 1.0, 1.5, 3.0, 5.0, 10.0, and 40.0~GeV$^2$, and
achieved $\chi^2_\nu= 2151.72/1866 = 1.153$.

\subsection{Comparison of recent experiments and models}
\label{ssec:compEM}

\begin{figure}[tb]
\begin{center}
\includegraphics[width=0.75\columnwidth]{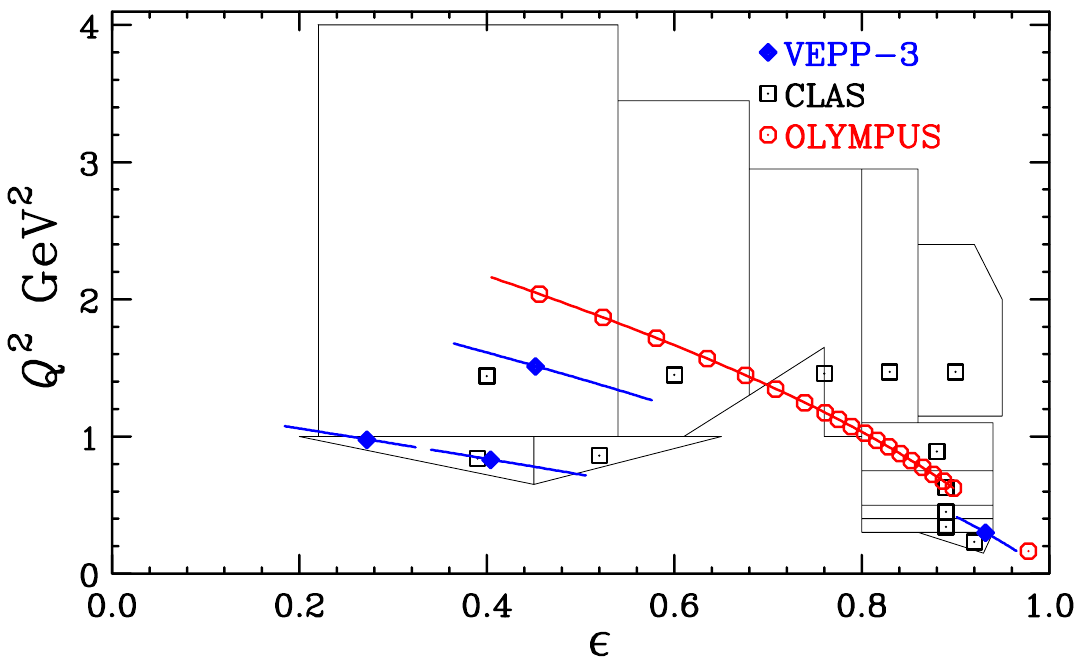}
\caption{
	Kinematic regions probed by the three two-photon experiments showing the
	$Q^2$ and $\epsilon$ plane.  Symbols indicate values at which data points
	were reported by the respective experiments. The boxed regions show the bins
	over which the CLAS data are summed and the blue curves indicate the
	kinematic region over which the VEPP-3 data points are summed to obtain the
	results at the data points shown by the symbols.  The binning of the OLYMPUS
	data are binned such that the gaps between bins are not visible in the red
	curve.}
\label{fig:KinAll}
\end{center}
\end{figure}

The kinematic coverage in $Q^2$ versus $\eps$ of the three new experiments,
VEPP-3, CLAS, and OLYMPUS is shown in Fig.~\ref{fig:KinAll}, which shows the
binning used by each experiment. Both VEPP-3 and OLYMPUS ran with monoenergetic
beams so their bins are sums over angle ranges leading to a correlated variation
over $Q^2$ and $\eps$.  CLAS had a range of beam energies at much lower
integrated luminosity and thus summed bins over ranges in both $Q^2$ and
$\eps$.  The combined data sets are shown in Table~\ref{tab:AllData}.

\begin{table}[thb]
\centering
\begin{tabular}{l @{\qquad} S[table-format=1.3] S[table-format=1.3] S[table-format=1.4]
S[table-format=1.4] S[table-format=1.3] c l @{\qquad}
S[table-format=1.3] S[table-format=1.3] S[table-format=1.4]
S[table-format=1.4] S[table-format=1.3]}
\toprule
&{$Q^2$}	& {$\eps$}	& {$R_{2\gamma}$} 	& {$\delta R_{2\gamma}^{\rm total}$} &
  {$\delta R_{2\gamma}^{\rm scale}$} & &
	& {$Q^2$}	& {$\eps$}	& {$R_{2\gamma}$} 	& {$\delta R_{2\gamma}^{\rm total}$}	&
	  {$\delta R_{2\gamma}^{\rm scale}$} \\
\cmidrule{1-6} \cmidrule{8-13}
\multicolumn{2}{l}{{\em VEPP-3}}& & & & & &\multicolumn{2}{l}{{\em OLYMPUS}} \\
&1.510		& 0.452		& 1.0332			& 0.0116			& {--} &
	&& 0.165	& 0.978		& 0.9967			& 0.0046			& 0.0036 \\						
&0.298		& 0.932		& 1.0002			& 0.0023			& {--} &
	&& 0.624	& 0.898		& 0.9948			& 0.0042			& 0.0045 \\
&0.976		& 0.272		& 1.0174			& 0.0052			& {--} &
	&& 0.674	& 0.887		& 0.9913			& 0.0047			& 0.0045 \\
&0.830		& 0.931		& 1.0133			& 0.0038			& {--} &
	&& 0.724	& 0.876		& 0.9927			& 0.0064			& 0.0045 \\
\multicolumn{2}{l}{{\em CLAS}} & & & & &
	&& 0.774	& 0.865		& 0.9921			& 0.0056			& 0.0045 \\
&0.84 		& 0.39 		& 1.0070 			& 0.0226 			& 0.003 &
	&& 0.824	& 0.853		& 0.9918			& 0.0049			& 0.0045 \\
&0.86 		& 0.52 		& 0.9897 			& 0.0167 			& 0.003 &
	&& 0.874	& 0.841		& 0.9952			& 0.0053			& 0.0045 \\
&1.44 		& 0.40 		& 1.0287			& 0.0146 			& 0.003 &
	&& 0.924	& 0.829		& 0.9967			& 0.0049			& 0.0045 \\
&1.45 		& 0.60 		& 1.0047			& 0.0156 			& 0.003 &
	&& 0.974	& 0.816		& 0.9998			& 0.0051			& 0.0045 \\
&1.46 		& 0.76 		& 0.9943			& 0.0136 			& 0.003 &
	&& 1.024	& 0.803		& 0.9969			& 0.0059			& 0.0045 \\
&1.47 		& 0.83 		& 0.9956			& 0.0138 			& 0.003 &
	&& 1.074	& 0.789		& 0.9955			& 0.0069			& 0.0045 \\
&1.47  		& 0.90 		& 0.9965			& 0.0133 			& 0.003 &
	&& 1.124	& 0.775		& 0.9960			& 0.0066			& 0.0045 \\
&0.23 		& 0.92 		& 0.9921 			& 0.0057 			& 0.003 &
	&& 1.174	& 0.761		& 1.0011			& 0.0085			& 0.0045 \\
&0.34 		& 0.89 		& 0.9888 			& 0.0052 			& 0.003 &
	&& 1.246	& 0.739		& 1.0007			& 0.0072			& 0.0045 \\
&0.45 		& 0.89 		& 0.9974 			& 0.0049 			& 0.003 &
	&& 1.347	& 0.708		& 0.9985			& 0.0073			& 0.0045 \\
&0.63 		& 0.89 		& 1.0025 			& 0.0068 			& 0.003 &
	&& 1.447	& 0.676		& 0.9912			& 0.0080			& 0.0045 \\
&0.89 		& 0.88 		& 1.0097 			& 0.0066 			& 0.003 &
	&&1.568	& 0.635		& 1.0126			& 0.0084			& 0.0045 \\
& & & & & & & & 1.718	& 0.581	& 1.0063	& 0.0123			& 0.0045 \\
& & & & & & & & 1.868	& 0.524	& 1.0055	& 0.0151			& 0.0045 \\
& & & & & & & & 2.038	& 0.456	& 1.0212	& 0.0150			& 0.0045 \\
\bottomrule
\end{tabular}
\caption{
	Recent data for $R_{2\gamma}$, with $Q^2$ in GeV$^2$.  $\delta R_{2\gamma}^{\rm total}$ is the
	quadrature sum of statistical and uncorrelated systematic uncertainties.
	$\delta R_{2\gamma}^{\rm scale}$ is the scale-type uncertainty for each
	experiment.}
\label{tab:AllData}
\end{table}

\begin{figure}[tb]
\centering
\begin{minipage}{0.5\textwidth}
\centering
\includegraphics[width=0.73\linewidth]{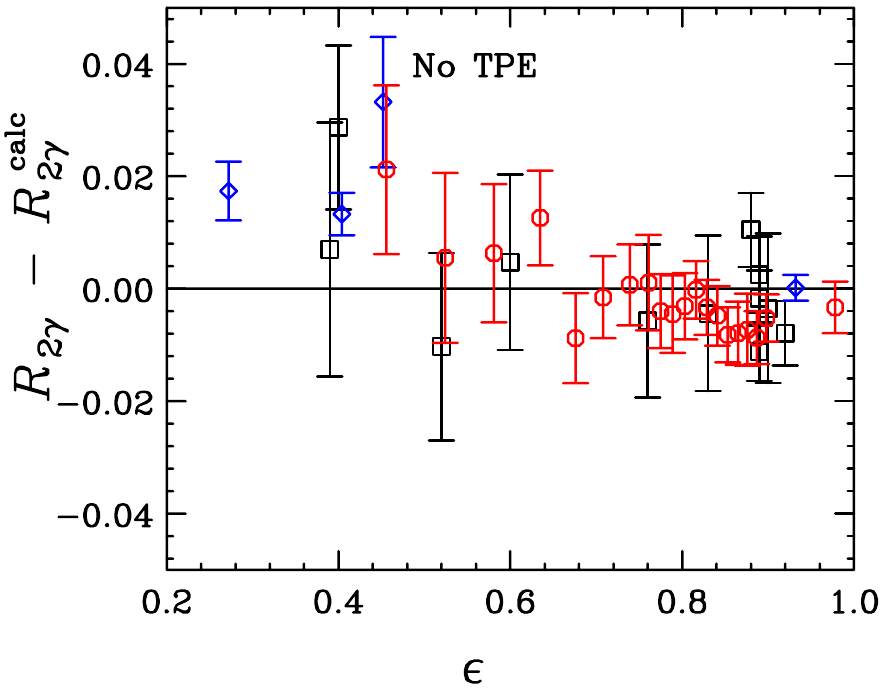}
\end{minipage}%
\begin{minipage}{0.5\textwidth}
\centering
\includegraphics[width=0.73\linewidth]{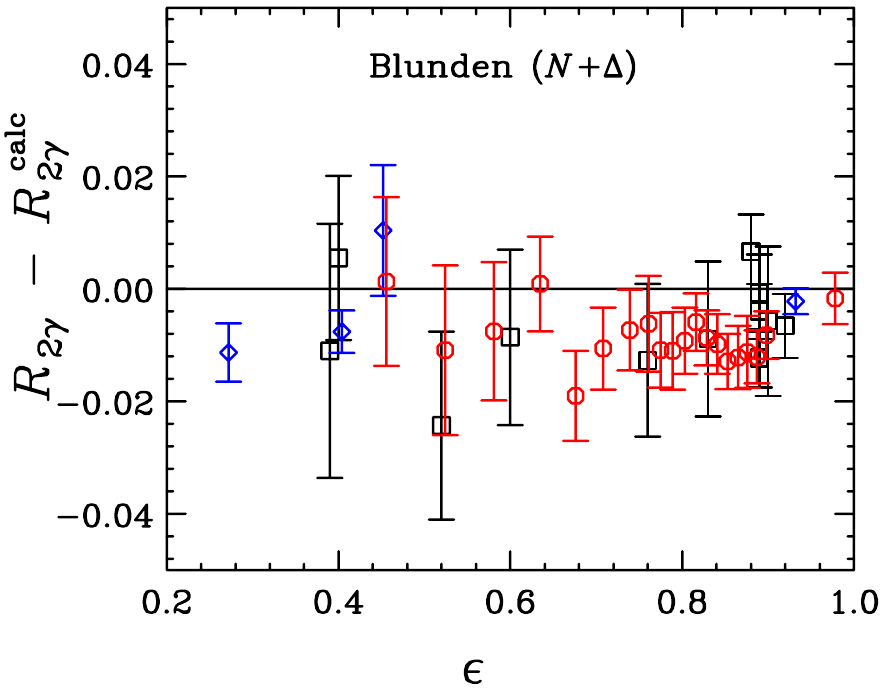}
\end{minipage}
\begin{minipage}{0.5\textwidth}
\centering
\includegraphics[width=0.73\linewidth]{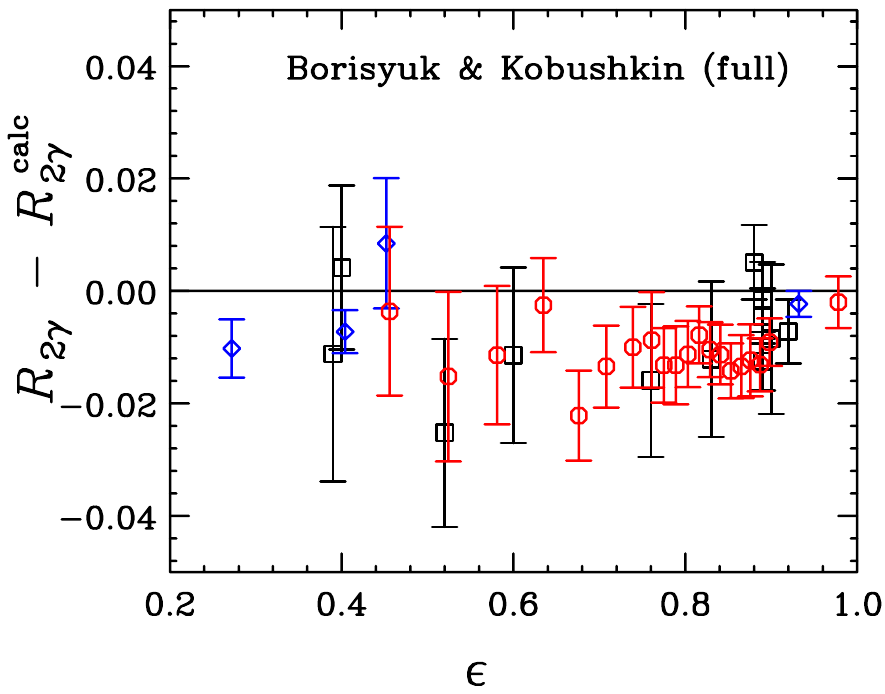}
\end{minipage}%
\begin{minipage}{0.5\textwidth}
\centering
\includegraphics[width=0.73\linewidth]{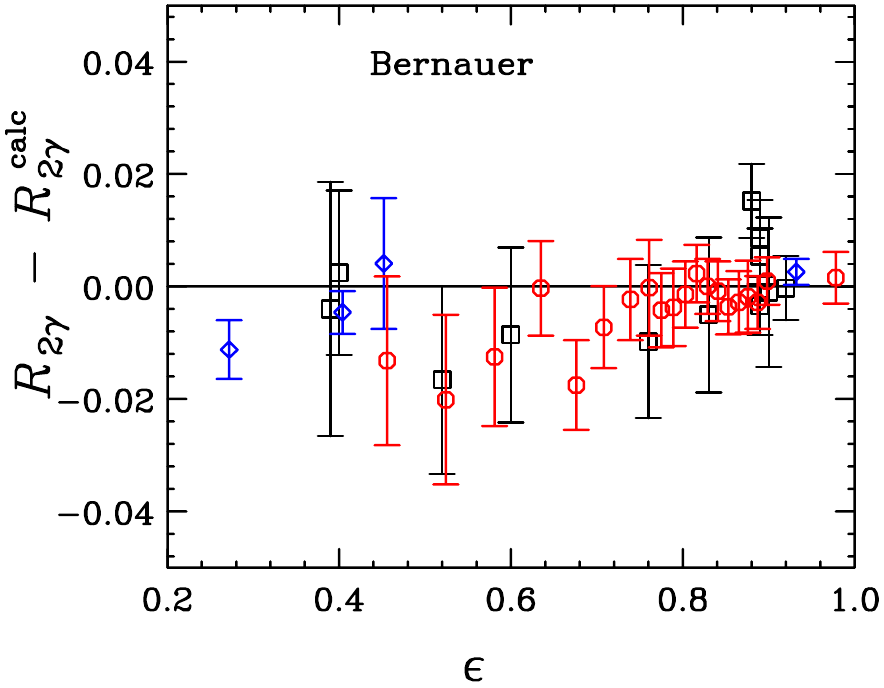}
\end{minipage}
\caption{
	Difference between $R_{2\gamma}$ and model predictions as a function of
	$\eps$. The blue diamonds are VEPP-3, the black boxes are from CLAS,
	and the red circles are from OLYMPUS. Error bars reflect the quadrature
	sum of statistical and uncorrelated systematic uncertainties.}
\label{fig:Diffeps}
\end{figure}

\begin{figure}[!tb]
\centering
\begin{minipage}{0.5\textwidth}
\centering
\includegraphics[width=0.73\linewidth]{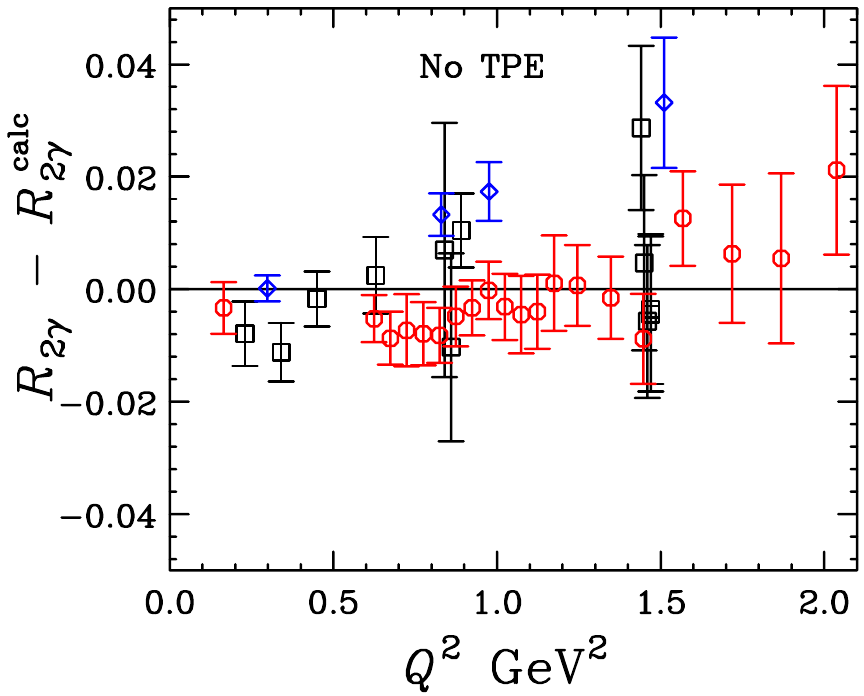}
\end{minipage}%
\begin{minipage}{0.5\textwidth}
\centering
\includegraphics[width=0.73\linewidth]{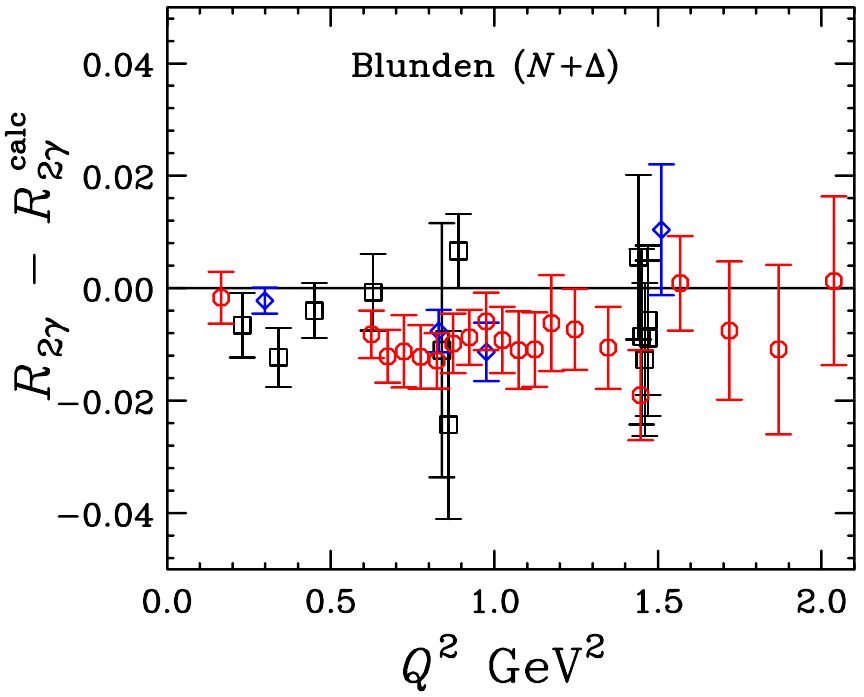}
\end{minipage}
\begin{minipage}{0.5\textwidth}
\centering
\includegraphics[width=0.73\linewidth]{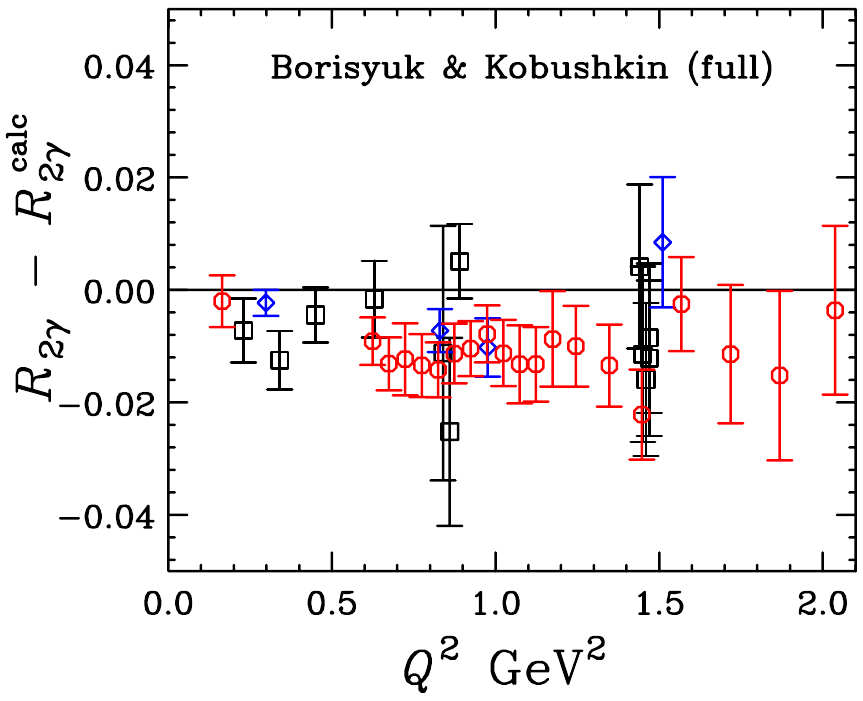}
\end{minipage}%
\begin{minipage}{0.5\textwidth}
\centering
\includegraphics[width=0.73\linewidth]{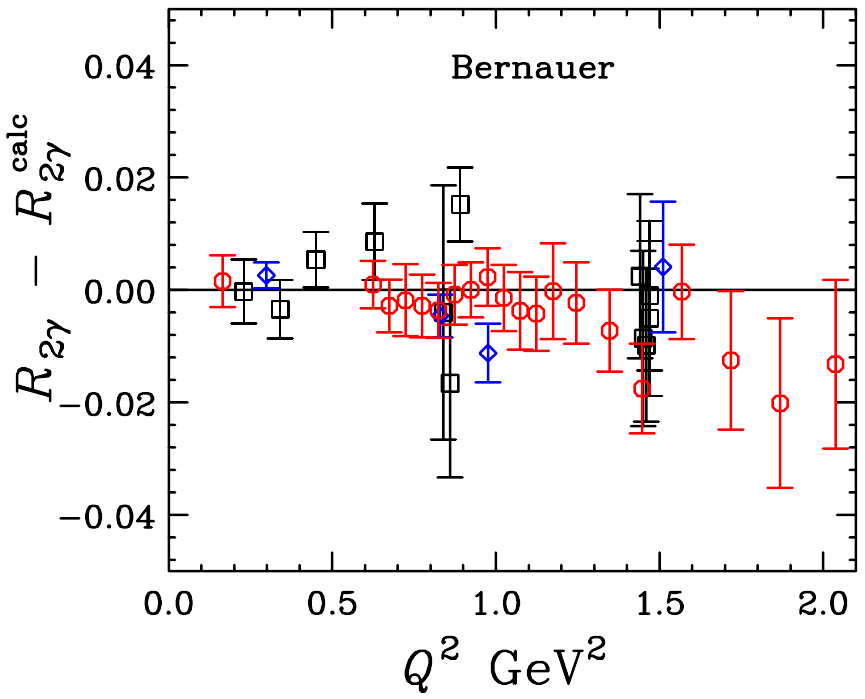}
\end{minipage}
\caption{
	 Difference between $R_{2\gamma}$ and model predictions as a function of
	 $Q^2$. Data symbols are the same as in Fig.~\ref{fig:Diffeps}.}
\label{fig:DiffQ2}
\end{figure}

The experiments measured few points with the same kinematics so a direct
comparison of all of the the data simultaneously would not be appropriate. A
plot of $R_{2\gamma}$ versus $\eps$ or versus $Q^2$ would hide the dependency
on the other variable.  One way to compare the data is to plot each data point's
difference from a given model, $R_{2\gamma}^{\rm data}-R_{2\gamma}^{\rm calc}$,
since the model can be calculated at the specific kinematics of the measured
data. This is shown in Figs.~\ref{fig:Diffeps} and \ref{fig:DiffQ2} at all
kinematics compared to the no TPE limit of $R_{2\gamma}=1$, the $N+\Delta$
models of both Refs.~\cite{Blunden:2016, Borisyuk:2015xma}, and the
parametrization of Bernauer~\etal~\cite{Bernauer:2013tpr}. The VEPP-3
data have been normalized to the model prediction at their luminosity
normalization points in each case and we have used the OLYMPUS data with
radiative corrections to all orders with the Mo and Tsai method. In the no TPE
limit we see that the OLYMPUS data is systematically below zero at large
$\eps$.  In the comparisons to Blunden and
Melnitchouk~\cite{Blunden:2016} and Borisyuk and
Kobushkin~\cite{Borisyuk:2015xma} calculations, the CLAS and VEPP-3 data are
evenly scattered about zero while the OLYMPUS data systematically fall below
zero at nearly all values of $\eps$.  The difference of all of the data
to the Bernauer parametrization is small except for, perhaps, at central values
of $\eps$ where there is a slight systematic negative difference. The
plot of $R_{2\gamma}=1$ versus $Q^2$ for the no TPE limit shows a more clear
difference between OLYMPUS and the other two data sets.


This does not present the entire picture since the CLAS and OLYMPUS data sets
both have scale-type, or normalization uncertainties that would move the entire
data sets up or down. To account for the normalization uncertainty we have
preformed two separate statistical analyses of the data. In the first we have
added the normalization uncertainties of CLAS and OLYMPUS in quadrature to the
statistical and uncorrelated systematic uncertainties.  This over inflates the
error bars on individual data points but provides an upper limit on the
confidence-level agreement between the data and models. We also took into
account the fact that not all of the CLAS data presented in
Figs.~\ref{fig:RatioEpsDep} and~\ref{fig:RatioQ2Dep} are independent. As was
done in Ref.~\cite{Rimal:2016toz} we selected the 12 independent data points
with the best discriminatory power.  These are the ones shown in
Fig.~\ref{fig:KinAll}. The VEPP-3 paper does not report a separate normalization
uncertainty, rather the normalization depends upon the model to which the data
are being compared. A standard $\chi^2_\nu$ has been calculated for each data set
separately and as a whole.

The results of this analysis are shown in the columns labelled ``No normalization'' of
Table~\ref{tab:global}. With this treatment of the normalization uncertainties,
the data exclude the the no-TPE hypothesis at the 98\% confidence level, though
the OLYMPUS data alone verify this hypothesis at the 89\% confidence level.
There is excellent agreement between the collective data set and the Bernauer
parametrization, and both the hadronic models tested are excluded at greater
than the 96\% confidence level. We stress that one should not read too much into
these confidence levels because the error bars on the data points are inflated.

\begin{table}[tbh]
\centering
\begin{tabular}{l @{\qquad} l S[table-format=1.2] S[table-format=1.0]
l S[table-format=1.2] S[table-format=1.0] S[table-format=1.4] S[table-format=1.2]}
\toprule
 & & \multicolumn{2}{l}{{No normalization}} &\phantom{ab}& \multicolumn{2}{l}{{With normalization}} \\
\cmidrule{3-4} \cmidrule{6-9}
& Data set 	& {$\chi^2_\nu$}	& {$\nu$}	& & {$\chi^2_\nu$}	& {$\nu$}	& {${\cal N}$} &
{$\left(\frac{{\cal N}-1}{\delta R_{2\gamma}^{\rm norm}}\right)$} \\[2pt]
\midrule
\multicolumn{5}{l}{{\em Model:} $\delta_{\g2}=0$}	\\
&VEPP-3	& 7.97		& 4		&& 7.97		& 4		& {--}			& {--}	\\
&CLAS	& 0.99		& 12	&& 1.25		& 11	& 1.0012		& 0.40 	\\
&OLYMPUS & 0.64		& 20 	&& 0.68		& 19	& 1.0034  		& 0.76  \\
&All	& 1.57		& 36	&& 1.73		& 34	& {--}			& {--}	\\[5pt]
\multicolumn{5}{l}{{\em Model:} Blunden \& Melnitchouk~\cite{Blunden:2016}} 	\\
&VEPP-3	& 2.62		& 4		&& 2.62		& 4		& {--}			& {--}	\\
&CLAS	& 0.90		& 12	&& 0.91		& 11	& 1.0032		& 1.07	\\
&OLYMPUS & 1.57		& 20	&& 0.64		& 19	& 1.0082  		& 1.82	\\
&All	& 1.46		& 36	&& 0.96		& 34	& {--}			& {--}	\\ [5pt]
\multicolumn{5}{l}{{\em Model:} Borisyuk \& Kobushkin~\cite{Borisyuk:2015xma}} \\ 
&VEPP-3	& 2.28		& 4		&& 2.28		& 4		& {--}			& {--}	\\
&CLAS	& 1.02		& 12	&& 0.94		& 11	& 1.0038		& 1.27	\\
&OLYMPUS & 2.15		& 20	&& 0.75		& 19	& 1.0097  		& 2.16	\\
&All	& 1.79		& 36	&& 1.00		& 34	& {--}			& {--}	\\ [5pt]
\multicolumn{5}{l}{{\em Model:} Bernauer~\etal~\cite{Bernauer:2013tpr}} \\ 
&VEPP-3	& 1.90		& 4		&& 1.90		& 4		& {--}			& {--}	\\
&CLAS	& 0.74		& 12	&& 0.90		& 11	& 0.9985		& -0.40	\\
&OLYMPUS & 0.46		& 20	&& 0.51		& 19	& 1.0019  		& \ 0.42 \\
&All	& 0.71		& 36	&& 0.80		& 34	& {--}			& {--}	\\ 
\bottomrule
\end{tabular}
\caption{
	Comparison of VEPP-3, CLAS, OLYMPUS, and the combined data set (All) to
	various TPE calculations showing the reduced $\chi^2$ value and the
	normalization factor ${\cal N}$ derived from the fit. The ``No normalization'' column
	represents a comparison when the normalization uncertainties of CLAS and
	OLYMPUS are added in quadrature. The column labelled ``With normalization'' is when the
	CLAS and OLYMPUS normalizations are allowed to float, as described in the
	text.}
  \label{tab:global}
\end{table}

In our second treatment of the normalization uncertainties we have allowed the
normalization of the CLAS and OLYMPUS data to float independently but with a
penalty determined by the normalization uncertainty of each data set.  We select
the normalization, ${\cal N}$, that minimizes a modified $\chi^2$ defined by
\be
\label{eq:modchi}
	\chi^2=\sum_n \left(\frac{R_{2\gamma}\, {\cal N}-R_{2\gamma}^{\rm calc}}
	{\delta R_{2\gamma}^{\rm total}}\right)^2 +
		\left(\frac{{\cal N}-1}{\delta R_{2\gamma}^{\rm norm}}\right)^2\, ,
\ee
where $R_{2\gamma}$ is the value reported by the experiments, $\delta
R_{2\gamma}^{\rm total}$ is the quadrature sum of the statistical and
uncorrelated systematic uncertainties, $R_{2\gamma}^{\rm calc}$ is the
calculated value for a particular model, and $\delta R_{2\gamma}^{\rm norm}$ is
the normalization uncertainty. The number of degrees of freedom, $\nu$, is then
number of data points, $n$, in the set minus one.  For CLAS $\nu=11$, and for
OLYMPUS $\nu=19$. The analysis for the VEPP-3 data does not change from the ``No
normalization'' analysis.

\begin{figure}[tb]
\centering
\begin{minipage}{0.5\textwidth}
\centering
\includegraphics[width=0.75\linewidth]{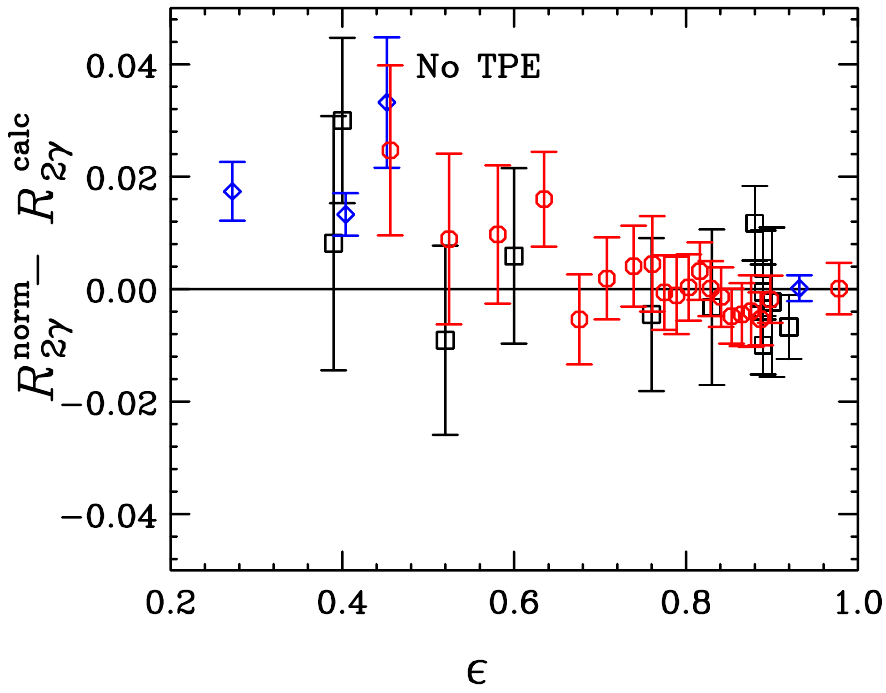}
\end{minipage}%
\begin{minipage}{0.5\textwidth}
\centering
\includegraphics[width=0.75\linewidth]{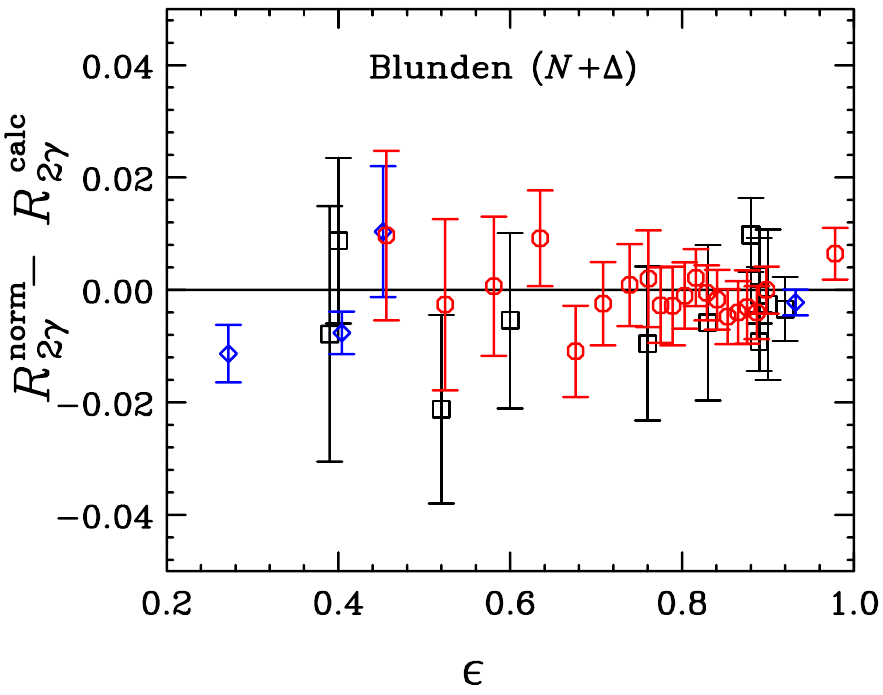}
\end{minipage}
\begin{minipage}{0.5\textwidth}
\centering
\includegraphics[width=0.75\linewidth]{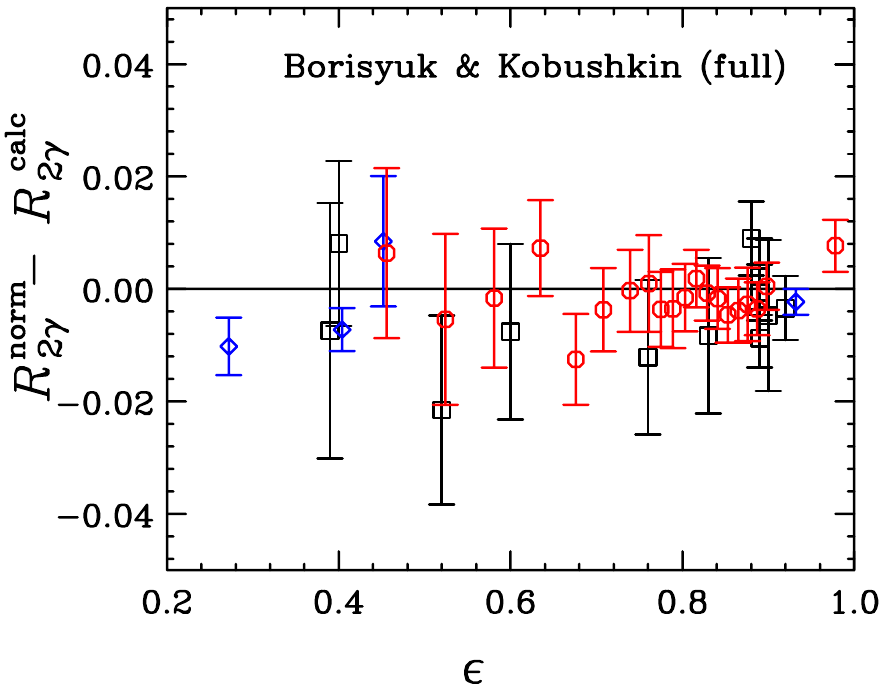}
\end{minipage}%
\begin{minipage}{0.5\textwidth}
\centering
\includegraphics[width=0.75\linewidth]{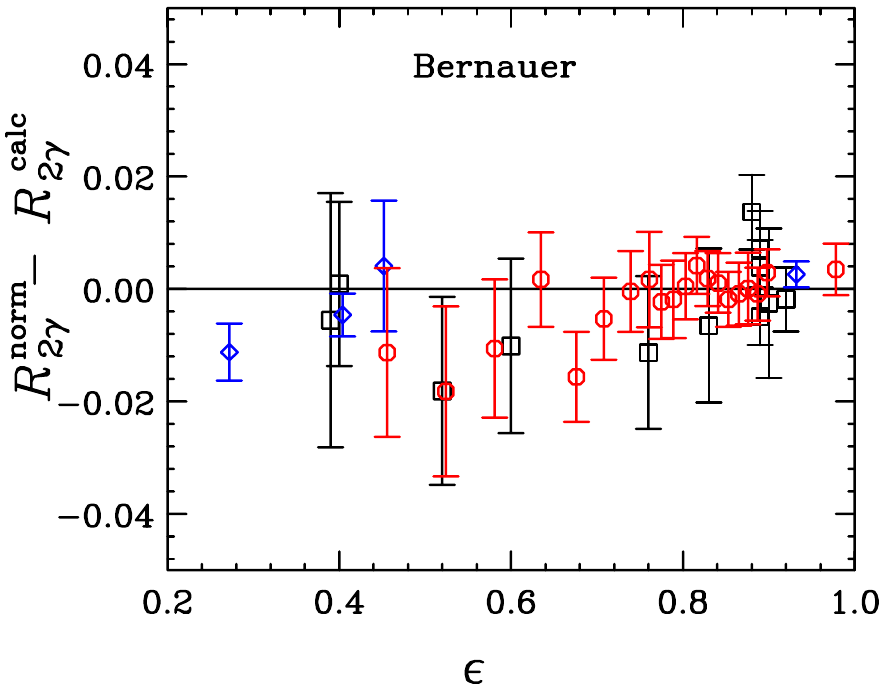}
\end{minipage}
\caption{
	Difference between {\it normalized} $R_{2\gamma}$ and model predictions as a
	function of $\eps$. Data symbols are the same as in
	Fig.~\ref{fig:Diffeps}.}
\label{fig:Nfits}
\end{figure}

The statistical results are shown in Table~\ref{tab:global}, and
Fig.~\ref{fig:Nfits} shows the difference between the normalized data values
$R_{2\gamma}^{\rm norm}=R_{2\gamma}{\cal N}$ and the model predictions
$R_{2\gamma}^{\rm calc}$ with total uncertainties also scaled by the
normalization factor.  The combined data still excludes the no-TPE hypothesis
but now at the 99.5\% confidence level and there is good agreement with the
hadronic models of Refs.~\cite{Blunden:2016, Borisyuk:2015xma} with confidence
levels of 53\% and 48\%, respectively. However, in both cases a large upward
normalization is required for the OLYMPUS data that is different from one by
nearly $2\delta R_{2\gamma}^{\rm norm}$.  The Bernauer parametrization agrees
with the data at the 79\% confidence level.

\section{Conclusions and outlook}
\label{sec:conclusion}
New $e^+p/e^-p$ data from three experiments are now available for
$Q^2<2.1$~GeV$^2$. These data are in reasonable agreement with each other except
for a steeper $Q^2$ dependence in the VEPP-3 results, which largely disappears
when compared to calculations that also increase with $Q^2$.  Collectively, the
data sets rule out $\delta_{\gamma\gamma}=0$ at greater than the 95\% confidence
level.

Allowing a renormalization of the CLAS and OLYMPUS results achieves reasonable
agreement with the calculations of Blunden and Melnitchouk~\cite{Blunden:2016}
and Borisyuk and Kobushkin~\cite{Borisyuk:2015xma} and the parametrization of
Bernauer~\etal~\cite{Bernauer:2013tpr}.  The calculations largely reconcile the
Rosenbluth polarization transfer discrepancy of the electromagnetic form factors
of the proton.  However, to achieve this agreement with the calculations, the
CLAS and OLYMPUS results must be shifted by approximately one and two times
their respective correlated uncertainties. Without the renormalization only the
Bernauer prediction remains in good agreement with the three experimental
results and the theoretical calculations are systematically higher than the
results.  However the $\eps$ and $Q^2$ dependence is generally followed by the
calculations.

The results of these experiments are by no means definitive. The majority of the
data are well below where the form factor discrepancy is significant
($Q^2>2$~GeV$^2$), so questions regarding the source of this discrepancy remain
largely unanswered.  There is a clear need for similar experiments at larger
$Q^2$, and perhaps more importantly, at $\eps<0.5$. Figure~\ref{fig:R2gBlunden}
shows that $R_{2\gamma}$ remains small, even at $Q^2=2.5$~GeV$^2$, for
$\eps>0.5$.  This, of course, poses significant experimental difficulties due to
the rapid drop in the elastic cross section at large lepton scattering angles. 
At the present time no new experiments have been approved for studies in the
high-$Q^2$ region, so the question may remain unanswered for several years. 
There have been discussions in the community to produce an $e^+$ beam at
Jefferson Lab~\cite{Voutier:2014kea}, but any such facility is uncertain and
many years in the future.

An upcoming MUSE experiment~\cite{Gilman:2013eiv} at Paul Scherrer Institute
(PSI) will address the problem of the proton radius \cite{Pohl:2010zza,
Pohl:2013yb} via precision measurements at small transferred momenta. MUSE will
provide a comparison of electron and positron scattering on the proton, as well
as positive and negative muons, directly constraining TPE for these processes at
very low $Q^2$.

Effects due to TPE have been sought in experiments on polarization observables.
Polarization measurements~\cite{Meziane:2010xc}, where the real part of TPE
could alter the angular dependence of the recoil proton polarization, have not
observed substantial deviations of the polarization ratio, but reported
deviations in the individual polarization components.  Single-spin asymmetries
(SSA) caused by normal (with respect to scattering plane) polarization provide a
probe of the imaginary part of TPE amplitude. The data on beam SSA
\cite{Abrahamyan:2012cg, Waidyawansa:2016znm} are in good agreement with
unitarity-based calculations~\cite{Afanasev:2004hp, Afanasev:2004pu,
Gorchtein:2005za, Gorchtein:2008dy} for the proton and light nuclei, but
disagree with the data on a high-$Z$ target $^{208}$Pb both in sign and
magnitude, possibly due to Coulomb distortion effects. The measurements of
single-spin target asymmetry~\cite{Zhang:2015kna} in quasi-elastic scattering on
a transversely polarized $^3$He target showed a TPE effect that agreed with GPD
predictions at high momentum transfer. The data both on target and beam SSA show
evidence of inelastic excitations of the intermediate hadronic state and provide
valuable input for theoretical constraints of TPE.

On the theoretical front, there has been significant progress in calculations of
TPE based on the use of dispersion relations~\cite{Borisyuk:2015xma,
Tomalak:2014sva, Blunden:2016}. The use of spin-\sfrac{1}{2} and
spin-\sfrac{3}{2} helicity amplitudes from electroproduction data throughout the
resonance region is a notable advance~\cite{Borisyuk:2015xma}. At forward angles
and low $Q^2$ the dispersive approach allows one to use total photonucleon cross
section data to constrain hadronic uncertainties~\cite{Gorchtein:2014hla,
Tomalak:2015aoa}.  Connecting the low to moderate $Q^2$ hadronic models with the
high $Q^2$ QCD-based models studied in Refs.~\cite{Afanasev:2005mp,
Borisyuk:2008db, Kivel:2009eg, Kivel:2012vs} remains an elusive goal.

Another area where progress might be made is regarding higher order radiative
corrections.  The large difference between exponentiated and non-exponentiated
radiative corrections that increase with decreasing $\eps$ suggests higher order
corrections may be warranted. In addition, a reanalysis of the existing form
factor and polarization data to uniformly apply and update the radiative
corrections might provide further insight into the TPE process and the role it
has in lepton-nucleon scattering.

\section*{Acknowledgements}

This material is based upon work supported by NSERC (Canada) and the
U.S.~Department of Energy, Office of Science, Office of Nuclear Physics under
contracts DE-AC05-06OR23177, DE-SC0013620, and DE-FG02-94ER40818. AA
acknowledges support of the Gus Weiss endowment at George Washington University.
PGB thanks Jefferson Lab for support during a sabbatical leave, where part of
this work was completed.

\section*{References}
\raggedright


\end{document}